\pdfoutput=1

\documentclass{article}
\usepackage[a4paper]{geometry}

\usepackage{graphicx}
\usepackage{subcaption}
\usepackage{color}
\usepackage{natbib}
\usepackage{bm}
\usepackage{amsmath,amssymb}
\usepackage{amsfonts,amstext}
\usepackage{tdhmacro}
\usepackage{a4}

\begin{document}

\newcommand{\kcp}{k}
\newcommand{\kcpi}{k_i}
\newcommand{\kbar}{\overline{k}}
\newcommand{\hstar}{\nsegs^{\star}}
\newcommand{\model}{\mathcal{M}}
\newcommand{\modelA}{\mathcal{M}_1}
\newcommand{\modelB}{\mathcal{M}_2}
\newcommand{\ups}{\upsilon}
\newcommand{\1}{1\mskip-5mu\relax\mathrm{l}}
\newcommand{\G}{\mathcal{G}}
\newcommand{\prob}{P}
\newcommand{\eps}{\varepsilon}
\newcommand{\Ndim}{M}
\newcommand{\nsegs}{h}
\newcommand{\nobs}{n}
\newcommand{\iter}{m}
\newcommand{\neww}{_{\text{new}}}
\newcommand{\old}{_{\text{old}}}
\newcommand{\AllGraph}{\mathbb{M}}
\newcommand{\data}{D}
\newcommand{\graph}{\mathcal{G}}
\newcommand{\allgraph}{\mathbb{M}}
\newcommand{\allmodels}{{\mathcal{M}\in\mathbb{M}}}
\newcommand{\newone}{_{0\text{new}}}
\newcommand{\newtwo}{_{1\text{new}}}
\newcommand{\oldone}{_{0\text{old}}}
\newcommand{\oldtwo}{_{1\text{old}}}
\newcommand{\parent}{\pi_\model}
\newcommand{\subnet}{\mathcal{N}}
\newcommand{\nrodata}{K}  % I --> K
\newcommand{\chpoint}{b}
\newcommand{\edge}{e}
\newcommand{\edgeIN}{e_{j}^h}
\newcommand{\nonedgeIN}{\overline{e_{ij}^h}}
\newcommand{\edgeProb}{\theta_{ij}}
\newcommand{\edgeHyper}{\alpha_{ij}}
\newcommand{\edgeHyperComp}{\overline{\alpha_{ij}}}
\newcommand{\CompEdgeSetIN}{\{e_{ij}^{\tilde{h}}\}_{\tilde{h}\neq h}}
\newcommand{\Nnets}{K-1}
\newcommand{\Nedges}{B^h_{ij}}
\newcommand{\Nnonedges}{\overline{B^h_{ij}}}
\newcommand{\Nsegs}{K}
\newcommand{\binomA}{a}
\newcommand{\binomB}{b}
\newcommand{\betaHypAa}{\alpha}
\newcommand{\betaHypAb}{\overline{\alpha}}
\newcommand{\betaHypBa}{\gamma}
\newcommand{\betaHypBb}{\overline{\gamma}}
\newcommand{\Naa}{N_1^1}
\newcommand{\Nab}{N_1^0}
\newcommand{\Nba}{N_0^1}
\newcommand{\Nbb}{N_0^0}
\newcommand{\edgeHIGH}{e\hspace{-2mm}\uparrow}
\newcommand{\edgeLOW}{e\hspace{-2mm}\downarrow}
\newcommand{\Ndata}{T}
\newcommand{\seg}{h}
\newcommand{\equationsize}{\scriptsize}
\newcommand{\parents}{\pi_i}
\newcommand{\parentsA}{\pi^{(1)}_i}
\newcommand{\parentsB}{\pi^{(2)}_i}
\newcommand{\vecparents}{\pi}
\newcommand{\Npart}{Z}
\newcommand{\nodeOther}{\tilde{n}}
\newcommand{\autoc}{A}
\newcommand{\Nhood}{\mathcal{N}}
\newcommand{\state}{k}
\newcommand{\covMatGP}{{\bf C_{\phi_k}}}
\newcommand{\covMatGPel}{C_{\phi_k}}
\newcommand{\covMatGPinv}{{\bf C}_{\phi_k}^{-1}}
\newcommand{\icovMatGP}{{\bf 'C_{\phi_k}}}
\newcommand{\covMatGPi}{{\bf C'_{\phi_k}}}
\newcommand{\covMatGPii}{{\bf C''_{\phi_k}}}
\newcommand{\matDer}{{\bf A}}
\newcommand{\matDerBar}{{\bf \tilde{A}}}
\newcommand{\hatX}{{\bf \hat{X}}}
\newcommand{\hatx}{{\bf \hat{x}}}
\newcommand{\dotX}{{\bf \dot{X}}}
\newcommand{\tildeX}{{\bf \tilde{X}}}
\newcommand{\bfX}{{\bf X}}
\newcommand{\Jacobian}{{\bf J}}
\newcommand{\covMatGPint}{{\bf C^*_{\phi_k}}}
\newcommand{\dotxsc}{\dot{x}}
\newcommand{\dotxsctilde}{\dot{\tilde{x}}}
\newcommand{\dotx}{{\bf \dot{x}}}
\newcommand{\dotxtilde}{{\bf \dot{\tilde{x}}}}
\newcommand{\kernel}{K}
\newcommand{\kernelInt}{K^*}
\newcommand{\kernelpar}{\boldsymbol\phi}
\newcommand{\ODEpar}{\boldsymbol\theta}
\newcommand{\mBar}{{\bf\tilde{m}}}
\newcommand{\hypIGa}{a}
\newcommand{\hypIGb}{b}
\newcommand{\invT}{\tau}
\newcommand{\bfsigmaTwo}{\boldsymbol\sigma^2}
\newcommand{\tif}{f}
\newcommand{\tiparall}{\boldsymbol\Psi}
\newcommand{\potentialFun}{\Psi}
\newcommand{\normConst}{C}
\newcommand{\ZC}{Z_C}
\newcommand{\expect}{\mathbb{E}}
\newcommand{\var}{\mathbb{V}}
\newcommand{\eqname}{{Eq.}}    % singular
\newcommand{\eqnames}{{Eqns.}} % plural
\newcommand{\Vmat}{\boldsymbol\theta}
\newcommand{\Dmat}{{\bf{D}}}
\newcommand{\DmatA}{{\bf{D}}^{(1)}}
\newcommand{\DmatB}{{\bf{D}}^{(2)}}
\newcommand{\byi}{{\bf y}}
\newcommand{\sigi}{\sigma^2}
\newcommand{\deli}{\delta^2}
\newcommand{\Kmat}{{\bf{K}}_i}
\newcommand{\SigmaTilde}{\tilde{\boldsymbol\Sigma}_i}
\newcommand{\bftildetheta}{\mbox{{\boldmath $\tilde{\theta}$}}}
\newcommand{\shortciteA}{\citet}   %   \citet{key} ==>>                Jones et al. (1990)
\newcommand{\shortcite}{\citep}    %   \citep{key} ==>>                (Jones et al., 1990)
\newcommand{\delay}{\Delta}
\newcommand{\Ndelays}{k}
\newcommand{\spreadFactor}{sf}
\newcommand{\figurenames}{Figs.}
\newcommand{\vareps}{\eps}
\newcommand{\Sect}{Section}
\newcommand{\Fig}{Figure}
\newcommand{\EE}{E}
\newcommand{\EET}{\tilde{E}}
\newcommand{\MAD}{\mathbb{A}}
\newcommand{\VAR}{\mathbb{V}}

\title{Targeting Bayes factors with direct-path non-equilibrium thermodynamic integration\footnote{Accepted for publication in Computational Statistics (doi:10.1007/s00180-017-0721-7). }}

\author{Marco Grzegorczyk$^1$ \and Andrej Aderhold$^2$ \and Dirk Husmeier$^2$}
\date{%
    $^1$Johann Bernoulli Institute (JBI), Groningen University, The Netherlands\\%
    $^2$School of Mathematics and Statistics, Glasgow University, UK \\[2ex]%
    \today
}

\maketitle

\begin{abstract}
Thermodynamic integration (TI) for computing marginal likelihoods is based on an inverse annealing path from the prior to the posterior distribution. 
In many cases, the resulting estimator suffers from high variability, which particularly stems from the prior regime.  
When comparing complex models with differences in a comparatively small number of parameters, 
intrinsic errors from sampling fluctuations may outweigh the differences in the log marginal likelihood estimates. In the present article, we propose a thermodynamic integration scheme that directly targets the log Bayes factor. The method is based on a modified annealing path between the posterior distributions of the two models compared, which systematically avoids the high variance prior regime. We combine this scheme with the concept of non-equilibrium TI to minimise discretisation errors from numerical integration.  Results obtained on Bayesian regression models applied to standard benchmark data, and a complex hierarchical model applied to biopathway inference,  demonstrate a significant reduction in estimator variance over state-of-the-art TI methods.
%\keywords{Bayesian inference, marginal likelihood, temperature ladder, variance reduction, Jarzynski's theorem, benchmark studies, biopathway}
\end{abstract}

\section{Introduction}
\label{sec:introduction}
A central quantity in Bayesian statistics is the marginal likelihood 
\begin{equation}
\label{eq:MargLhood}
p(\data|\model) 
=
\int 
p(\data,\bftheta|\model)
 d\bftheta 
=
\int 
p(\data|\bftheta,\model)
p(\bftheta|\model)
 d\bftheta 
\end{equation}
where  $\data$ are the data, and $\model$ represents a given statistical model with parameter vector $\bftheta$.
The difficulty in practically computing the marginal likelihood is exemplified by
considering the Monte Carlo sum
\begin{equation}
\label{eq:MCsum}
X
\; = \;
\frac{1}{M}
\sum_{i=1}^M
p(\data|\bftheta_i,\model)
\end{equation}
where $\{\theta_i\}$ is an iid sample from 
$p(\bftheta|\model)$.  Under fairly general regularity conditions the estimator $X$ converges almost surely to $p(\data|\model)$,
by the strong law of large numbers, and is asymptotically efficient, with asymptotic variance 
$C/\sqrt{N}$ (where $N$ is the size of $\data$), by the central limit theorem.
However, even for modestly complex systems, the constant in the numerator, $C$, can reach exorbitant magnitudes, rendering the scheme
not viable for practical applications.
The practical shortcomings of a variety of alternative numerical methods, 
like the harmonic mean estimator \shortcite{HarmonicMean1994}, bridge sampling
\shortcite{bridgeSampling}, or Chib's method \shortcite{Chib2001marginal},
have been discussed in the statistics and machine learning literature (e.g. \shortciteA{MurphyBook}).
The most widely used and robust method appears to be 
thermodynamic integration (TI). This method was originally proposed by
\shortciteA{Kirkwood} and further developed in statistical physics
for the mathematically equivalent problem of computing free energies; see e.g.
\shortciteA{Schlitter1991} and \shortciteA{HusmeierSchlitter1992}. 
 \shortciteA{bridgeSampling}  adapted TI to the computation
of marginal likelihoods,  \shortciteA{Lartillot2006} demonstrated the application of TI to complex systems, 
and  \shortciteA{Friel} and \shortciteA{CalderheadGiroloamiTI2009} 
popularised TI more widely  in the statistics community
by demonstrating a computationally powerful combination with
parallel tempering \shortcite{parallelTempering}.

Thermodynamic integration is based on an integral of the expected log likelihood
along an inverse annealing path from the prior to the posterior distribution.
The resulting estimator typically suffers from high variability, which particularly stems from the parameter prior regime. 
When comparing complex models with differences in a comparatively small number of parameters, 
these intrinsic errors from sampling fluctuations may 
outweigh the differences in the log marginal likelihood estimates. 
The objective of the present study is to explore the scope for variance reduction by
directly targeting the log Bayes factor via a modified transition path between the two models
such that the high-variance prior domain is avoided.
This idea is not new. In statistical physics it is well known \shortcite{Schlitter1991,HusmeierSchlitter1992} that applying TI to the 
computation of a reaction free energy, which is mathematically equivalent to the log Bayes factor, is computationally more efficient than the separate computation of the standard free energies for the two reaction states involved (educt versus product states); the latter is mathematically equivalent to the difference of the log marginal likelihoods of two statistical models to be compared.
Also in the statistics literature, the direct targeting of the log Bayes factor has been discussed before.
For instance,  path sampling \shortcite{bridgeSampling} and annealed importance sampling \shortcite{Neal_AIS} have been conceived in a way to allow the direct computation of the ratio of two partition functions, $Z_1$ and $Z_2$, associated with two models $\model_1$ and $\model_2$. However, in the work of
\shortciteA{Neal_AIS}, $Z_1$ is set to the normalisation factor of the prior distribution, and the method thus reduces to the computation of the log marginal likelihood\footnote{
Note that in the regression example presented by \shortciteA{Neal_AIS}, where the objective is model selection between a Gaussian and a Cauchy distribution for the noise, the log marginal likelihoods are computed separately with annealed importance sampling, and then combined to produce the log Bayes factor. Unlike the scheme proposed in the present article, the log Bayes factor is \emph{not} targeted directly. 
}. 
\shortciteA{bridgeSampling} do consider a direct comparison between two alternative models: a homoscedastic versus a heteroscedastic linear regression model. Rather than computing the Bayes factor, the authors apply their path sampling approach to infer the posterior distribution of the entire spectrum of intermediate models. While this is a more ambitious approach than model selection with Bayes factors, it will be computationally onerous beyond the one-dimensional regime considered in their example.  

To the best of our knowledge, the present article presents the first systematic study of the variance reduction that can be achieved with a thermodynamic integration path that directly targets the log Bayes factor by transiting between the posterior distributions of the two models involved. The mathematical exposition and implementation of this scheme is combined with a  comprehensive comparative performance assessment based on a set of standard benchmark data to quantify the improvement in variance reduction, accuracy  and computationally efficiency that can be achieved over state-of-the-art established TI methods, in particular 
the recent improvement proposed by  \shortciteA{frielTIcorrection}.

This article is organised as follows. 
In Section~\ref{sec:rationale} we provide a brief rationale for 
targeting Bayes factors directly rather than indirectly via  the marginal likelihood.
Section~\ref{sec:TI} reviews standard 
thermodynamic integration.
In Section~\ref{sec:NETI} we discuss a modified numerical integration and sampling scheme 
from statistical physics, termed non-equilibrium TI (NETI), to reduce numerical discretisation errors. 
Section~\ref{sec:TInew} describes NETI-DIFF, the proposed new TI scheme along an alternative 
integration path between two posterior distributions.
Sections~\ref{sec:MHS}-\ref{sec:Gibbs} describe practical numerical implementations based on 
Metropolis-Hastings and Gibbs sampling,
and Section~\ref{sec:Tladder}
proposes a new improved inverse temperature ladder.
Section~\ref{sec:data} provides an overview of a set of benchmark problems on which 
we have evaluated the methods, and Section~\ref{sec:results} presents our empirical findings.
We conclude this article in Section~\ref{sec:discussion} with a discussion, a comparison with the
controlled thermodynamic integral of \shortciteA{Oates2015TI},
and an outlook
on future work.

% ========================================
\section{Rationale}
\label{sec:rationale}
% ========================================
Consider two alternative models, $\model_1$ and $\model_2$, and define 
$\EE_i= -\log p(\data|\bftheta,\model_i)$, the negative log likelihood of model $i$.
Further, define the log likelihood ratio $\Delta \EE= \EE_2-\EE_1$, the negative unnormalised log posterior 
$\EET_i= \EE_i+\log p(\bftheta|\model_i)$, the negative log posterior ratio
$\Delta \EET= \EET_2-\EET_1$, and let
$\langle\ldots\rangle_{i}$ denote the posterior average with respect
to the posterior distribution
\begin{math}
p(\bftheta|\data,\model_i)
\end{math}. 
We can then adapt Jarzynski's theorem from statistical physics 
\shortcite{Jarzynski}
to show that 
\begin{equation}
\label{eq:Jarzynski}
p(\data|\model_i) =  \Big\langle \exp(\EE_i[\bftheta])\Big\rangle_i^{-1}, \quad\quad
\frac{p(\data|\model_2)}{p(\data|\model_1)}
 =  \Big\langle \exp(-\Delta\EET[\bftheta])\Big\rangle_1
\end{equation}
A proof is given in the Appendix. 
In real applications with non-trivial models, the negative log likelihood is typically in the order of a two to three digit figure, which when put into the argument of the exponential function will lead to an astronomically large number. An estimator aiming to approximate $p(\data|\model_i)= \langle \exp(\EE_i[\bftheta])\rangle_i^{-1} $ from a limited sample drawn from the posterior distribution will inevitably suffer from substantial variation. For nested models or models with sufficient parameter overlap, on the other hand, $\Delta\EET(\bftheta)$ will typically be small, 
\begin{math}
|\Delta\EET(\bftheta)|
\ll 
\min\{|\EE_1(\bftheta)|,|\EE_2(\bftheta)|\}
\end{math}.
We can therefore reduce the intrinsic estimation uncertainty considerably by computing the Bayes factor 
directly rather than indirectly via two separate marginal likelihood estimations.

\section{Methodology}
\label{sec:method}
% ========================================
\subsection{Thermodynamic integration for marginal likelihoods}
\label{sec:TI}
%\vspace{-3mm}
% ========================================
Thermodynamic integration is based on an inverse annealing path from 
the prior to the posterior distribution, and computing the expectation 
of the log likelihood with respect to the following annealed posterior distributions at inverse temperatures $\invT\in[0,1]$:
\begin{equation}
\label{eq:TI1}
p_{\invT}(\bftheta|\data,\model)
\; =\;
\frac{1}{Z(\data|\invT,\model)}
p(\data|\bftheta,\model)^{\invT}
p(\bftheta|\model), \quad\quad Z(\data|\invT,\model)
\; = \;
\int p(\data|\bftheta,\model)^{\invT}
p(\bftheta|\model) d\bftheta
\end{equation}
Taking the derivative of $\log Z(\data|\invT,\model)$ gives:
\begin{eqnarray}
\frac{d}{d\invT} \log Z(\data|\invT,\model) & = & \frac{1}{Z(\data|\invT,\model)} \frac{d}{d\invT}  Z(\data|\invT,\model)
\nonumber  
\\
& = & 
\frac{1}{Z(\data|\invT,\model)}
\int 
\frac{d}{d\invT} 
p(\data|\bftheta,\model)^{\invT}
p(\bftheta|\model)
d\bftheta  
\nonumber  
\\
& = & 
\int 
\log p(\data|\bftheta,\model)
\frac{
p(\data|\bftheta,\model)^{\invT}
p(\bftheta|\model)}
{Z(\data|\invT,\model)}
d\bftheta  
\nonumber  
\\
& = &
\int 
p_{\invT}(\bftheta|\data,\model)
\log p(\data|\bftheta,\model)
d\bftheta  
\nonumber\\
& = & 
\mathbb{E}_{\invT}\Big[
\log p(\data|\bftheta,\model)
\Big]
\label{eq:exPowerPost}
\end{eqnarray}
From \eqname~(\ref{eq:exPowerPost}) we get:
\begin{eqnarray}
\log p(\data|\model) & = &  
\log Z(\data|\invT=1) - \log Z(\data|\invT=0)
\nonumber
\\
& = &
 \int_0^1  \frac{d}{d \invT}\log Z(\data|\invT) d \invT  =  
 \int_0^1  \mathbb{E}_{\invT}\Big[\log p(\data|\bftheta,\model)\Big] d\invT
 \label{eq:pipi}
\end{eqnarray}
This one-dimensional integral can be solved numerically, e.g. with the trapezoid rule: 
{

\begin{equation}
\label{eq:TI_trapezoid_standard} 
\log(p(\data|\model)) \approx
\sum_{k=2}^K \frac{\invT_k-\invT_{k-1}}{2}  
\left\{
\expect_{\invT_k}\big[\log p(\data|\bftheta,\model)\big] + \expect_{\invT_{k-1}}\big[\log p(\data|\bftheta,\model) \big]
\right\}  
\end{equation}
}
Some care has to be taken with respect to the choice of discretisation points $\invT_k, k=\{0,1,2,\ldots,K\}$, as the major contributions to the integral usually come from a small region around $\invT \rightarrow 0$. This motivates the form
\begin{equation}
\label{eq:powerLaw}
\invT_k \; = \; \left(\frac{k-1}{K-1}\right)^{\alpha}; \quad k \in \{1,2,\ldots,K\}
\end{equation}
for $\alpha > 1$. Theoretical results for the optimal choice of $\alpha$ can be found in 
\shortciteA{Schlitter1991}, but require knowledge that is usually not available in practice
(like the functional dependence of $\mathbb{E}_{\invT}[\log p(\data|\bftheta,\model)]$ on $\invT$). In practice, 
$\alpha=5$ is widely used, as e.g. in \shortciteA{frielTIcorrection}, and we have used this value in the present study.
A potentially numerically more stable alternative 
was proposed in \cite{frielTIcorrection}. The authors show that:
 \begin{equation}
	\label{eq_pre_approx}
	\frac{d}{d\invT} \{ \expect_{\invT}[\log(p(\data|\bftheta,\model))] \}_{\invT} 
	\;=\;
	\var_{\invT}(\log(p(\data|\bftheta,\model))	
\end{equation} 
where $\var_{\invT}(.)$ is the variance w.r.t. the power posterior in \eqname~(\ref{eq:TI1}). The second derivative of $\expect_{\invT}[\log(p(\data|\bftheta,\model))]$ at a point $\invT\in[\invT_{k-1},\invT_{k}]$ can then be approximated by the difference quotient of the first derivative of $\expect_{\invT}[\log p(\data|\bftheta,\model)]$
\eqname~(\ref{eq_pre_approx}):
\begin{equation}
	\label{eq_approx} 	
	\frac{d^2}{dt^2} \{ \expect_{t}[\log(p(\data|\bftheta,\model))] \}_{t=\invT} 
	\nonumber 
	 \approx 
	\frac{\var_{\invT_{k}}(\log(p(\data|\bftheta,\model))-\var_{\invT_{k-1}}(\log(p(\data|\bftheta,\model))}{\invT_{k}-\invT_{k-1}}
\end{equation}
\shortciteA{frielTIcorrection} then employ the corrected trapezoid rule\footnote{$\int_a^b f(x) dx = {(b-a)} \frac{f(b)+f(a)}{2} - \frac{(b-a)^3}{12} f''(c)$ for some $c\in[a,b]$.} to compute each sub-integral
\\
$\int_{\invT_{k-1}}^{\invT_{k}}\expect_{\invT}[\log(p(\data|\bftheta,\model))] d\invT$. This yields:
{
%\footnotesize
\begin{eqnarray}
%\footnotesize
\log(p(\data|\model)) & = &  \int_{0}^{1} \expect_{\invT} \big[ \log p(\data|\bftheta,\model) \big] d \invT 
= \sum_{k=2}^K \int_{\invT_{k-1}}^{\invT_k} \expect_{\invT} \big[ \log p(\data|\bftheta,\model) \big] d \invT
\nonumber\\
&\approx& 
\sum_{k=2}^K \frac{\invT_k-\invT_{k-1}}{2}  
\Bigg\{
\expect_{\invT_k}\big[\log p(\data|\bftheta,\model)\big] 
 + \expect_{\invT_{k-1}}\big[\log p(\data|\bftheta,\model) \big]
\Bigg\}  
\nonumber\\
& & 
-\sum_{k=2}^K \frac{(\invT_k-\invT_{k-1})^2}{12}  
\Big\{ \var_{\invT_k}\big[\log p(\data|\bftheta,\model)\big] 
- \var_{\invT_{k-1}}\big[\log p(\data|\bftheta,\model) \big]
\Big\}  
\label{eq:TI_trapezoid_correction} 
\end{eqnarray}
}

% ========================================
\subsection{Nonequilibrium thermodynamic integration}
\label{sec:NETI}
%\vspace{-3mm}
% ========================================
The computation of the expectation values $\expect_{\invT_k}\big[\log p(\data|\bftheta,\model)\big] $ is expensive and limits the number of discretisation points $K$ that can be practically applied. An alternative scheme we use in the present work is to approximate
\begin{eqnarray}
\log p(\data|\model)  & = &
 \int_0^1  \mathbb{E}_{\invT}\Big[\log p(\data|\bftheta,\model)\Big] d\invT
  \approx 
 \int_0^1  \log p(\data|\bftheta^{(\invT)},\model) d\invT
 \nonumber\\
 & \approx &
 \sum_{k=2}^K \frac{\invT_k-\invT_{k-1}}{2}  
\Big\{
\log p(\data|\bftheta^{(\invT_k)},\model)
+ \log p(\data|\bftheta^{(\invT_{k-1})},\model) 
\Big\}  
\end{eqnarray}
where $\theta^{(\invT)}$ is a single draw from the power posterior defined in \eqname~(\ref{eq:TI1}),
and $0=\invT_1<\invT_2<\ldots<\invT_K=1$.  The computational resources gained are used to choose $K$ 
orders of magnitude larger than in equilibrium TI,\footnote{Note that $K$ can be set equal to the total number of MCMC iterations $N_{iter}$, which otherwise would have to be subdivided onto $K$ discretisation points.} with the implication that $(\invT_k-\invT_{k-1}) \rightarrow 0$ and discretisation errors in numerical integration are avoided.
 This scheme was originally proposed in statistical physics \shortcite{HusmeierSchlitter1992} under the name \emph{non-equilibrium thermodynamic integration} (NETI), and is conceptionally similar to annealed importance sampling \shortcite{Neal_AIS}.
 The underlying rationale is as follows: rather than use the computational resources for the computation of the expectation value at a limited number of discretisation points - and incur discretisation errors  - spread the computational resources over the whole "temperature" range and use as fine a discretisation as possible. %We will provide empirical evidence that this scheme compares favourably with a computation of the marginal likelihood based on equations~(\ref{eq:pipi}) and (\ref{eq:TI_trapezoid_correction}).
%For further details and an empirical evaluation, see \shortciteA{HusmeierSchlitter1992}.
This avoids the problem that had to be addressed in \cite{frielTIcorrection}: how to select the inverse temperatures and minimise the numerical integration error in standard TI.
The price to pay is a relaxation error as a consequence of the non-equilibrium nature of the method, as discussed by
\shortciteA{HusmeierSchlitter1992}. The authors proposed a scheme for correcting this relaxation error, by running simulations over different simulation lengths $N_{iter}$, regressing the estimates against an approximate upper bound on the relaxation error $\mathcal{R}$, and then  
extrapolating for $\mathcal{R}\rightarrow 0$. 
In preliminary investigations omitted from the present article, we found that a single simulation with an increased value of $N_{iter}$ matching the total computational costs of the extrapolation scheme achieved similar results, and we used this conceptionally simpler approach in all our studies\footnote{The extrapolation scheme proposed by \shortciteA{HusmeierSchlitter1992} can reduce actual computation  time by parallelisation, but this was not an issue for the simulations carried out in the present work.}.

% ========================================
\subsection{Novel thermodynamic integration for Bayes factors}
\label{sec:TInew}
% ========================================
When comparing two models, we are typically interested in the Bayes factor $p(\data|\model_2)/p(\data|\model_1)$. The standard approach is to apply thermodynamic integration to both models $\model_1$ and $\model_2$ separately, by independently inversely annealing the prior distributions to the respective posterior distributions. This approach ignores the fact that both models usually have many aspects in common and share certain parameters. This applies particularly to nested models, where all the parameters of the less complex model are also included in the more complex model. One would expect to reduce the estimation uncertainty
by following a direct transition path from the posterior distribution of the less complex model to that of the more complex model, rather than transiting through the uninformative prior distribution twice.
Consider two models $\model_1$ and $\model_2$ with joint parameter vector $\bftheta$ and a joint parameter 
prior $p(\bftheta|\model_1,\model_2)$ defined such that it reduces to the parameter priors for the separate models 
by marginalisation:
\begin{equation}
p(\bftheta|\model_1) \; = \; \int_{\model_2/\model_1} p(\bftheta|\model_1,\model_2) d\bftheta
\label{eq:A1}, \quad\quad
p(\bftheta|\model_2) \; = \; \int_{\model_1/\model_2} p(\bftheta|\model_1,\model_2) d\bftheta
\end{equation}
where $\model_2/\model_1$ is the subset of parameters contained in model $\model_2$, but not in
model $\model_1$, and $\model_1/\model_2$ is the subset of parameters contained in model $\model_1$, but not in
model $\model_2$. A mathematically more accurate notation would split $\bftheta$ into three subsets, 
$\bftheta=\{\bftheta_1,\bftheta_2,\bftheta_{12}\}$ such that $\bftheta_1 \in \model_1/\model_2$, 
$\bftheta_2 \in \model_2/\model_1$ and $\bftheta_{12} \in \model_1\cap\model_2$. 
\eqname~(\ref{eq:A1}) implies that $p(\bftheta|\model_1)=p(\bftheta_1,\bftheta_{12}|\model_1)$ and $p(\bftheta|\model_2)=p(\bftheta_2,\bftheta_{12}|\model_2)$. For that reason we can use a mathematically 
redundant but less opaque notation that does not make the partition $\bftheta=\{\bftheta_1,\bftheta_2,\bftheta_{12}\}$ explicit.
Define the tempered posterior distribution
\begin{equation}
\label{eq:tempPost}
p_{\invT}(\bftheta|\data,\model_1,\model_2)
 =
\frac{p(\data|\bftheta,\model_2)^{\invT}
p(\data|\bftheta,\model_1)^{1-\invT}
p(\bftheta|\model_1,\model_2)}
{Z(\data|\invT,\model_1,\model_2)}
\end{equation}
where
\begin{equation}
\label{eq:TIZ2}
Z(\data|\invT,\model_1,\model_2)
 = 
\int
\left(\frac{p(\data|\bftheta,\model_2)}{p(\data|\bftheta,\model_1)}
\right)^{\invT}
p(\data|\bftheta,\model_1)
p(\bftheta|\model_1,\model_2) d\bftheta
\end{equation}
From \eqname~(\ref{eq:A1}) we get:
\begin{equation}
p(\data|\model_1) =  Z(\data|\invT=0,\model_1,\model_2), \quad\quad
\label{eq:B1}
p(\data|\model_2)  =  Z(\data|\invT=1,\model_1,\model_2)
\end{equation}
Taking the derivative of the partition function in \eqname~(\ref{eq:TIZ2}) gives:
\begin{eqnarray}
\frac{d}{d\invT} \log Z(\data|\invT,\model_1,\model_2) & = & 
\frac{1}{Z(\data|\invT,\model_1,\model_2)} \frac{d}{d\invT}  Z(\data|\invT,\model_1,\model_2)
\nonumber  \\
& = & 
\frac{1}{Z(\data|\invT,\model_1,\model_2)}
\int 
\frac{d}{d\invT} 
\left(\frac{p(\data|\bftheta,\model_2)}{p(\data|\bftheta,\model_1)}
\right)^{\invT}
p(\data|\bftheta,\model_1)
p(\bftheta|\model_1,\model_2)
d\bftheta  
\nonumber  \\
& = & 
\int 
\log 
\left(\frac{p(\data|\bftheta,\model_2)}{p(\data|\bftheta,\model_1)}
\right)
\frac{p(\data|\bftheta,\model_2)^{\invT}
p(\data|\bftheta,\model_1)^{1-\invT}
p(\bftheta|\model_1,\model_2)}
{Z(\data|\invT,\model_1,\model_2)}
d\bftheta  
\nonumber  \\
& = & 
\int 
p_{\invT}(\bftheta|\data,\model_1,\model_2)
\log 
\left(\frac{p(\data|\bftheta,\model_2)}{p(\data|\bftheta,\model_1)}
\right)
d\bftheta  
\nonumber\\
& = & 
\mathbb{E}_{\invT}\left[
\log \left(\frac{p(\data|\bftheta,\model_2)}{p(\data|\bftheta,\model_1)}\right)
\right]
\label{eq:CCC}
\end{eqnarray}
Combining \eqnames~(\ref{eq:B1}-\ref{eq:CCC}) gives the following thermodynamic integral for the 
log Bayes factor:
\begin{eqnarray}
\log \left(\frac{p(\data|\model_2)}{p(\data|\model_1)}\right) & = & 
\log Z(\data|\invT=1,\model_1,\model_2) - \log Z(\data|\invT=0,\model_1,\model_2)
\nonumber \\
& = & 
\int_0^1 \left[\frac{d}{d\invT} \log Z(\data|\invT,\model_1,\model_2)\right] d \invT
\nonumber
\\
& = & 
\int_0^1
\mathbb{E}_{\invT}\left[
\log \left(\frac{p(\data|\bftheta,\model_2)}{p(\data|\bftheta,\model_1)}\right)
\right] d \invT
\end{eqnarray}
Again, we follow the idea of non-equilibrium thermodynamic integration 
and make the approximation
{
%\footnotesize
\begin{eqnarray}
\log \left(\frac{p(\data|\model_2)}{p(\data|\model_1)}\right)
& \approx &
\int_0^1
\left[
\log \left(\frac{p(\data|\bftheta^{(\invT)},\model_2)}{p(\data|\bftheta^{(\invT)},\model_1)}\right)
\right] d \invT
\nonumber \\
& \approx & 
\sum_{k=2}^K \frac{\invT_k-\invT_{k-1}}{2}  
\Bigg\{
\log \left(\frac{p(\data|\bftheta^{(\invT_k)},\model_2)}{p(\data|\bftheta^{(\invT_k)},\model_1)}\right)
\nonumber
\\
& & + \log \left(\frac{p(\data|\bftheta^{(\invT_{k-1})},\model_2)}{p(\data|\bftheta^{(\invT_{k-1})},\model_1)}\right)
\Bigg\}  
\label{eq:ioio}
\end{eqnarray}
} % end footnote size
where $\bftheta^{(\invT)}$ is a single draw from the tempered posterior distribution defined in 
\eqname~(\ref{eq:tempPost}), $0=\invT_1<\invT_2<\ldots<\invT_K=1$, $K\gg1$, and $(\invT_{k}-\invT_{k-1})\ll1$.

In comparison with statistical physics, the proposed scheme corresponds to the direct computation of a free energy difference \shortcite{Schlitter1991,HusmeierSchlitter1992}, which is more efficient, in terms of reduced estimation variance for given computational costs, than computing the difference of two separately computed standard free energies. The analogy from classical statistics is model comparison via a paired test, which is known to have higher power than an unpaired test. 

In what follows, we refer to the estimator defined by  \eqname~(\ref{eq:ioio}) as NETI-DIFF. 
We describe how to compute the variance of this estimator in the Appendix~\ref{sec:Var}.

% ========================================
\subsection{Metropolis-Hastings scheme}
\label{sec:MHS}
%\vspace{-3mm}
% ========================================
The implementation of a Metropolis-Hastings scheme to target the distribution in (\ref{eq:tempPost}) is straightforward. Given the current parameters $\bftheta$, sample new parameters
$\bftildetheta$ from a proposal distribution $q(\bftildetheta|\bftheta)$, and accept these new parameters with the following acceptance probability:
\begin{equation}
a(\bftildetheta|\bftheta)= 
\min\left\{
\frac{p(\data|\bftildetheta,\model_2)^{\invT}
p(\data|\bftildetheta,\model_1)^{1-\invT}
p(\bftildetheta|\model_1,\model_2)q(\bftheta|\bftildetheta)}
{p(\data|\bftheta,\model_2)^{\invT}
p(\data|\bftheta,\model_1)^{1-\invT}
p(\bftheta|\model_1,\model_2)q(\bftildetheta|\bftheta)}
,1
\right\}
\end{equation}
Otherwise, set $\bftildetheta=\bftheta$, and follow this scheme iteratively.

% ========================================
\subsection{Gibbs sampling for linear models}
\label{sec:Gibbs}
% ========================================
Consider a standard  linear model with parameter vector $\Vmat$, design matrix $\Dmat$,
and prior distribution 
\begin{equation}
\label{reg_prior}
p(\Vmat|\delta^2,\sigma^2) = N(\boldsymbol\mu_0,\sigi\deli{\bf I})
\end{equation}
The data, 
%$\data=\{\by_1,\ldots,\by_{\Ni}\}$, 
$\data=\{y_1,\ldots,y_{\Ndata}\}$ 
or 
$\by= (y_1,\ldots,y_{\Ndata})^{\transp}$,
are assumed to be obtained under the assumption of independent and identically distributed normal noise, with
%model-specific 
variance $\sigi$:
\begin{equation}
p(\byi|\Vmat,\sigi)= N(\Dmat\Vmat,\sigi{\bf I})
\end{equation}
We want to compare two competing models $\model_1$ and $\model_2$,  represented by two alternative design matrices $\DmatA$ and $\DmatB$:
\begin{equation}
p(\byi|\Vmat,\sigi,\model_1)= N(\DmatA\Vmat,\sigi{\bf I}), \quad\quad
p(\byi|\Vmat,\sigi,\model_2)= N(\DmatB\Vmat,\sigi{\bf I})
\end{equation}
For notational compactness we choose a representation that leaves the dimension of $\Vmat$ invariant with respect to changing model dimensions by padding obsolete entries in the design matrix with zeros.
For instance, to compare the models 
$\model_1: y= \theta_1 x_1 + \theta_2 x_2$, and $\model_2: y= \theta_1 x_1 + \theta_3 x_3 + \theta_4 x_4$
based on a data set of $n$ observations 
$ \{y_t,x_{1,t},x_{2,t},x_{3,t},x_{4,t}\}$, $t=1,\ldots,n$,
we get the following design matrices:
{
\begin{equation*}
\DmatA 
=
\left(
\begin{array}{cccc}
x_{1,1} &x_{2,1} & 0 & 0  \\
x_{1,2} &x_{2,2} & 0 & 0 \\
\vdots & \vdots & \vdots & \vdots \\
x_{1,n} &x_{2,n} & 0 & 0 \\
\end{array}
\right),
\;
\DmatB 
=
\left(
\begin{array}{cccc}
x_{1,1} &0  & x_{3,1} & x_{4,1} \\
x_{1,2} &0  & x_{3,2} & x_{4,2} \\
\vdots & \vdots & \vdots & \vdots \\
x_{1,n} &0  & x_{3,n} & x_{4,n}  \\
\end{array}
\right)
\end{equation*}
From (\ref{eq:tempPost}) we get
{
%\footnotesize
\begin{eqnarray}
p_{\invT}(\bftheta|\data,\model_1,\model_2) & \propto &
p(\data|\bftheta,\model_2)^{\invT} p(\data|\bftheta,\model_1)^{1-\invT} p(\bftheta|\model_1,\model_2)
\nonumber
\\
& \propto & 
N(\DmatA\Vmat,\sigi{\bf I})^{1-\invT}
N(\DmatB\Vmat,\sigi{\bf I})^{\invT}
N(\bfmu_0,\sigi\delta^2{\bf I})
\nonumber
\\
& \propto & 
\exp\left(
\frac{-(1-\invT)}{2\sigma^2}
\left[
\DmatA\Vmat-\byi
\right]^{\transp}
\left[
\DmatA\Vmat-\byi
\right]
\right)
\nonumber
\\
& &
\exp\left(
\frac{-\invT}{2\sigma^2}
\left[
\DmatB\Vmat-\byi
\right]^{\transp}
\left[
\DmatB\Vmat-\byi
\right]
\right)
\nonumber
\\  
& & 
\exp\left(
\frac{-1}{2\sigma^2\delta^2}
[\Vmat-\boldsymbol\mu_0]^{\transp}
[\Vmat-\boldsymbol\mu_0]
\right)
%\nonumber
\\
& = & 
\exp\left(
\frac{-1}{2\sigma^2}
\Vmat^{\transp}
\left[
\invT\{\DmatB\}^{\transp}\DmatB
+
(1-\invT)\{\DmatA\}^{\transp}\DmatA
+
\delta^{-2}{\bf I}\right]
\Vmat
\right)
\nonumber
\\ 
& & 
\exp\left(
\frac{1}{\sigma^2}
\Vmat^{\transp}
\left(\left[
\invT\{\DmatB\}^{\transp} 
+
(1-\invT)\{\DmatA\}^{\transp} 
\right]
\by
+
\delta^{-2}\boldsymbol\mu_{0}\right)
\right) C(\by)
%\nonumber
\end{eqnarray}
}
where the factor $C(\by)$ does not depend on $\Vmat$. Comparing this with the identity
\begin{equation}
N(\Vmat|\boldsymbol\mu,\sigma^2{\bf H}^{-1})
\nonumber
\propto 
\exp\left(
\frac{-1}{2\sigma^2}
[\Vmat-\boldsymbol\mu]^{\transp}{\bf H}
[\Vmat-\boldsymbol\mu]
\right)
 = 
\exp\left(
\frac{-1}{2\sigma^2}
\Vmat^{\transp}{\bf H}\Vmat
\right)
\exp\left(
\frac{1}{\sigma^2}
\Vmat^{\transp}{\bf H}\boldsymbol\mu
\right)
C(\boldsymbol\mu)
\end{equation}
we get:
\begin{equation}
\label{eq:tempPostGibbs}
p_{\invT}(\bftheta|\data,\model_1,\model_2)
\; = \;
N(\bftheta|\boldsymbol\mu,\sigma^2{\bf H}^{-1})
\end{equation}
where 
{
%\footnotesize
\begin{eqnarray}
\label{eq:tempPostGibbs1}
{\bf H}
 = 
\invT\{\DmatB\}^{\transp} \DmatB
+
(1-\invT)\{\DmatA\}^{\transp} \DmatA
+ \delta^{-2} {\bf I}, \quad
\nonumber
\\
\boldsymbol\mu
 = 
{\bf H}^{-1}
\left(
\left[
\invT\{\DmatB\}^{\transp} 
+
(1-\invT)\{\DmatA\}^{\transp} 
\right]
\by
+
\delta^{-2} \bfmu_0
\right)
\end{eqnarray}
} 
Hence, we can directly sample $\bftheta$ from the tempered conditional distributions in a Gibbs sampling scheme without having to resort to Metropolis-Hastings.
For linear models where the variance $\sigma^2$ is not known and has to be sampled from the tempered posterior distribution too, we refer to Appendix~\ref{sec:fcd_var}.

% ========================================
\subsection{Sigmoid inverse temperature ladder}
\label{sec:Tladder}
% ========================================
Given a single model $\mathcal{M}$, conventional TI follows an inverse annealing path from the prior $p(\bftheta|\mathcal{M})$ to the posterior $p(\bftheta|\mathcal{M},\data)$, symbolically $p(\bftheta|\mathcal{M})\rightarrow p(\bftheta|\mathcal{M},\data)$. Unlike TI, NETI-DIFF is based on a direct transition from the posterior of one model $\mathcal{M}_1$ to the posterior of another model $\mathcal{M}_2$, $p(\bftheta|\mathcal{M}_1,\data)\rightarrow p(\bftheta|\mathcal{M}_2,\data)$. For nested models, e.g. $\mathcal{M}_1 \subset \mathcal{M}_2$, we start at the less complex model $\mathcal{M}_1$ and move towards the more complex model $\mathcal{M}_2$, e.g. using the power-law inverse temperature ladder, defined in \eqname~(\ref{eq:powerLaw}). For a power $\alpha>1$ the distances $\invT_{i+1}-\invT_{i}$ between neighbouring discretisation points $\invT_{k}$ and $\invT_{k+1}$ increase in $k$ and the discretisation points will be concentrated around the nested model, $\mathcal{M}_1$ ($\invT=0$), and fewer points will be set near $\mathcal{M}_2$ ($\invT=1$). However, in many applications non-nested models have to be compared, and it is then not clear which of the two models should be used as starting point. Imbalances can be avoided by choosing a sigmoid inverse temperature ladder, such that the discretisation points are mirrored at the midpoint $\invT^{\star}=0.5$ of the interval $[0,1]$. Every discretisation point $\invT<0.5$ closer to $\mathcal{M}_1$ then has its counterpart $\invT^{\star}=1-\invT$ with the same distance $\invT$ to $\mathcal{M}_2$, and vice-versa.

To obtain a sigmoid  inverse temperature ladder for NETI-DIFF 
%that is comparable to a TI power-law ladder with power $\alpha$, see \eqname~(\ref{eq:powerLaw}), 
we apply the following procedure.
%
%Suppose that TI is performed with $N_{iter}$ iterations in total, i.e. $N_{iter}/K$ iterations at each discretisation point $\invT_1,\ldots,\invT_K$, and that NETI-DIFF can also be run with $N_{iter}$ iterations (points). To implement a sigmoid inverse temperature ladder, 
We first specify $50\%$ of the discretisation points $\invT_1<\ldots<\invT_{\frac{N_{iter}}{2}}$ within the interval $[0,0.5]$, and then we mirror the ladder at the midpoint $\invT=0.5$.\footnote{For uneven $N_{iter}$, we fix one point at $\invT=0.5$ and apply the procedure to the remaining $N_{iter}-1$ points.} This yields the remaining $50\%$ of the discretisation points, $\invT_{\frac{N_{iter}}{2}+i}=1-\invT_{\frac{N_{iter}}{2}+1-i}$ ($i=1,\ldots,\frac{N_{iter}}{2}$).
As we want the first $50\%$ of the discretisation points to get as close as possible to the midpoint $\invT=0.5$ subject to a power law with power $\alpha$, we determine the minimal integer $N^{\star}$ such that
\begin{equation}
\invT_i := \left(\frac{i}{N^{\star}}\right)^\alpha < 0.5 \;\;\;\; (i=1,\ldots,\frac{N_{iter}}{2})
\end{equation}
The solution is: $ N^{\star} = \lfloor  x^{\star} \rfloor$, where 
\begin{equation}
\left(\frac{N_{iter}}{2x^{\star}}\right)^\alpha = 0.5 \Leftrightarrow x^{\star} = \frac{N_{iter}}{2} \cdot 0.5^{-\frac{1}{\alpha}}
\end{equation}

% ========================================
\section{Benchmark problems and data}
\label{sec:data}
%\vspace{-3mm}
% ========================================

We have evaluated the proposed method on four benchmark data sets. Given data $\data$ the goal is to estimate the log Bayes factor $\mathcal{B}$ between two models $\mathcal{M}_1$ and $\mathcal{M}_2$. We assume the models to be equally likely a priori, $p(\mathcal{M}_1)=p(\mathcal{M}_2)$, so that the Bayes factor is the ratio of marginal likelihoods:
\begin{equation}
\label{BAYES_FACTOR}
\mathcal{B}(\mathcal{M}_1,\mathcal{M}_2) = \log\left\{  \frac{ p(\mathcal{M}_2|\data)}{p(\mathcal{M}_1|\data)}    \right\} = \log\left\{  \frac{ p(\data|\mathcal{M}_2)}{p(\data|\mathcal{M}_1)} \right\}
\end{equation}
For nonuniform prior distributions, $p(\mathcal{M}_1)\neq p(\mathcal{M}_2)$, it is straightforward to add the correction factor 
$\log p(\model_2)/p(\model_1)$, which is computationally cheap compared to the marginal likelihood ratio.

For method evaluation, we need to compare with a ground truth. 
For a linear model, we have a proper ground truth, as the Bayes factor can be computed analytically. This applies to the Radiata pine data (Section~\ref{sec:radiata}) and the Radiocarbon data (Section~\ref{sec:carbon}). For the Pima Indian data (Section~\ref{sec:pima}), we use a generalised linear model, and for the biopathway data (Section~\ref{sec:Biopepa}), we use a nonlinear model. In these cases, a closed-form solution of the Bayes factor does not exist. 
For the Pima Indian data, we follow the method suggested in \shortciteA{frielTIcorrection} and use the numerical result from a very long MCMC run as an approximate gold standard. For the biopathway data, we use the knowledge of the true interaction structure of the system as a surrogate gold standard and assess the performance in terms of network reconstruction accuracy.  We think this provides an adequate balance between using linear models, for which a strong ground truth exists, and generalised linear/non-linear models, for which a strong ground truth is intrinsically unavailable, and a weaker surrogate ground truth has to be used instead. 

\subsection{Radiata pine}
\label{sec:radiata}	
The Radiata pine data have been used in \shortciteA{frielTIcorrection} and were originally published in \cite{Williams1959}. Like \shortciteA{frielTIcorrection} we focus on the log Bayes factor between two competing non-nested linear regression models for explaining the 'maximum compression strength' $y$ of $n=42$ Radiata pine specimens. Both linear models contain an intercept and one single covariate. The first model ($\mathcal{M}_1$) uses the 'density' $x_1$ and the second model ($\mathcal{M}_2$)  the 'adjusted density' $x_2$ of the specimen. After standardizing the observation vectors ${{\bf x}}_1$ and ${{\bf x}}_2$ of the two covariates to mean $0$, the likelihood of model $\mathcal{M}_k$ ($k=1,2$) is:
\begin{equation}
\mathcal{M}_k: \;\; p({{\bf y}}|\bftheta^{(k)},\sigma^2) = N({{\bf D}}^{(k)} \bftheta^{(k)}, \sigma^2 {{\bf I}})
\end{equation}
where ${{\bf y}}$ is the vector of the observed 'maximum compression strengths', ${{\bf D}}^{(k)}=({{\bf 1}},{{\bf x}}_k)$ is the $n$-by-$2$ design matrix and $\bftheta^{(k)}$ is the $2$-dimensional vector of regression coefficients of model $\mathcal{M}_k$. Both models share the intercept parameter $\theta_0$, but differ w.r.t. the second parameter, i.e. $\bftheta^{(k)}=(\theta_0,\theta_k)^{\top}$.
For comparability we use exactly the same Bayesian model as in \shortciteA{frielTIcorrection}, where an inverse Gamma prior is imposed on the noise variance: $p(\sigma^{-2})=  GAM(3,2\cdot300^2 )$
and Gaussian priors are used for the regression coefficient vectors:\footnote{Unlike the prior in \eqname~(\ref{reg_prior}), the prior from \shortciteA{frielTIcorrection} uses fixed variances in \eqname~(\ref{PRIOR_FROM_FRIEL}).}
\begin{equation}
\label{PRIOR_FROM_FRIEL}
\mathcal{M}_k: \;\;\; p(\bftheta^{(k)})    =N  \left(  \left( \begin{array}{r}3000\\185\\ \end{array} \right),  \left( \begin{array}{cc}0.06^{-1}&0\\0&6^{-1}\end{array} \right)  \right)
\end{equation}
This is a  model with fully conjugate priors, so that the marginal likelihoods $p({{\bf y}}|\mathcal{M}_k)$ can be computed in closed form \shortcite{frielTIcorrection}. With \eqname~(\ref{BAYES_FACTOR}) we obtain for the log Bayes factor $\mathcal{B}(\mathcal{M}_1,\mathcal{M}_2)=8.8571$.
Like \shortciteA{frielTIcorrection} we apply Gibbs sampling and re-sample the model parameters iteratively from their full conditional distributions: $p(\sigma^2|{{\bf y}},\bftheta^{(k)})$ and $p(\bftheta^{(k)}|{{\bf y}},\sigma^2)$.

\subsection{Pima Indians}
\label{sec:pima}
The Pima Indians data have also been used in \shortciteA{frielTIcorrection} and were originally published in \cite{Smith88}. Like \shortciteA{frielTIcorrection} we focus on the log Bayes factor between two nested logistic regression models for explaining the binary 'diabetes disease status' $y$ of $n=532$ female Pima Indians.
The first model ($\mathcal{M}_1$) contains an intercept and 4 covariates, namely 'the number of pregnancies', 'the plasma glucose concentration', 'the body mass index', and 'the diabetes pedigree function', while the second model ($\mathcal{M}_2$) extends model $\mathcal{M}_1$ by including one additional covariable 'age'. After standardizing all covariates to mean $0$ and variance $1$, the likelihood of model $\mathcal{M}_k$ ($k=1,2$) is:
\begin{equation}
\nonumber
\mathcal{M}_k: \;\;\;p({{\bf y}}| \bftheta^{(k)} ) = \prod_{i=1}^{n}  \frac{\left\{\exp(-{{\bf x}}_{i,k}^{\top}\bftheta^{(k)}) \right\}^{y_i} }{1+\exp(-{{\bf x}}_{i,k}^{\top}\bftheta^{(k)})} 
\end{equation}
where the $i$-th element of ${{\bf y}}$, $y_i\in\{0,1\}$, is the diabetes status of female $i$, ${{\bf x}}_{i,k}$ is the corresponding vector of covariates, including an initial $1$ for the intercept, and $\bftheta^{(k)}$ is the vector of regression coefficients of dimension $m=5$ ($\mathcal{M}_1$) or $m=6$ ($\mathcal{M}_2$). Again, we follow \shortciteA{frielTIcorrection} and impose the following Gaussian priors on the regression coefficient vectors: $p(\bftheta^{(k)}|\delta^2) = N({{\bf 0}}, \delta^2 {{\bf I}} )$, where $\delta^2=100$ gives rather uninformative priors. For the logistic regression neither the marginal likelihoods nor the full conditional distributions can be computed in closed form. We therefore use the Metropolis Hastings based Markov chain Monte Carlo (MCMC) sampling scheme from \shortciteA{frielTIcorrection}, which employs the following proposal mechanism: In each iteration a new candidate regression coefficient vector is obtained by adding a sample ${{\bf u}}$ from an $m$-dimensional multivariate Gaussian distribution to the current vector $\bftheta^{(k)}$. The Gaussian distribution of ${{\bf u}}$ has a zero mean vector and a diagonal covariance matrix, whose diagonal entries $d_1,\ldots,d_m$ depend on the inverse temperature $\invT\in[0,1]$ of the power posterior. For the TI approaches we set: $d_i=\min\{0.01\invT^{-1},100\}$, as in \shortciteA{frielTIcorrection}. For the proposed NETI-DIFF approch we use $d_6=\min\{0.01 \invT^{-1},100\}$, while we fix the first five diagonal entries $d_1,\ldots,d_5=0.01$. This modification is required, as the first five regression coefficients appear in both models $\mathcal{M}_1$ and $\mathcal{M}_2$. That is, they effectively appear constantly with inverse temperature $\invT=1$ throughout NETI-DIFF simulations. The marginal likelihoods cannot be computed in closed-form. We therefore use those values reported in \shortciteA{frielTIcorrection}, which were obtained from long TI simulations, as gold-standard: $\log\{p({{\bf y}}|\mathcal{M}_1)\}=-257.2342$ and $\log\{p({{\bf y}}|\mathcal{M}_2)\}=-259.8519$. \eqname~(\ref{BAYES_FACTOR}) yields the log Bayes factor: $\mathcal{B}(\mathcal{M}_1,\mathcal{M}_2)=-2.6177$.

\subsection{Radiocarbon dating}
\label{sec:carbon}
We use the Radiocarbon data from \cite{PearsonQua93} to compute the Bayes factors among 10 nested linear regression models. For predicting the 'true calendar age' $y$ of $n=343$ Irish oaks from one single covariable: 'the Radiocarbon dating process' $x$, we fit polynomial calibration curves $\mathcal{M}_i$ ($i=1,\ldots,10$) of the following type:
\begin{equation}
\mathcal{M}_i: \;\;\; y = \theta_0 + \sum_{j=1}^i \theta_j x^j 
\end{equation}
The likelihood of model $\mathcal{M}_i$ is then 
\begin{equation}
\mathcal{M}_i: \;\; p({{\bf y}}|\sigma^2,\bftheta^{(i)}) = N({{\bf D}}^{(i)} \bftheta^{(i)} , \sigma^2 {{\bf I}})
\end{equation}
where  ${{\bf y}}$ is the vector of calendar ages, $\bftheta^{(i)}=(\theta_0,\theta_1,\ldots,\theta_i)^{\top}$ is the vector of regression coefficients, and ${{\bf D}}^{(i)}$ is the $n$-by-$(i+1)$ design matrix. The first column of the design matrix consists entirely of ones (for the intercept), and the subsequent columns are built from the observation vector ${{\bf x}}$, ${{\bf D}}^{(i)}=({{\bf 1}},{{\bf x}}^1,\ldots,{{\bf x}}^i)$, where ${{\bf x}}^j$ denotes an element-wise power operation on ${{\bf x}}$. We impose conjugate priors on the parameters. For $\sigma^2$ we use an inverse Gamma distribution: $p(\sigma^{-2})= GAM(\frac{a}{2},\frac{b}{2})$, and on $\bftheta^{(i)}$ we impose Gaussian priors:
\begin{equation}
p(\bftheta^{(i)}|\sigma^2,\delta^2) = N({{\bf 0}},\sigma^2\delta^2 {{\bf I}})
\end{equation} 
For fixed hyperparameters $a$, $b$, and $\delta^2$ the marginal likelihood for a model $\mathcal{M}$ with design matrix ${{\bf D}}$ is given by:
\begin{equation}
\nonumber
p({{\bf y}}|\mathcal{M}) = \frac{  \Gamma(\frac{n+a}{2}) \cdot  b^{\frac{a}{2}} \cdot  (b + {{\bf y}}^{\top} ({{\bf I}}+\delta^2 {{\bf D}} {{\bf D}}^{\top})^{-1} {{\bf y}})^{-\frac{n+a}{2}  }  }{ \Gamma(\frac{a}{2}) \cdot \pi^{\frac{n}{2}} \cdot \det\left({{\bf I}} + \delta^2 {{\bf D}} {{\bf D}}^{\top}    \right)        } 
\end{equation}
so that the log Bayes factors $\mathcal{B}(\mathcal{M}_{i},\mathcal{M}_{l})$ for two models $\mathcal{M}_i$ and $\mathcal{M}_l$ can be computed in closed form with \eqname~(\ref{BAYES_FACTOR}).
For the Radiocarbon data we fix $a=b=0.2$, $\delta^2=1$, and we sample the parameters iteratively from their conditional distributions $p(\sigma^2|{{\bf y}},\bftheta^{(i)})$ and $p(\bftheta^{(i)}|{{\bf y}},\sigma^2)$ with Gibbs sampling.

\begin{figure}[tb]
\begin{minipage}[b]{.5\linewidth}
\centering \includegraphics[width=6.5cm]{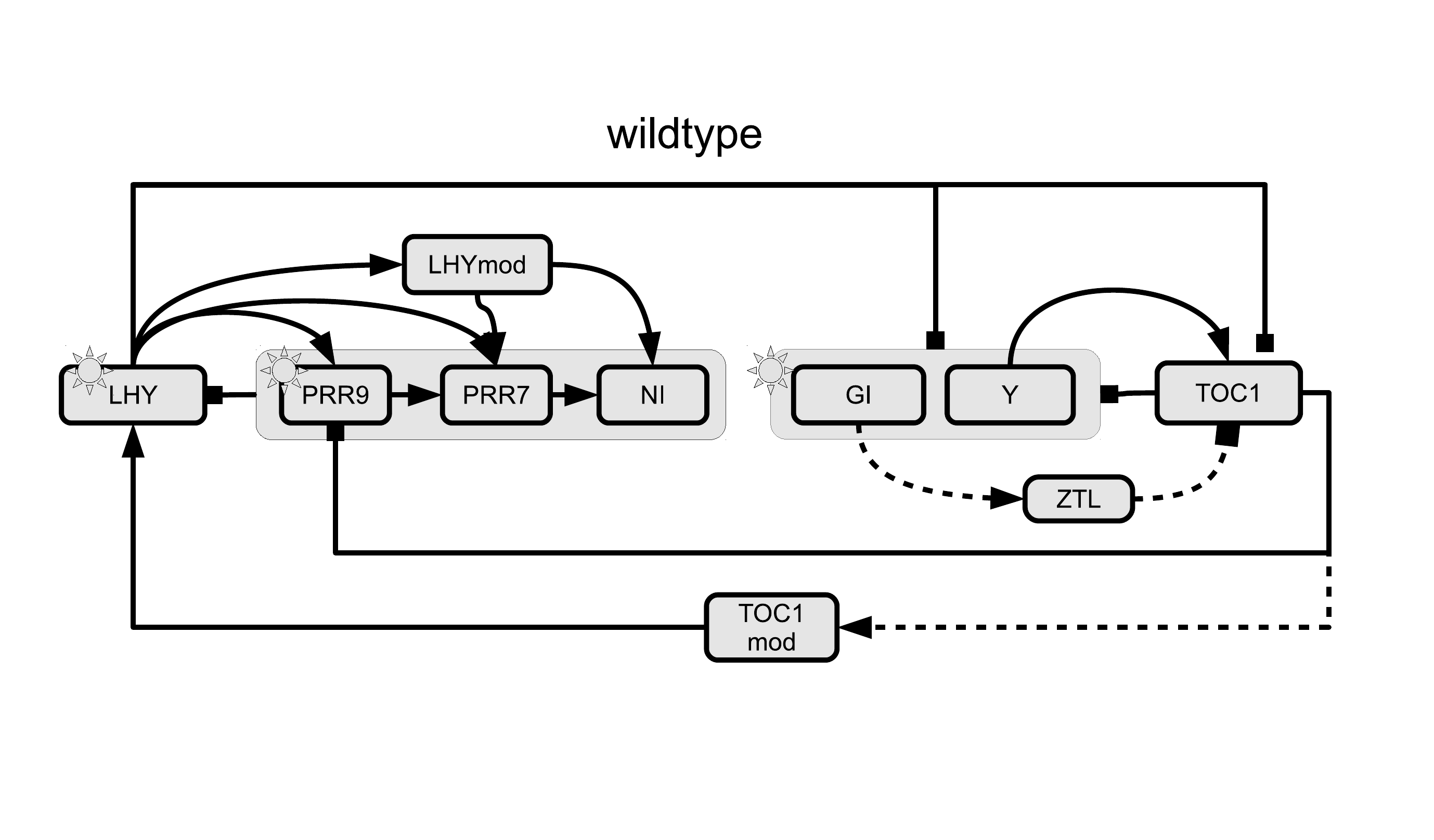}
\subcaption{}\label{fig:goldstd1_a}
\end{minipage}
\begin{minipage}[b]{.5\linewidth}
\centering \includegraphics[width=6.5cm]{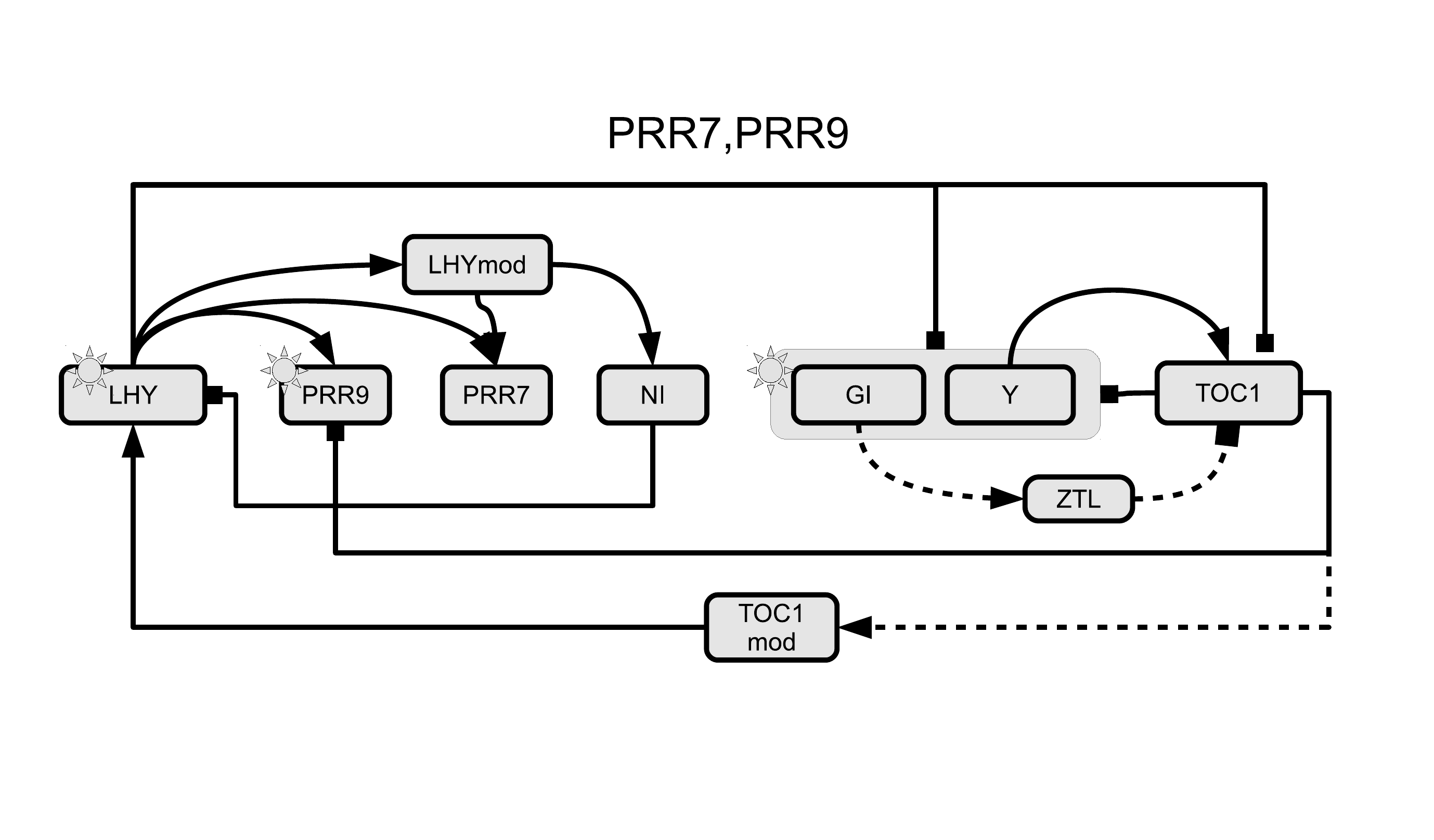}
\subcaption{}\label{fig:goldstd1_b}
\end{minipage}
\caption{\label{fig:goldstd1}
\textbf{Gene regulatory networks of the circadian clock in \emph{Arabidopsis  thaliana}: wildtype and mutant.}
The network displayed in panel \subref{fig:goldstd1_a} is the P2010 network proposed by \shortciteA{Pokhilko2010data}. Panel \subref{fig:goldstd1_b} shows a mutant network, in which the proteins PRR9 and PRR7 are dysfunctional and can no longer form a protein complex with NI. The nodes in the network represent proteins and genes, the edges indicate interactions. Arrows symbolize activations and bars inhibitions. Solid lines show protein-gene interactions; dashed lines show protein interactions.
The regulatory influence of light is symbolized by a sun symbol. 
Grey boxes group sets of regulators or regulated components.
Figure reproduced from \cite{Aderhold2014_SAGMB}.
} 
\end{figure}

\begin{figure}[tb]
\centering
\includegraphics[width=7cm]{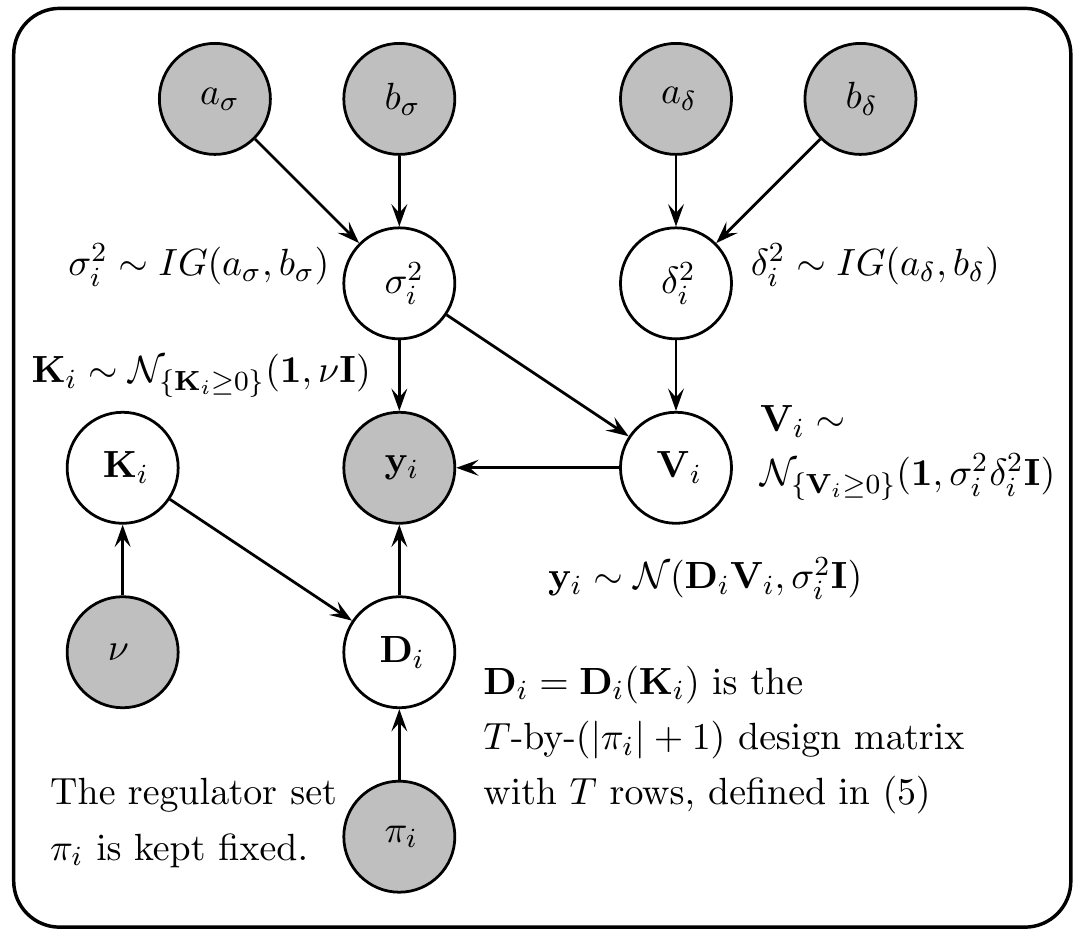}  
\vspace{-0.3cm}
\caption{
\label{fig:graphModelCheMA}
Hierarchical Bayesian model used for gene regulatory network reconstruction. Grey nodes refer to fixed quantities such as the observed response data or low-level hyperparameters. White nodes refer to quantities that can change, which includes the model parameters and high-level hyperparameters. Note that the design matrix ${\bf D}_i$ is not fixed because it depends on the Michaelis-Menten parameter vector ${\bf K}_i$.
}
\end{figure}

\subsection{Biopathway}
\label{sec:Biopepa}
The objective of the last application is model selection with respect to two alternative candidate interaction structures of ten genes in the circadian gene regulatory network  of 
\emph{Arabidopsis thaliana}, shown in Figure~\ref{fig:goldstd1}. 
The statistical model used for inference is a semi-mechanistic Bayesian hierarchical model for transcriptional regulation (\cite{STCO_2016}).
 Let $x_i(t)$ denote the mRNA concentration of gene $i$ at time $t$, and $\parents$ the set of its regulators. For instance, in the gene network of Figure~\ref{fig:goldstd1_a}, the regulators of gene \emph{PRR9} are two other genes, \emph{TOC1} and \emph{LHY}. So if $i=PRR9$, then $\parents=\{TOC1,LHY\}$. 
 A regulator can either act as activator or as repressor, and we represent that with the binary variable $I_{u,i}$, with $I_{u,i}=1$ indicating that gene $u$ is an activator for gene $i$, and $I_{u,i}=0$ indicating that gene $u$ is an inhibitor for gene $i$. For the example above, \emph{LHY} is an activator for \emph{PRR9}, hence $I_{u,i}=1$, while \emph{TOC1} is an inhibitor for \emph{PRR9}, hence $I_{u,i}=0$. From the fundamental equation of transcriptional regulation based on Michaelis-Menten kinetics we have for the gradient of $x_i$ \shortcite{Barenco2006}:
 \begin{equation}
\frac{d x_i(t)}{dt}|_{t=t^{\star}} = -v_{0,i} x_i(t^{\star}) + 
\sum_{u\in\parents}  v_{u,i} \frac{I_{u,i} x_u(t^{\star})+(1-I_{u,i})  k_{u,i}}{x_u(t^{\star})+k_{u,i}}
\label{EQ2_new}
\end{equation}
where the sum is over all genes $u$ that are in the regulator set of $\parents$ of gene $i$. 
The first term, $-v_{0,i} x_i(t^{\star})$, takes the degradation of $x_i$ into account, while $v_{u,i}$ and $k_{u,i}$ are the \emph{maximum reaction rate} and \emph{Michaelis-Menten} parameters for the regulatory effect of gene $u\in\parents$ on gene $i$, respectively.  
See the supplementary material of \shortciteA{Pokhilko2010data,Pokhilko2012clock} for similar examples in the mathematical biology literature. 
Without loss of generality, we now assume that $\parents$ is given by $\parents = \{x_1,\ldots,x_s\}$. 
\eqname~(\ref{EQ2_new}) can then be written in vector notation: 
\begin{equation}
\frac{d x_i(t)}{dt}|_{t=t^{\star}}   = {{\bf D}}_{i,t^{\star}}^{\top} {{\bf V}}_i
\end{equation}
where ${{\bf V}}_i=(v_{0,i},v_{1,i}\ldots,v_{s,i})^{\top}$ is the vector of the maximum reaction rate parameters, and the vector ${{\bf D}}_{i,t^{\star}}$ depends on the measured concentrations $x_u(t^{\star})$ and the Michaelis-Menten parameters $k_{u,i}$ ($u\in \parents$) via \eqname~(\ref{EQ2_new}):
\begin{equation}
\label{rows_of_Di}
{{\bf D}}_{i,t^{\star}}^{\top} = 
\Big(-x_i(t^{\star}),\frac{I_{1,i} x_1(t^{\star})+(1-I_{1,i}) k_{1,i}}{x_1(t^{\star})+k_{1,i}},\ldots,
\frac{I_{s,i} x_{s}(t^{\star})+(1-I_{s,i}) k_{s,i}} {x_s(t^{\star})+k_{s,i}}
\Big)
\end{equation}

 We combine the $s$ Michaelis-Menten parameters $k_{u,i}$ in a vector ${{\bf K}}_i =(k_{1,i}\ldots,k_{s,i})^{\top}$. For $n$ time points $t^{\star}\in\{t_1,\ldots,t_n\}$ we obtain $n$ row vectors from \eqname~(\ref{rows_of_Di}), and  we can arrange them successively in an $n$-by-$(|\parents|+1)$ design matrix ${{\bf D}}_i={{\bf D}}_i({{\bf K}}_i)$. The corresponing gradient vector is given by ${{\bf y}}_i:=(y_{i,1},\ldots,y_{i,n})^{\top}$, where $y_{i,j}$ is the gradient of $x_i$ at time point $t_j$. With  ${{\bf y}}_i$ being the response vector the likelihood is:
\begin{equation}
\nonumber
p({{\bf y}}_i|{{\bf K}}_i,{{\bf V}}_i,\sigma_i^2)=
	(2\pi \sigma_i^2)^{-\frac{n}{2}} e^{-\frac{1}{2\sigma^2_{i}} ({{\bf y}}_i - {{\bf D}}_i {{\bf V}}_i)^{\top} ({{\bf y}}_i - {{\bf D}}_i {{\bf V}}_i)}
\end{equation}
where  ${{\bf D}}_i={{\bf D}}_i({{\bf K}}_i)$ is the design matrix, given the Michaelis-Menten parameter vector ${{\bf K}}_i$. To ensure non-negative Michaelis-Menten parameters, truncated Normal prior distributions are used:  
\begin{equation}
\label{PriorK}
{{\bf K}}_i \sim \mathcal{N}_{\left\{{{\bf K}}_i\geq0 \right\}}({{\bf 1}},\nu{{\bf I}})
\end{equation} 
where $\nu>0$ is a hyperparameter, and the subscript, $\left\{{{\bf K}}_i\geq0 \right\}$, indicates the truncation condition, i.e. that each element of ${{\bf K}}_i$ has to be non-negative. 
For the maximum reaction rates, we use a  truncated ridge regression prior:
 \begin{equation}
\label{ridgePrior}
 {{\bf V}}_i|\sigma^2_i,\delta^2_i \sim \mathcal{N}_{\left\{{{\bf V}}_i\geq0\right\}}(  {{\bf 1}}, \delta^2_i \sigma_i^2 {{\bf I}}   )
 \end{equation}
 where $\delta^2_i$ is a hyperparameter that regulates the prior strength. For $\sigma_i^2$ and $\delta^2_i$ we use inverse Gamma priors, $\sigma_i^{2}\sim IG(a_{\sigma},b_{\sigma})$ and $\delta^2_i \sim IG(a_{\delta},b_{\delta})$. A graphical model representation can be found in \figurename~\ref{fig:graphModelCheMA}. 
 
 The posterior distribution of the parameters and hyperparameters has no closed-form solution, and we therefore resort to an MCMC scheme to sample from it. From the graphical model in \figurename~\ref{fig:graphModelCheMA} it can be seen that with the sole exception of the Michaelis-Menten parameters ${\bf K}_i$, the conditional distribution of each parameter conditional on its Markov blanket\footnote{Conditional on its Markov blanket, a node is independent of the rest of the graph; so the Markov blanket shields a node from the remaining graph. The Markov blanket of a node is the set of nodes in the graph that consists of the parents, the co-parents, and the children. In a graph A $\rightarrow$ B $\leftarrow$ C, we have: A is a parent of B (it has a directed edge from A to B), B is a child of both A and C, C is a co-parent of A, and A is a co-parent of C.} is of standard form (due to conjugacy) and can be sampled from directly. The MCMC scheme is therefore of the form of a Gibbs sampler, in which all parameters are sampled directly from their conditional distributions, except for ${\bf K}_i$, which is sampled via a Metropolis-Hastings within Gibbs step. 
The conditional distribution of the maximum rate parameter vector ${\bf V}_i$ is obtained from \eqnames~(\ref{eq:tempPostGibbs}-\ref{eq:tempPostGibbs1}) by replacing
\begin{math}
\bftheta 
\end{math}
by
\begin{math}
 {\bf V}_i, 
\end{math}
and adding an index $i$, for association with gene $i$, to all other quantities except for the identity matrix ${\bf I}$ and the inverse temperature $\invT$.
The derivation of the other conditional distributions is straightforward.
Pseudo code of the standard MCMC algorithm can be found in \cite{STCO_2016}. Pseudo code of the modified MCMC algorithm integrated into the proposed NETI-DIFF scheme is provided in the Appendix, Table~\ref{pseudocode_mcmc_TI}.

The data used for inference were obtained from \shortciteA{Aderhold2014_SAGMB}.
These are synthetic gene expression time series,  %from \shortciteA{Aderhold2014_SAGMB},
which were generated from a biologically realistic simulation of the  molecular interactions in these networks,
using the mathematical framework described in \shortciteA{Guerriero2012stochastic}  and implemented in the Biopepa software package  \shortcite{Ciocchetta2009biopepa}.
These time series correspond to gene expression measurements in 2 hour intervals over 24 hours, repeated 11 times for different experimental conditions related to various gene knockouts.
We repeated the simulations twice, for both of the two networks shown in Figure~\ref{fig:goldstd1}. Hence, the true interaction network is known, which  can be used to evaluate the accuracy of 
Bayesian model selection based on the modelling framework described above.

\begin{figure}[tbhp]
	\centering		
		\includegraphics[width=0.8\textwidth]{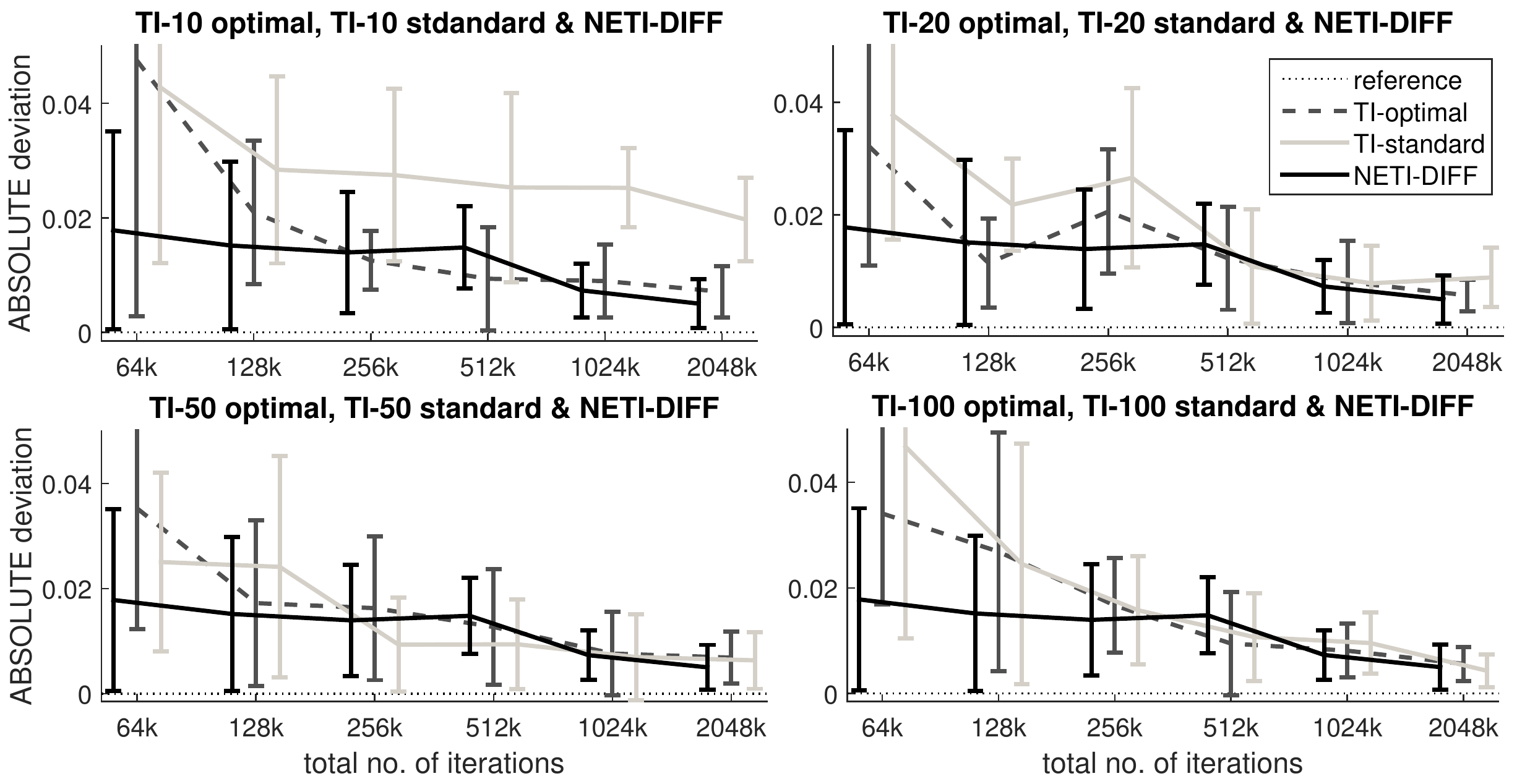}
		\caption{\label{fig:RADIATA_1}
		\textbf{Average absolute error on the Radiata pine data: TI versus NETI-DIFF.}  The figure shows the average absolute deviation between the estimated and the true log Bayes factor in dependence on the total number of MCMC iterations $N_{iter}$. In each panel the same NETI-DIFF results are shown, while the two TI approaches (TI-standard and TI-optimal) were applied with different numbers of discretisation points ($10$, $20$, $50$ and $100$). The error bars represent standard deviations. The horizontal axes give the total number of  (power posterior) MCMC iterations, $N_{iter}$.}
\end{figure}

\begin{figure}[tbhp]
	\begin{center}
		\includegraphics[width=.8\textwidth]{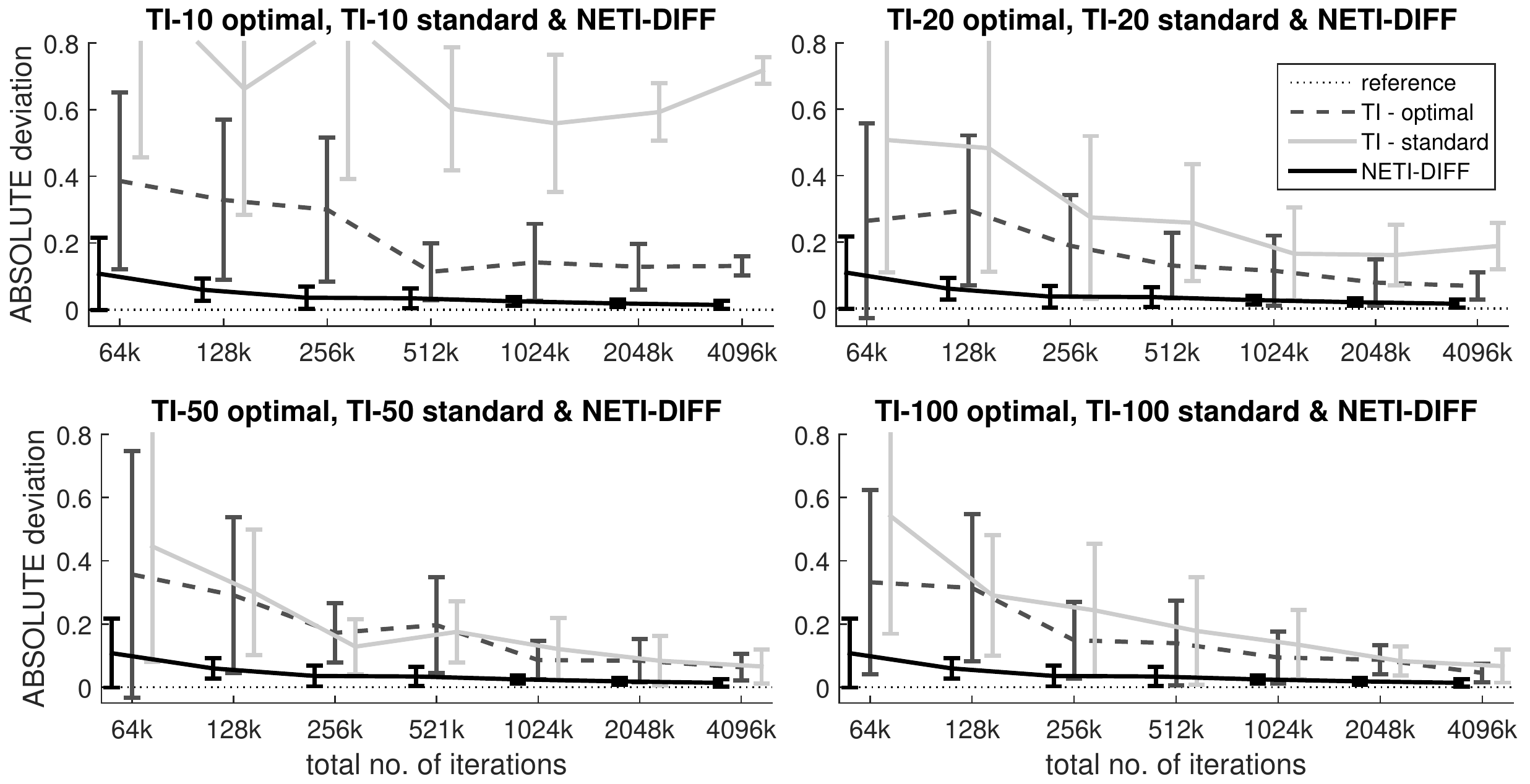}
		\caption{\label{fig:PIMA_1}{\bf  Average absolute error on the Pima Indians data: TI versus NETI-DIFF.} The figure shows the average absolute deviation between the estimated and the true log Bayes factor in dependence on the total number of MCMC iterations $N_{iter}$. In each panel the same NETI-DIFF results are shown, while the two TI approaches (TI-standard and TI-optimal) were applied with different numbers of discretisation points ($10$, $20$, $50$ and $100$). The error bars represent standard deviations. The horizontal axes give the total number of  (power posterior) MCMC iterations, $N_{iter}$.}
	\end{center}
\end{figure}

\begin{figure}[tbhp]
	\begin{center}
		\includegraphics[width=.59\textwidth]{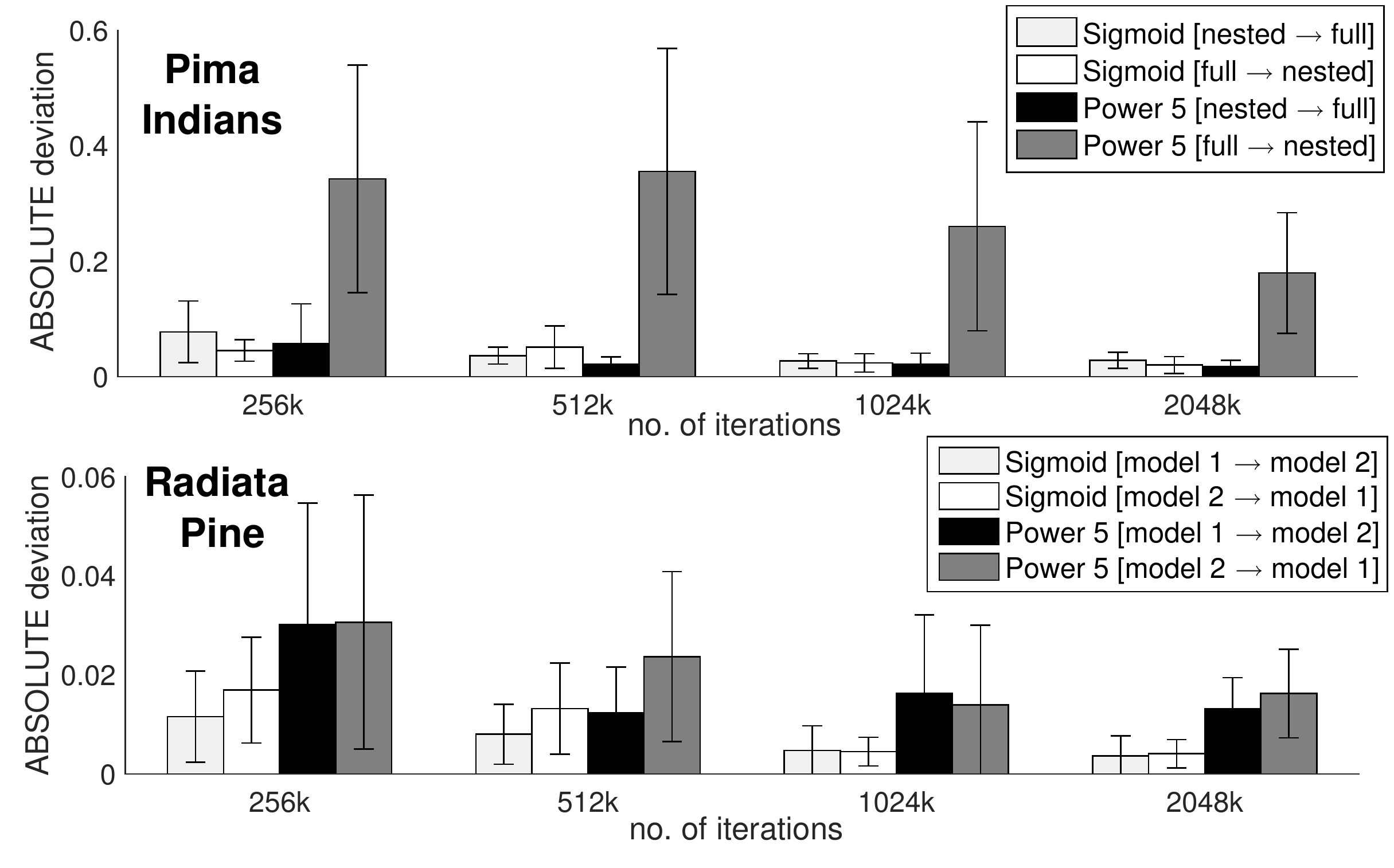}
		\caption{\label{fig:POWER_SIGMOID}{\bf Comparison of two inverse temperature ladders and two NETI-DIFF paths.} The vertical bars show the average absolute deviations between the estimated and true log Bayes factor, with error bars representing standard deviations. The horizontal axes give the total number of MCMC iterations $N_{iter}$. 
		The two inverse temperature ladders compared are the power law, \eqname~(\ref{eq:powerLaw}), versus the sigmoid function, defined in Section~\ref{sec:Tladder}.
		The alternative NETI-DIFF path swaps the initial model at $\invT=0$ with the final model at $\invT=1$.
		\emph{Top row:} Radiata pine data. \emph{Bottom row:} Pima Indians data. }
	\end{center}
\end{figure}

\begin{figure}[tbp]
    \begin{center}
        \includegraphics[width=.73\textwidth]{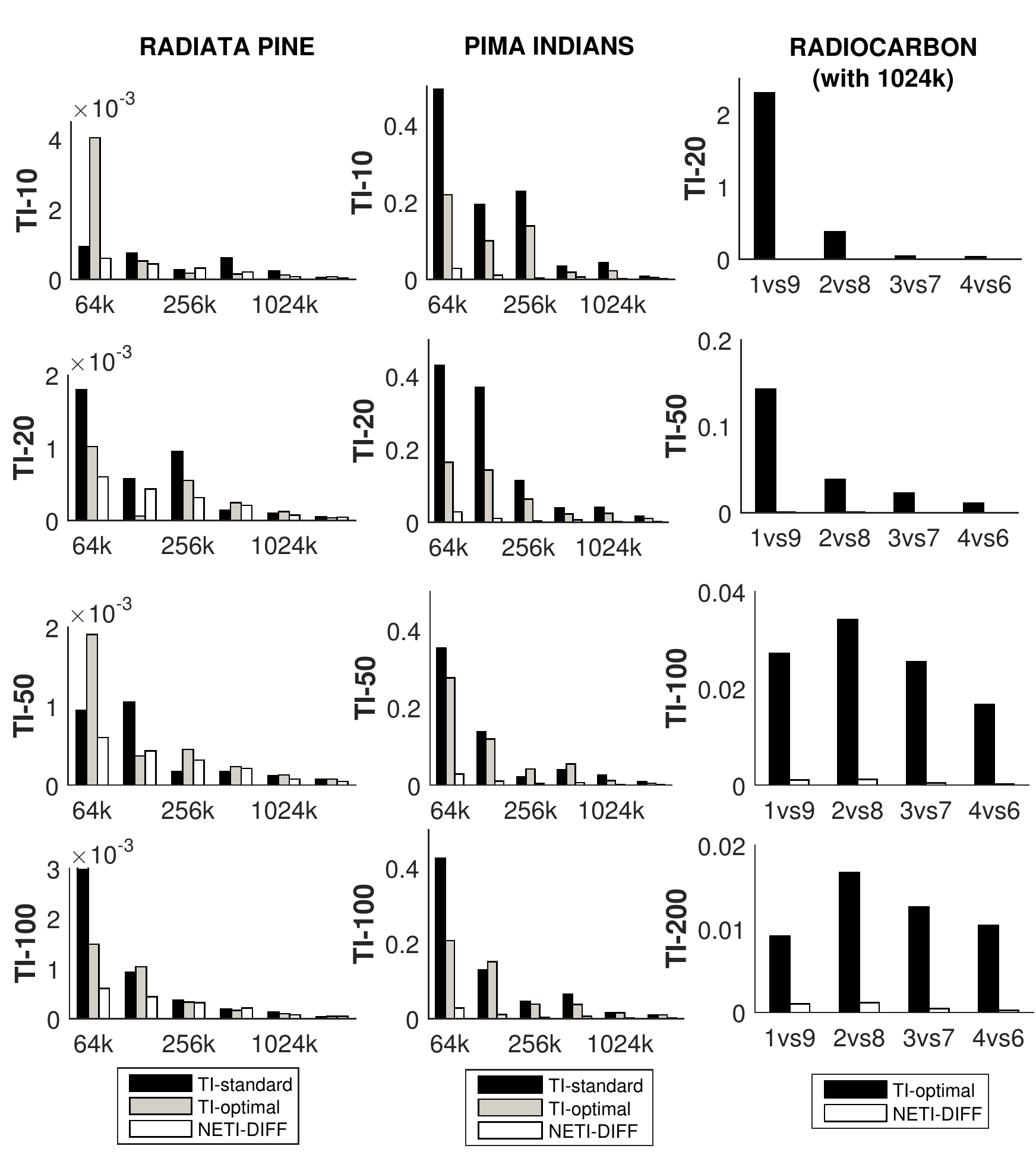}\\
        \caption{\label{fig:VARIANCES_1}{\bf Variance of log Bayes factor estimators.}
\emph{Left panel:} Radiata pine data, \emph{centre panel:} Pima
Indians data, \emph{right panel:}  Radiocarbon data. The vertical bars
show the variance $\VAR$, \eqname~(\ref{eq:VAR}),  for the TI-standard,
TI-optimal and NETI-DIFF
estimators of the log Bayes factor. For the Radiata pine and the
Pima Indians data we varied the number of total MCMC iteration
(horizontal axes). For the Radiocarbon data we performed $N_{iter}=1024k$
iterations and considered four different pairwise model comparisons
(horizontal axis). 
The rows represent different
numbers of discretisation points for TI (NETI-DIFF is unaffected). The three columns refer to the four panels
in Figure~\ref{fig:RADIATA_1} (right), Figure~\ref{fig:PIMA_1} (center)
and Figure~\ref{fig:RADIOC_1} (right)). The corresponding ratios of the variances are
shown in Figure~\ref{fig:VARIANCES_2}.}
    \end{center}
\end{figure}

\begin{figure}[tbp]
    \begin{center}
        \includegraphics[width=.73\textwidth]{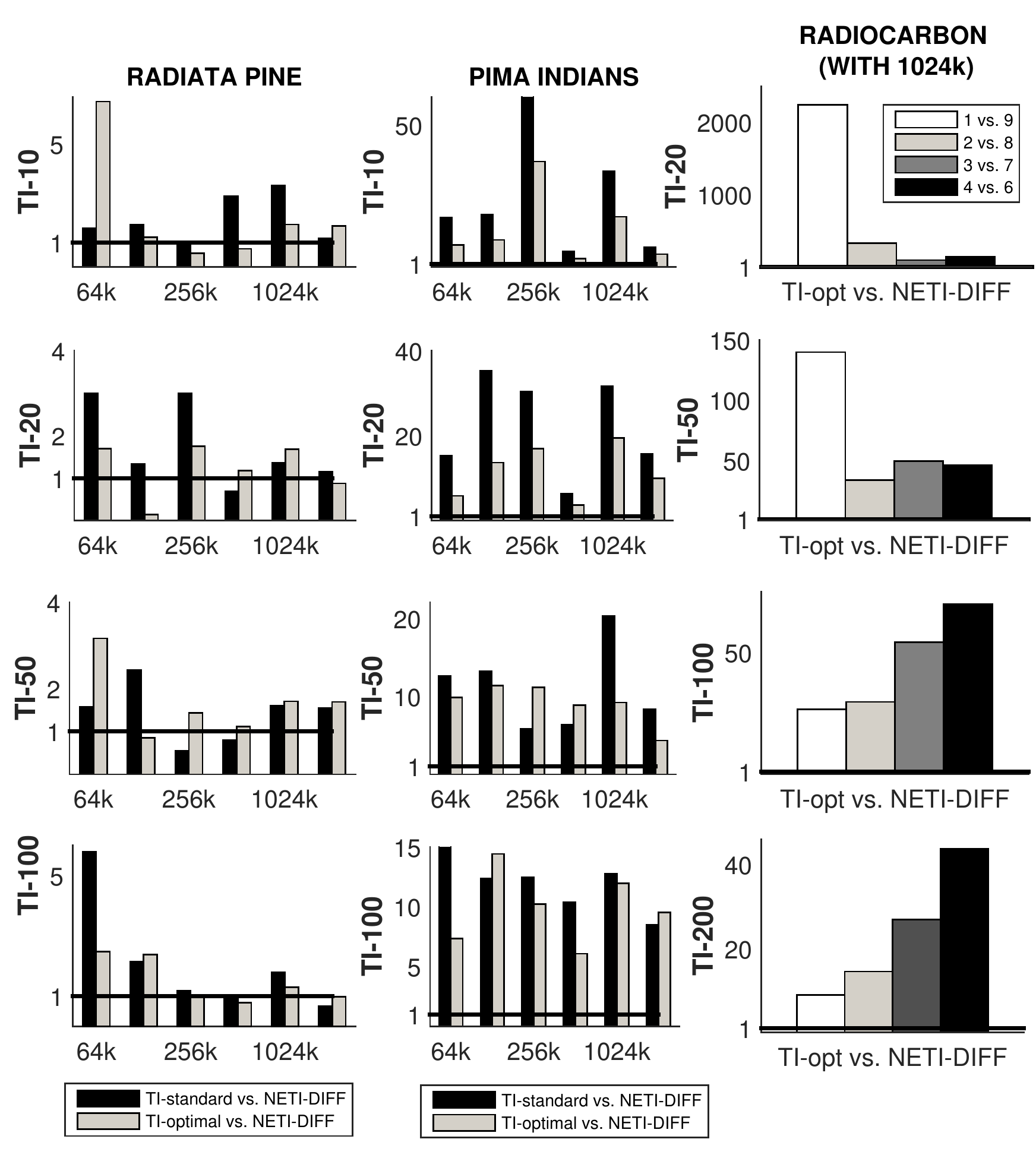}\\
        \caption{\label{fig:VARIANCES_2}{\bf Variance ratios of
log Bayes factors estimators.} 
\emph{Left panel:} Radiata pine data, \emph{centre panel:} Pima
Indians data, \emph{right panel:}  Radiocarbon data. 
The vertical bars show the variance ratios of the log Bayes factor
estimators:  TI-standard versus NETI-DIFF, and TI-optimal versus NETI-DIFF
(obtained from the variances in Figure~\ref{fig:VARIANCES_1}).
The horizontal reference line at value 1 indicates equal performance;
values above 1 indicate that NETI-DIFF achieves a variance reduction 
over the established TI schemes.
 For the Radiata pine and the Pima Indians
data we varied the number of total MCMC iterations $N_{iter}$ (horizontal axes). For
the Radiocarbon data we performed $N_{iter}=1024k$ iterations and considered four
different pairwise model comparisons (horizontal axis). 
The rows represent different
numbers of discretisation points for TI (NETI-DIFF is unaffected). 
The three columns refer to the four
panels in Figure~\ref{fig:RADIATA_1} (left), Figure~\ref{fig:PIMA_1}
(center) and Figure~\ref{fig:RADIOC_1} (right).}
    \end{center}
\end{figure}

\begin{figure}[tbhp]
	\begin{center}
		\includegraphics[width=.6\textwidth]{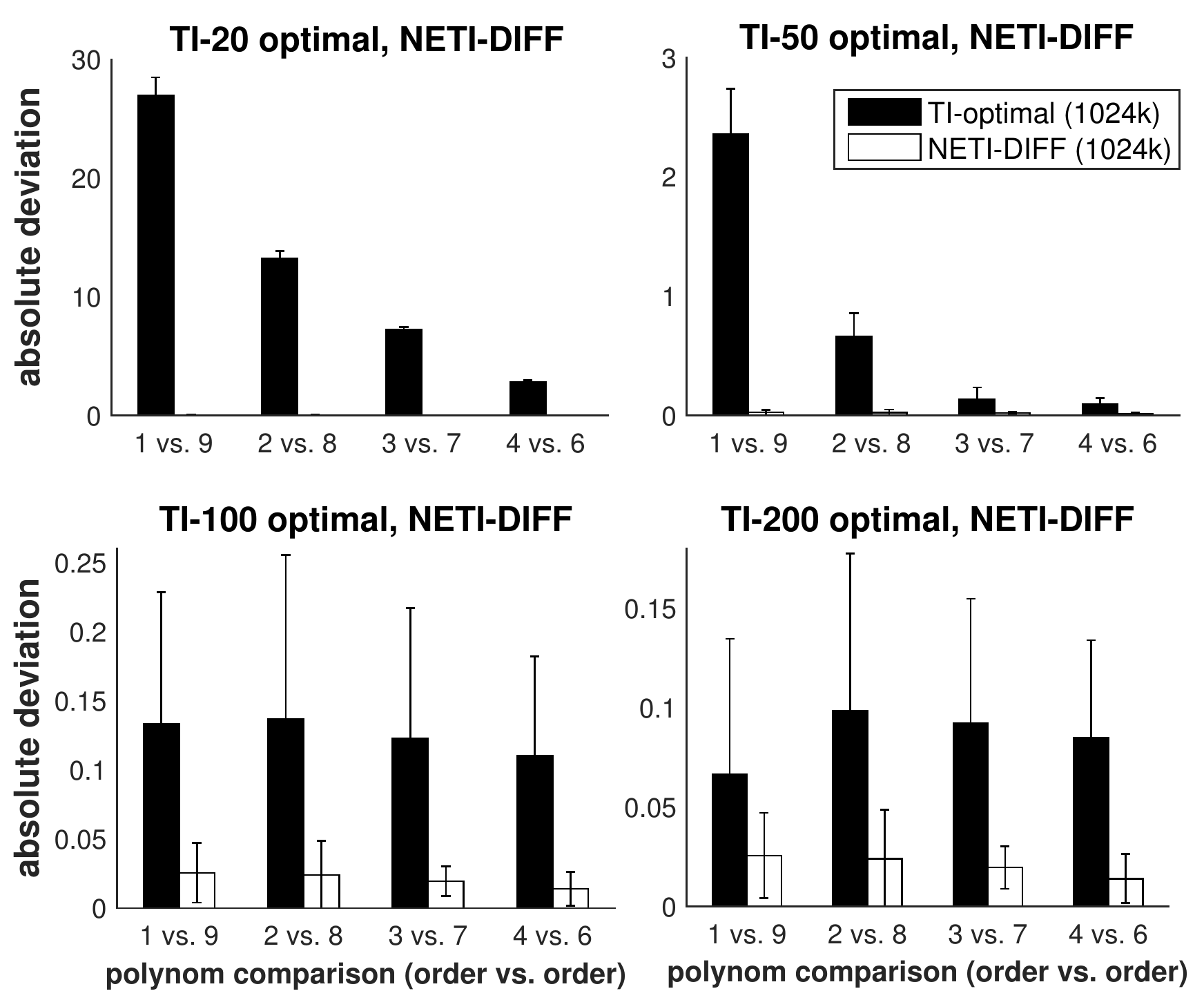}
		\caption{\label{fig:RADIOC_1}{\bf Average absolute error on the Radiocarbon data: NETI versus TI-optimal.} The figure shows the average absolute deviation between the estimated and the true log Bayes factor. In each panel the same NETI-DIFF results are shown, while TI-optimal was applied with different numbers of discretisation points ($20$, $50$, $100$ and $200$). The bars represent standard deviations and the horizontal axes indicate different model comparisons (polynomials of orders $i$ vs. $j$). The total number of  (power posterior) MCMC iterations was kept fixed at $N_{iter}=1024k$.}
	\end{center}
\end{figure}

\begin{figure}[tbhp]
	\begin{center}
		\includegraphics[width=.6\textwidth]{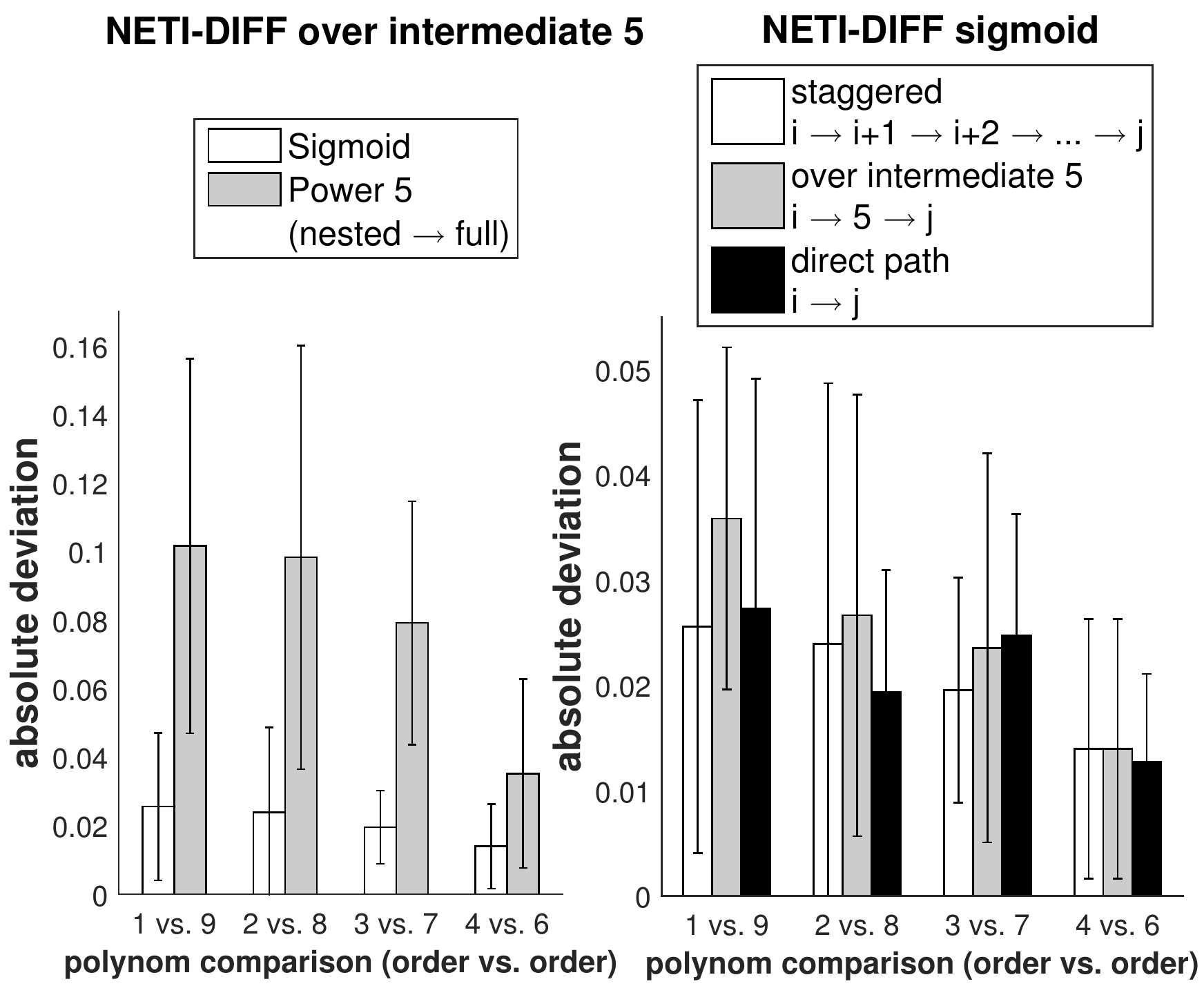}
		\caption{\label{fig:RADIOC_2}{\bf Influence of the inverse temperature ladder and the transition path.} The bars show the average absolute deviation $\MAD$, \eqname~(\ref{eq:MAD}), between the estimated and true log Bayes factor, computed with NETI-DIFF for the Radiocarbon data. The error bars show standard deviations. The horizontal axes indicate different model comparisons (polynomials of orders $i$ vs. $j$). The total number of (power posterior) MCMC iterations was kept fixed at $N_{iter}=1024k$. \emph{Left panel:} Comparison of two NETI-DIFF inverse temperature ladders (sigmoid vs. power 5). \emph{Right panel:} Comparison of three NETI-DIFF transition strategies (staggered vs. intermediate vs. direct).}
	\end{center}
\end{figure}

\begin{figure}[tbhp]
	\begin{center}
		\includegraphics[width=.64\textwidth]{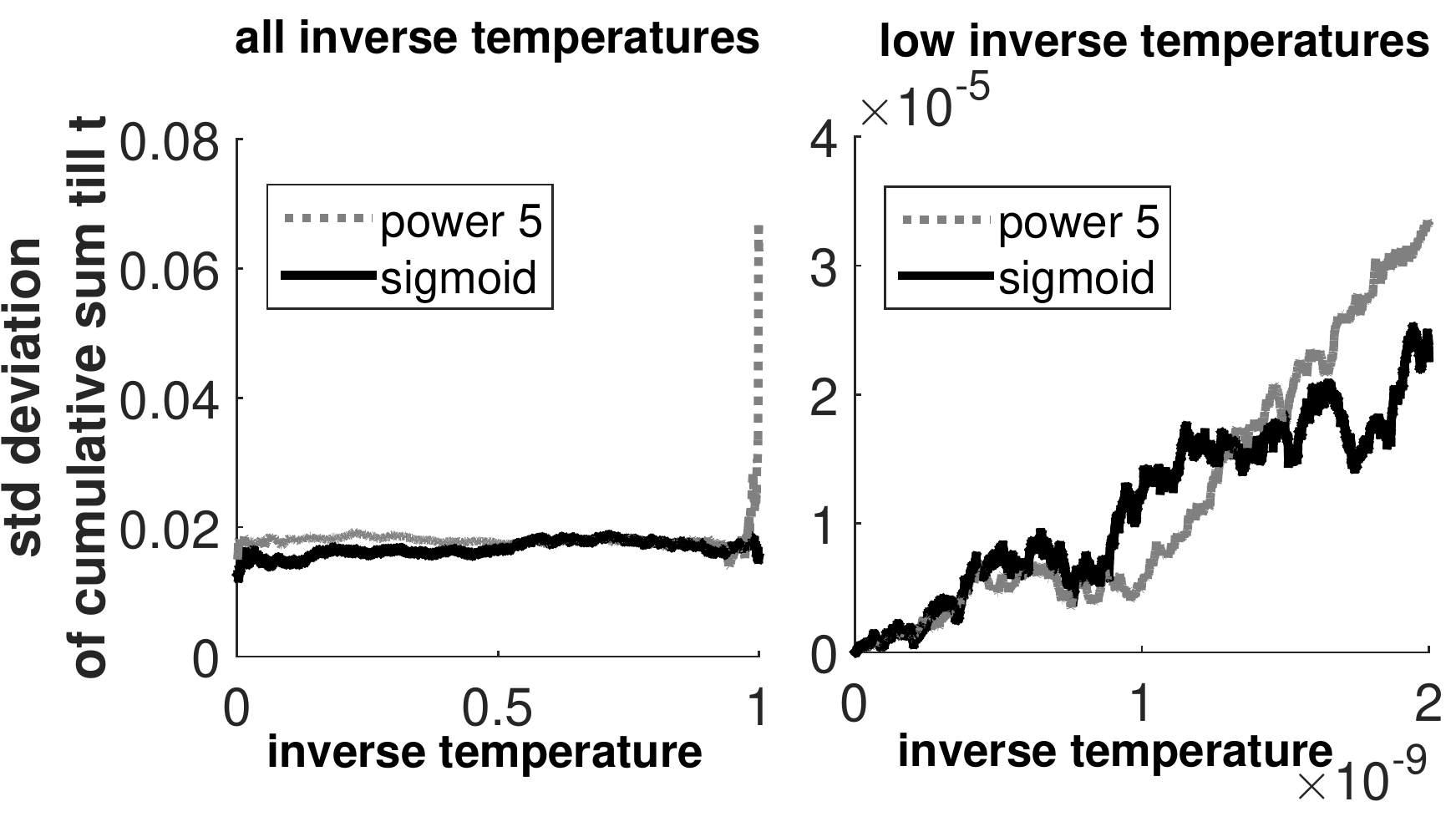}
		\caption{\label{fig:RADIOC_3}{\bf Comparison of the two inverse temperature ladders -- Radiocarbon data.} 
		The figures show the standard deviation of the partial NETI-DIFF integral, \eqname~(\ref{eq:ioio}), over the partial inverse temperature range $[0,\invT]$, obtained from five independent NETI-DIFF simulations.  The right panel shows a section of the left panel at higher resolution. \emph{Dashed line:} power law, \eqname~(\ref{eq:powerLaw}). \emph{Solid line:} sigmoid function, defined in Section~\ref{sec:Tladder}.
		 }
	\end{center}
\end{figure}

\begin{figure}[thbp]
\begin{minipage}[b]{.5\linewidth}
\centering 
\includegraphics[width=0.9\textwidth]{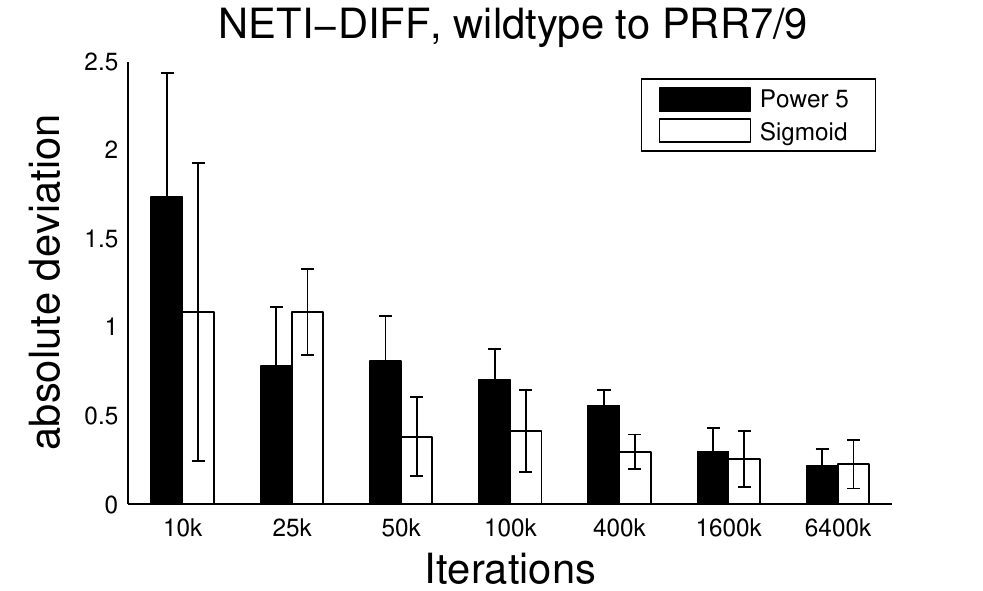}
\subcaption{}\label{fig:netidiff_pow_sigm_biasvar_a}
\end{minipage}
\begin{minipage}[b]{.5\linewidth}
\centering 
\includegraphics[width=0.9\textwidth]{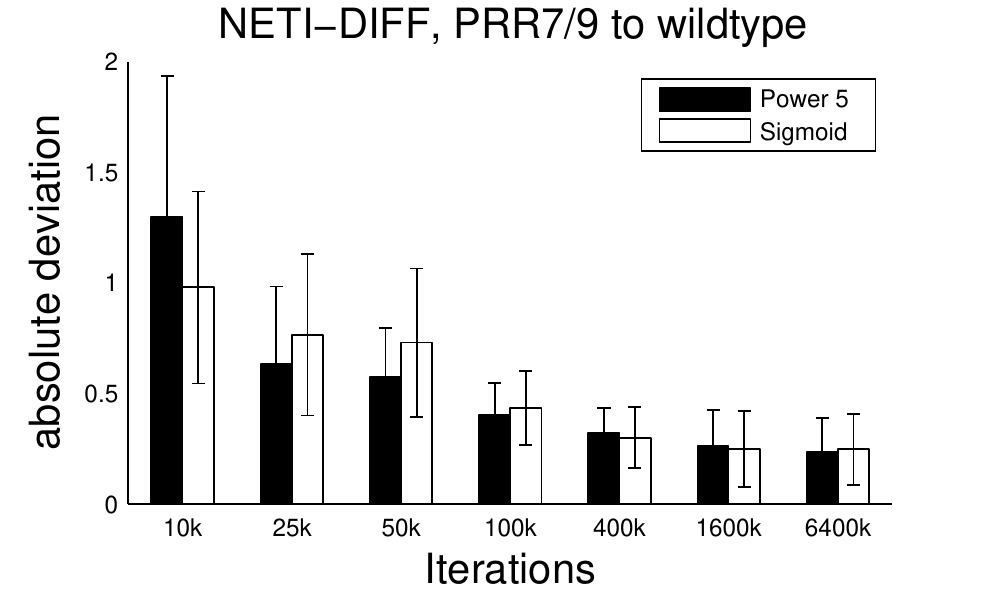}
\subcaption{}\label{fig:netidiff_pow_sigm_biasvar_b}
\end{minipage}
\begin{minipage}[b]{.5\linewidth}
\centering 
\includegraphics[width=0.9\textwidth]{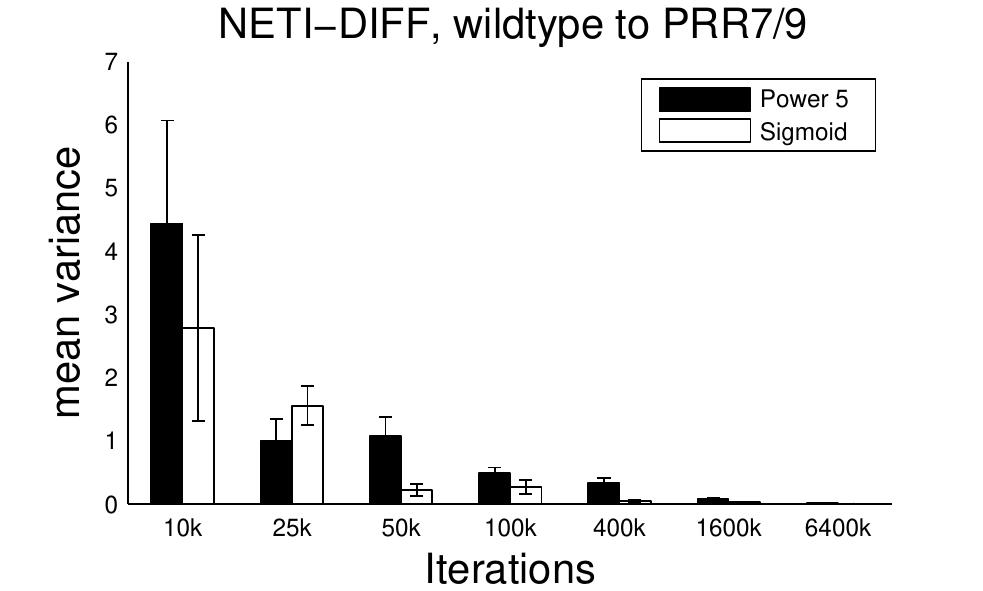}
\subcaption{}\label{fig:netidiff_pow_sigm_biasvar_c}
\end{minipage}
\begin{minipage}[b]{.5\linewidth}
\centering 
\includegraphics[width=0.9\textwidth]{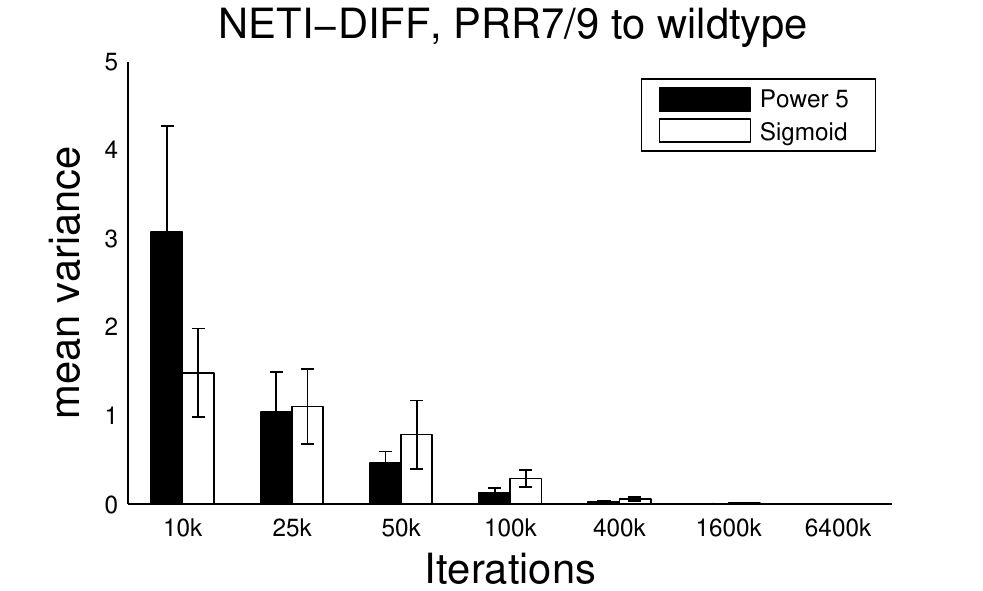}  
\subcaption{}\label{fig:netidiff_pow_sigm_biasvar_d}
\end{minipage}
\caption{\label{fig:netidiff_pow_sigm_biasvar} 
\textbf{Mean absolute error and mean variance for different inverse temperature ladders and biopathway data.} 
 Panels \subref{fig:netidiff_pow_sigm_biasvar_a} and \subref{fig:netidiff_pow_sigm_biasvar_b} show a comparison of the mean absolute error $\MAD$ (\eqname~\ref{eq:MAD}) and panels \subref{fig:netidiff_pow_sigm_biasvar_c} and \subref{fig:netidiff_pow_sigm_biasvar_d}  show a comparison of the mean variance $\VAR$ (\eqname~\ref{eq:VAR}) between two inverse temperature ladders: the power law from \eqname~(\ref{eq:powerLaw}) as black boxes, and the sigmoid form from Subsection~\ref{sec:Tladder} as white boxes. 
Results were obtained from 5 independent data instantiations from the wildtype biopathway of Figure~\ref{fig:goldstd1_a}, and 5 independent data instantiations from the PRR7/PRR9 mutant biopathway of Figure~\ref{fig:goldstd1_b}. Histogram height: average. Error bars: standard deviation. 
}
\end{figure}

\begin{figure*}[tbhp]
  \center
	\begin{minipage}[b]{1.0\linewidth}
	\centering 
	\includegraphics[width=0.33\textwidth]{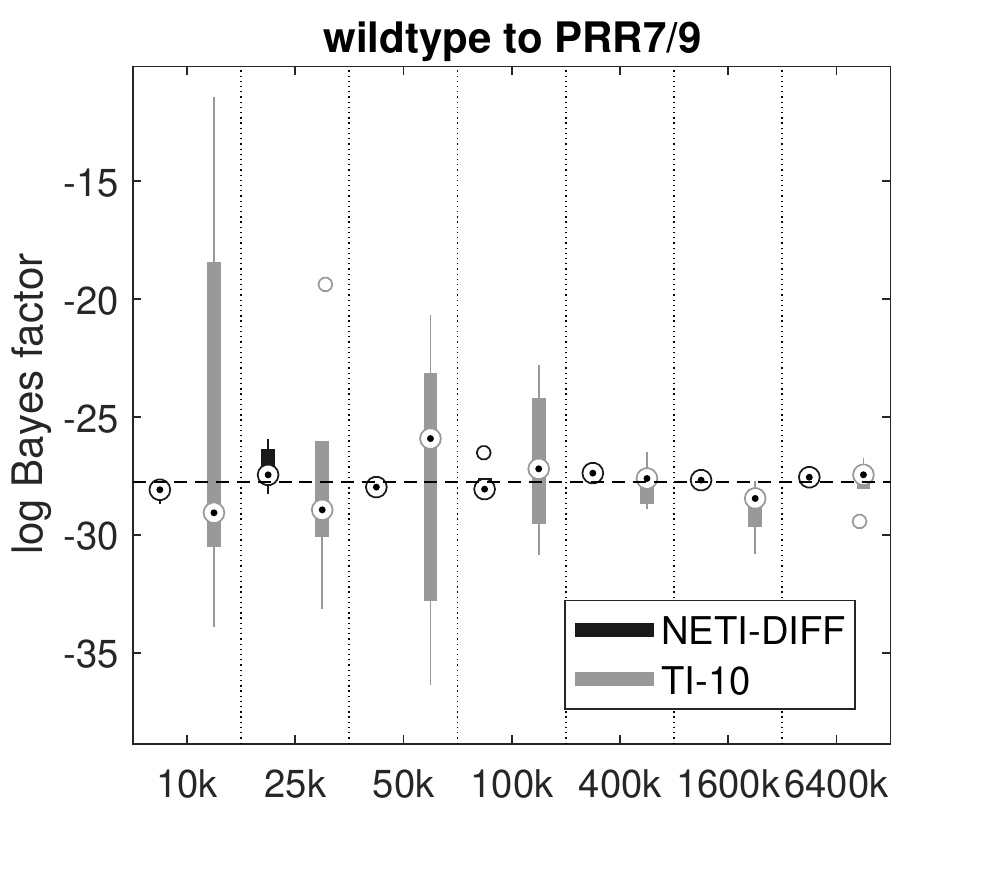} 
	\includegraphics[width=0.33\textwidth]{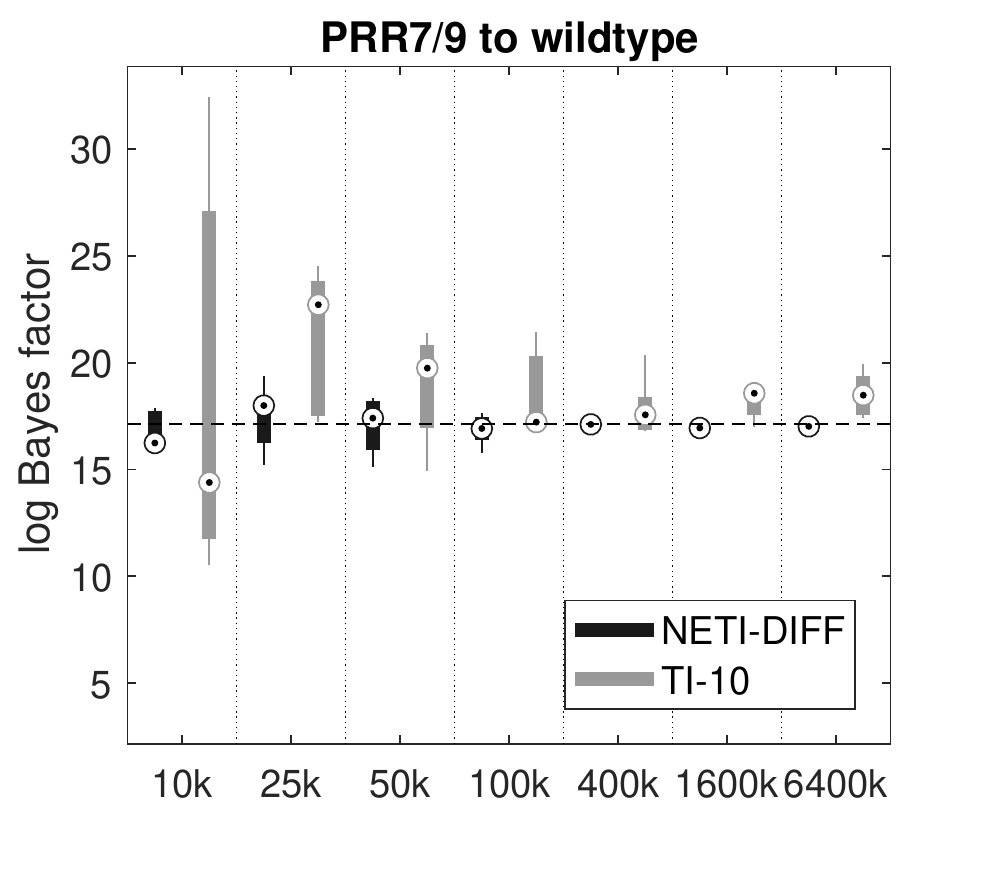} \\
	\subcaption{TI with $K=10$ inverse temperatures.}\label{fig:netidiff_ti_seq_boxplot_r1_a}
	\end{minipage}
	\begin{minipage}[b]{1.0\linewidth}
	\centering 
	\includegraphics[width=0.33\textwidth]{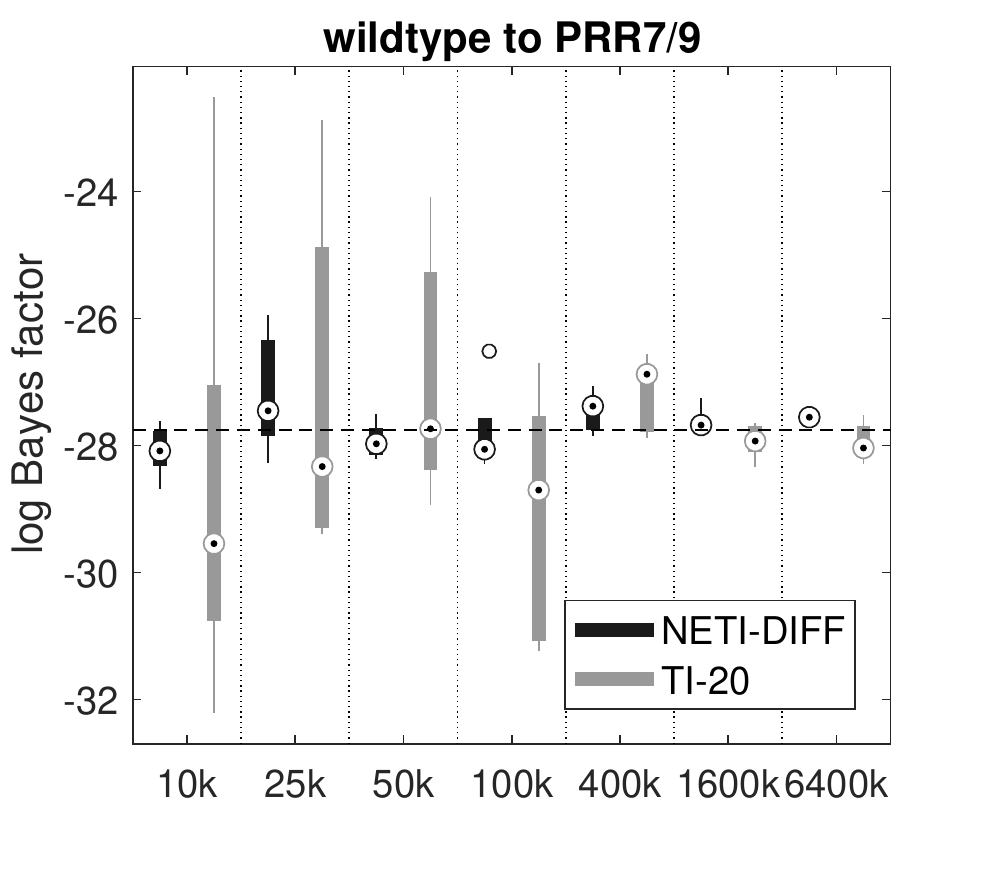} 
	\includegraphics[width=0.33\textwidth]{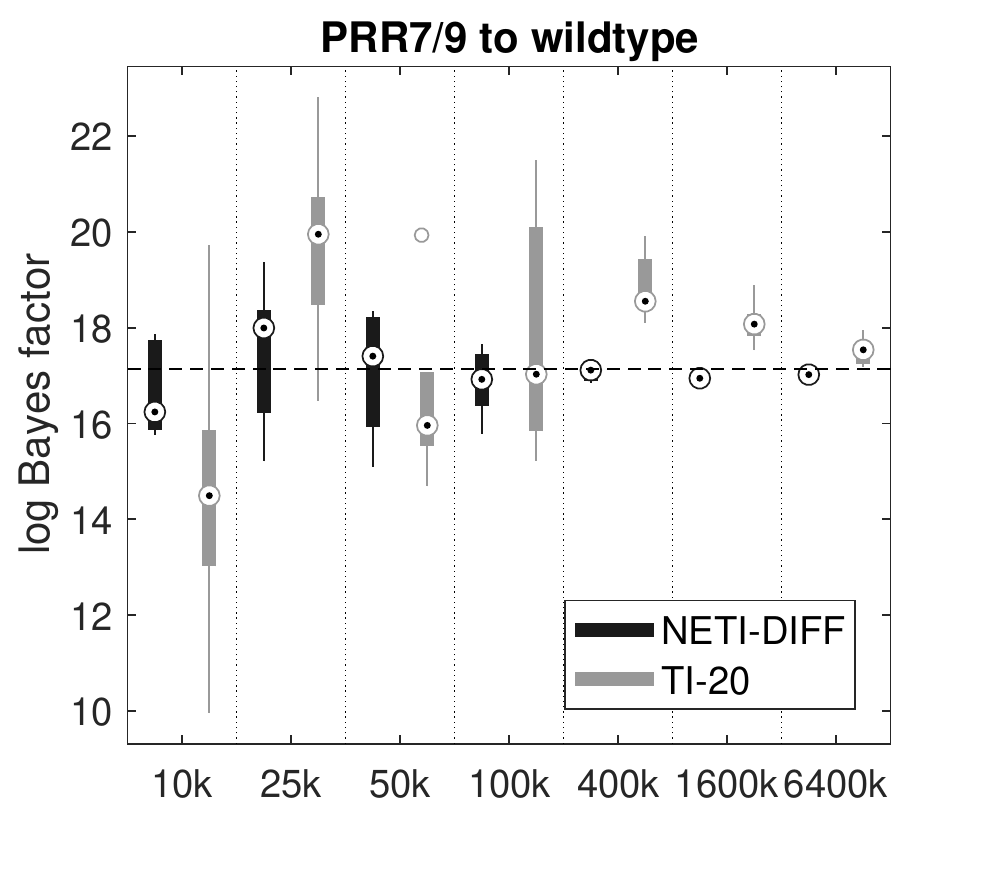} \\
	\subcaption{TI with $K=20$ inverse temperatures.}\label{fig:netidiff_ti_seq_boxplot_r1_b}
	\end{minipage}
	\begin{minipage}[b]{1.0\linewidth}
	\centering 
	\includegraphics[width=0.33\textwidth]{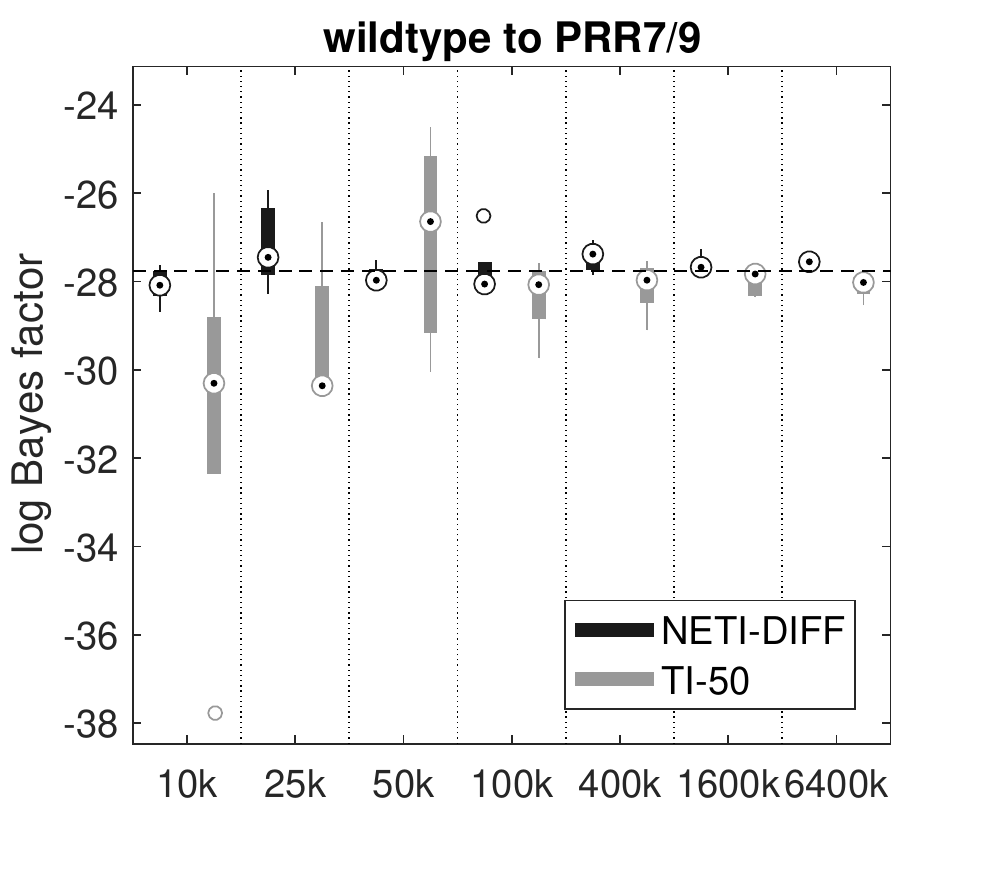} 
	\includegraphics[width=0.33\textwidth]{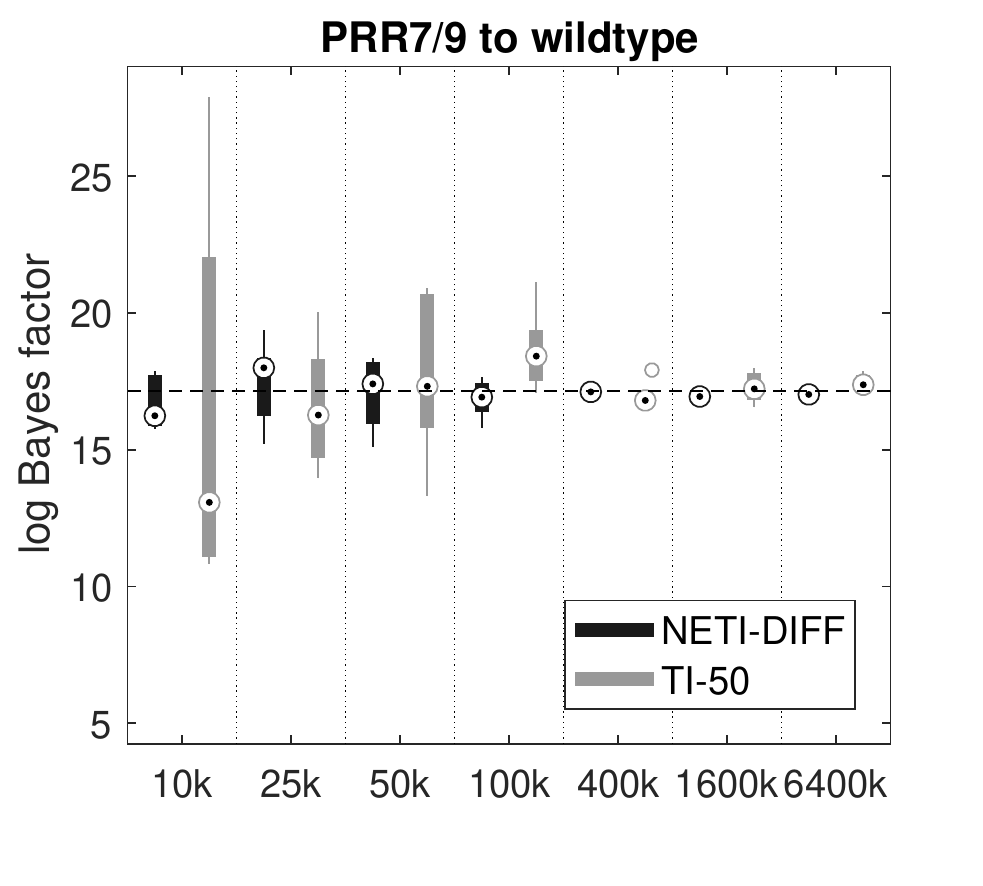} \\
	\subcaption{TI with $K=50$ inverse temperatures.}\label{fig:netidiff_ti_seq_boxplot_r1_c}
	\end{minipage}
	\begin{minipage}[b]{1.0\linewidth}
	\centering 
	\includegraphics[width=0.33\textwidth]{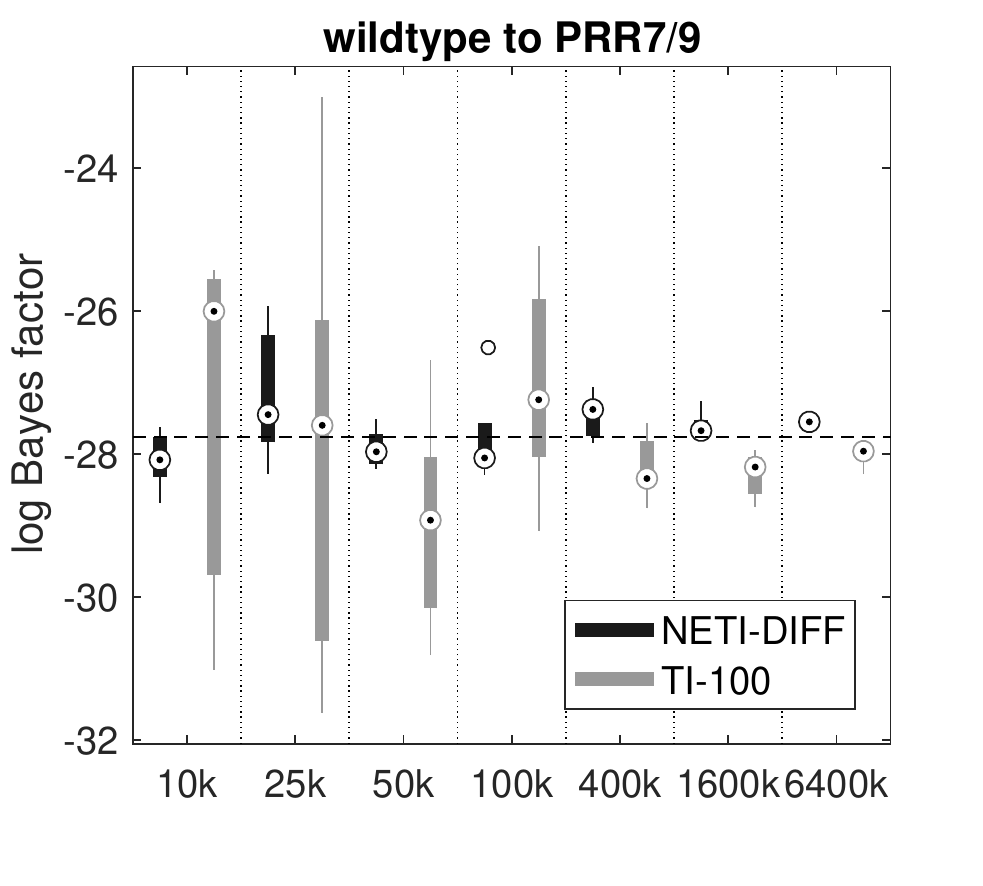} 
	\includegraphics[width=0.33\textwidth]{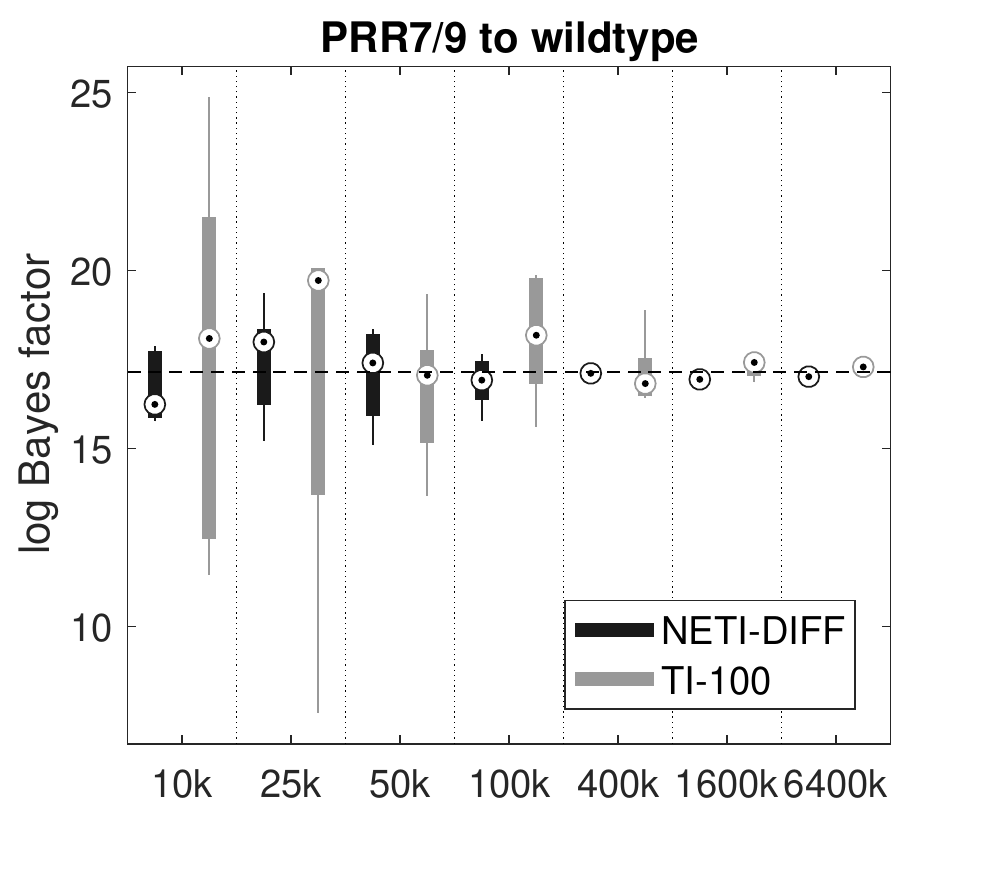} \\
	\subcaption{TI with $K=100$ inverse temperatures.}\label{fig:netidiff_ti_seq_boxplot_r1_d}
	\end{minipage}
\caption{\label{fig:netidiff_ti_seq_boxplot_r1} 
\footnotesize\textbf{Log Bayes factors for the biopathway data: comparison between NETI-DIFF and TI.}
The figure shows the distribution of the log Bayes factor $\log p(\data|\model_2)/p(\data|\model_1)$, where $\model_1$ is the biopathway from Figure~\ref{fig:goldstd1_a} (wildtype), and $\model_2$ is the biopathway from Figure~\ref{fig:goldstd1_b} (PRR7/PRR9 mutant). NETI-DIFF is the same for all four rows. 
Left column: data generated from $\model_1$; negative log Bayes factors select the correct model. Right column: data generated from $\model_2$; positive log Bayes factors select the correct model. The horizontal line shows the `true' value of the log Bayes factor (in the sense defined in the text).
The box plots show distributions over 5 independent MCMC runs. The horizontal axis shows $N_{iter}$, the total number of iterations, ranging from $10k$ to $6400k$.
}
\end{figure*}

\begin{figure*}[tbhp]
	\begin{minipage}[b]{1.0\linewidth}
	\centering 
	\includegraphics[width=0.45\textwidth]{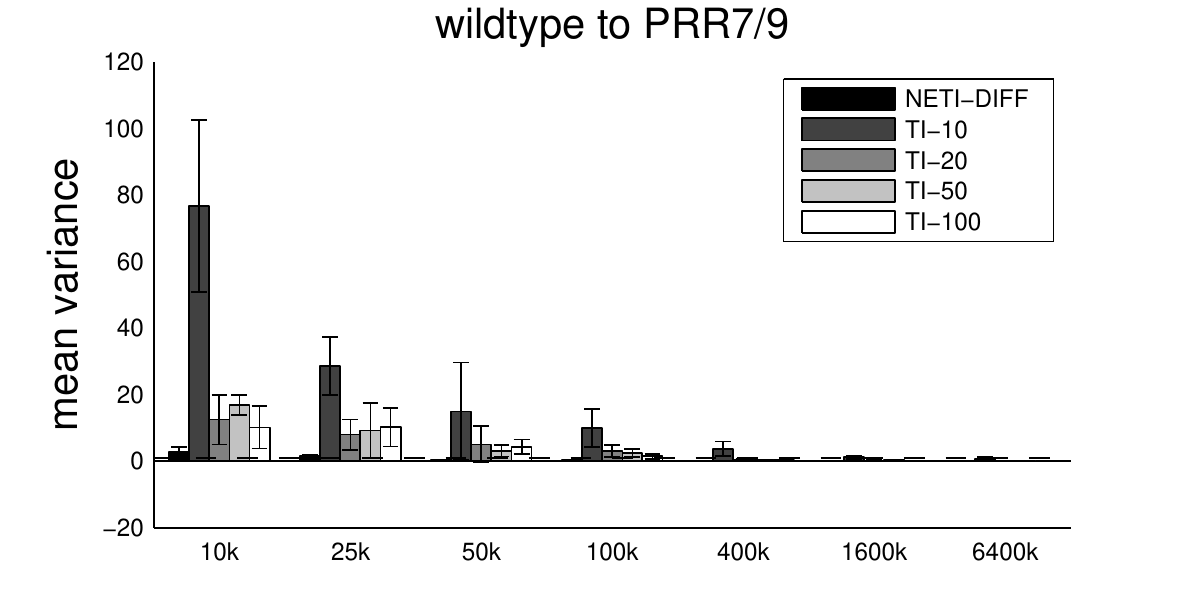}
	\includegraphics[width=0.45\textwidth]{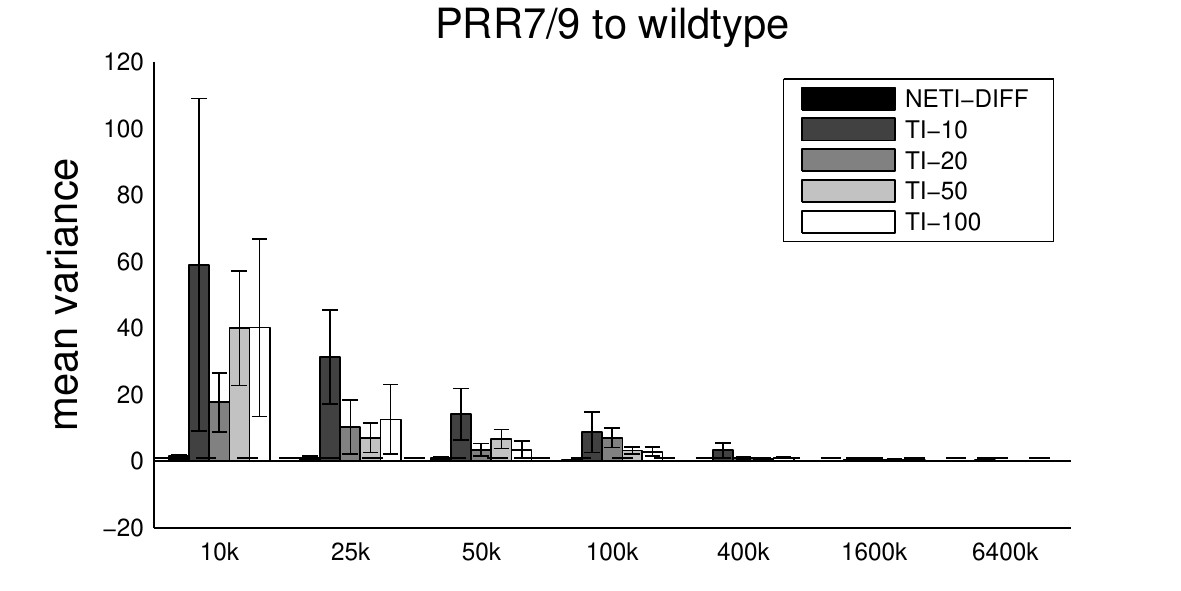}\\
	\subcaption{Mean of variance $\VAR$, defined in \eqname~(\ref{eq:VAR}) with standard deviations (error bars).}\label{fig:netidiff_ti_seq_meanvar_a}
	\end{minipage}
	\begin{minipage}[b]{1.0\linewidth}
	\centering 
	\includegraphics[width=0.45\textwidth]{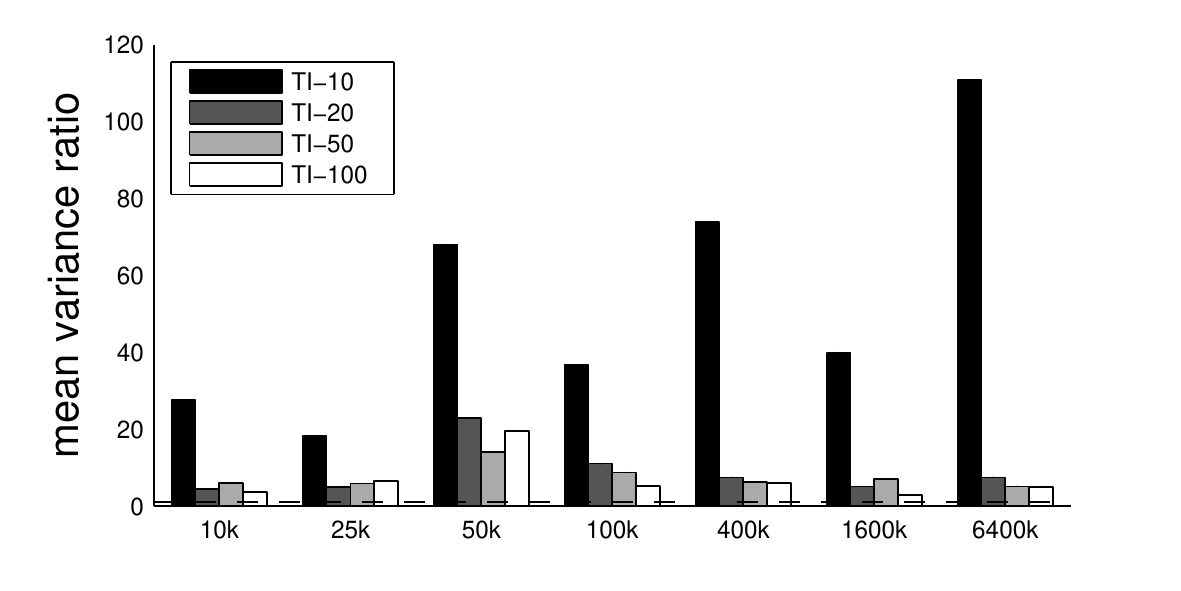}
	\includegraphics[width=0.45\textwidth]{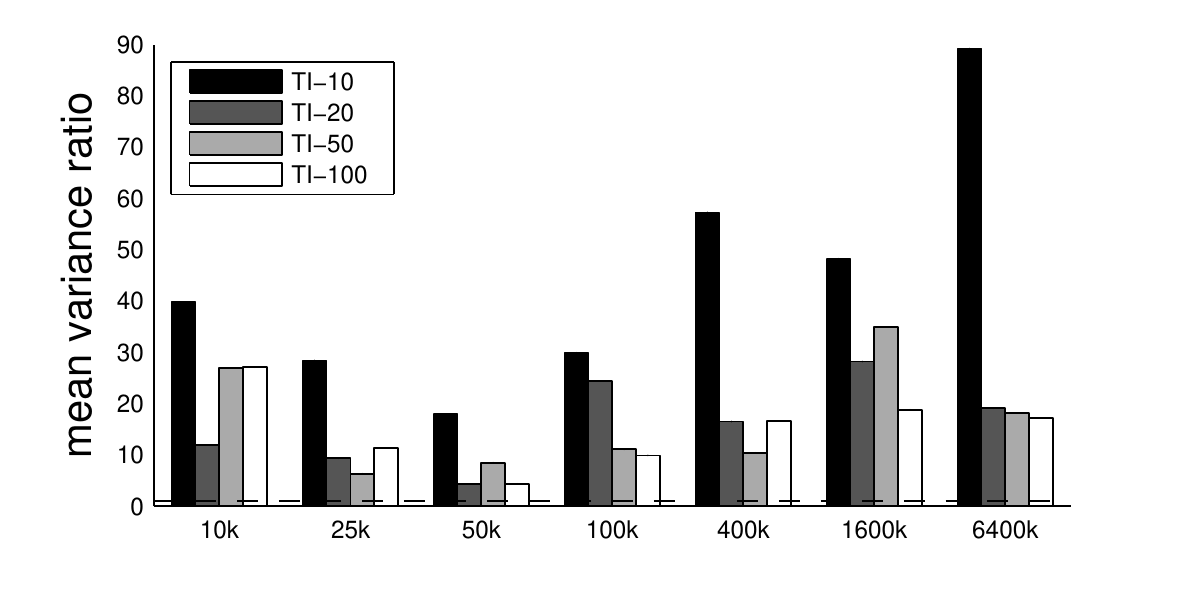} \\
	\subcaption{Ratio of the mean variance obtained with TI, divided by the average variance obtained with NETI-DIFF: $\overline{\VAR}(\mathrm{TI})/\overline{\VAR}(\mathrm{NETI-DIFF})$. Values above 1 indicate a performance improvement with NETI-DIFF over TI.}\label{fig:netidiff_ti_seq_meanvar_b}
	\end{minipage}
	\begin{minipage}[b]{1.0\linewidth}
	\centering 
	\includegraphics[width=0.45\textwidth]{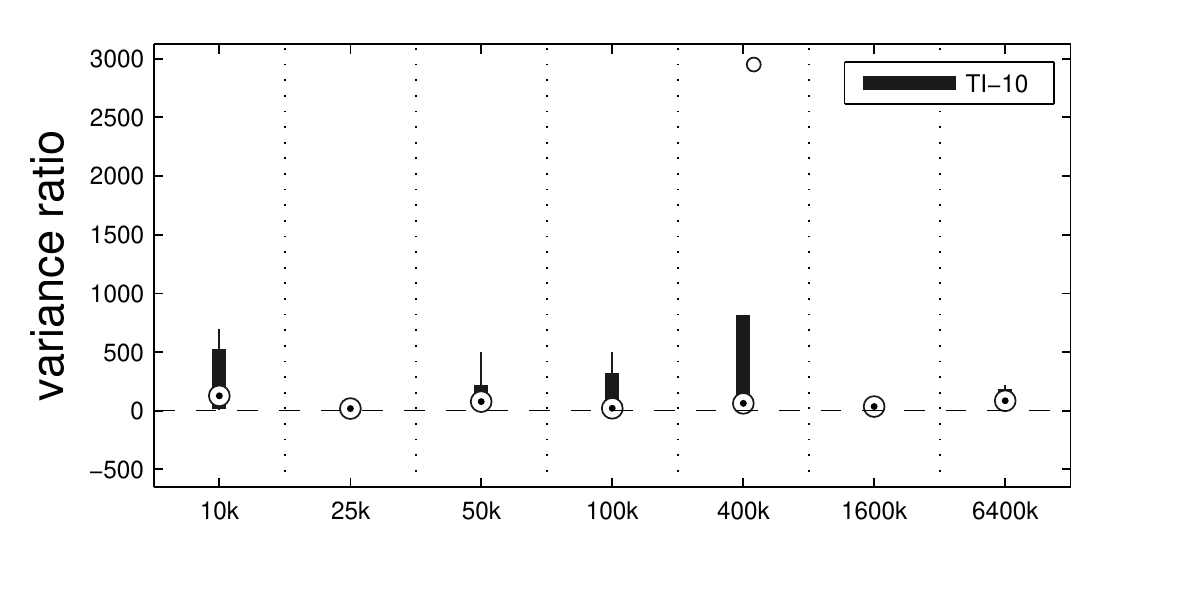}
	\includegraphics[width=0.45\textwidth]{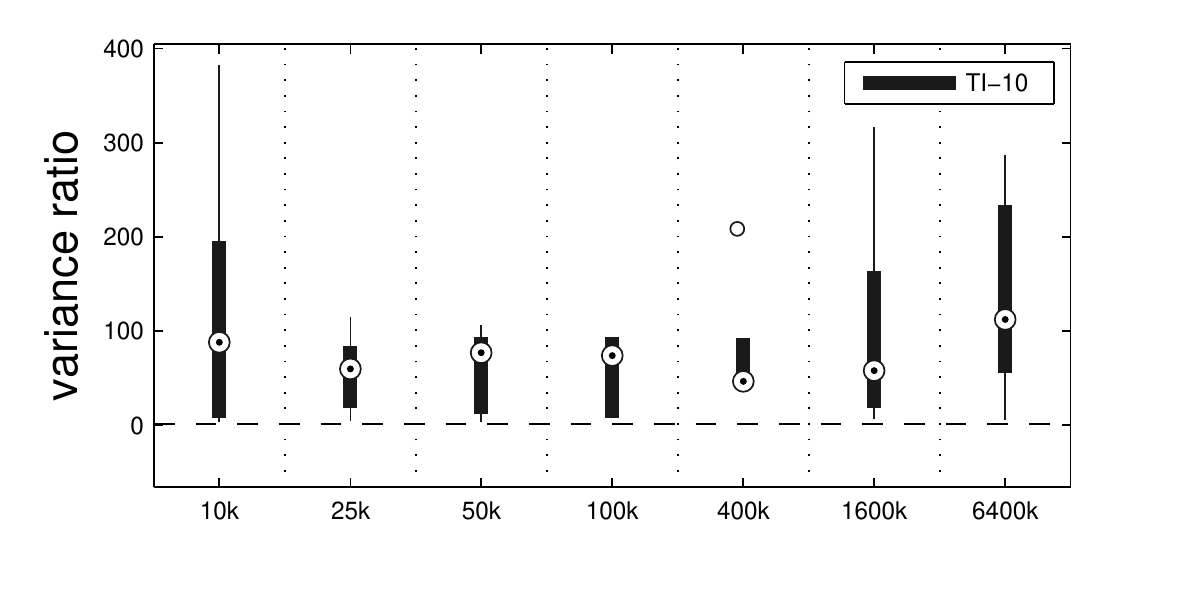} \\
	\subcaption{Distribution of the variance ratios $\VAR(\mathrm{TI})/\VAR(\mathrm{NETI-DIFF})$ for TI with $K=10$ inverse temperatures. Values above 1 indicate a performance improvement with NETI-DIFF over TI.}\label{fig:netidiff_ti_seq_meanvar_c}
	\end{minipage}
	\begin{minipage}[b]{1.0\linewidth}
	\centering 
	\includegraphics[width=0.45\textwidth]{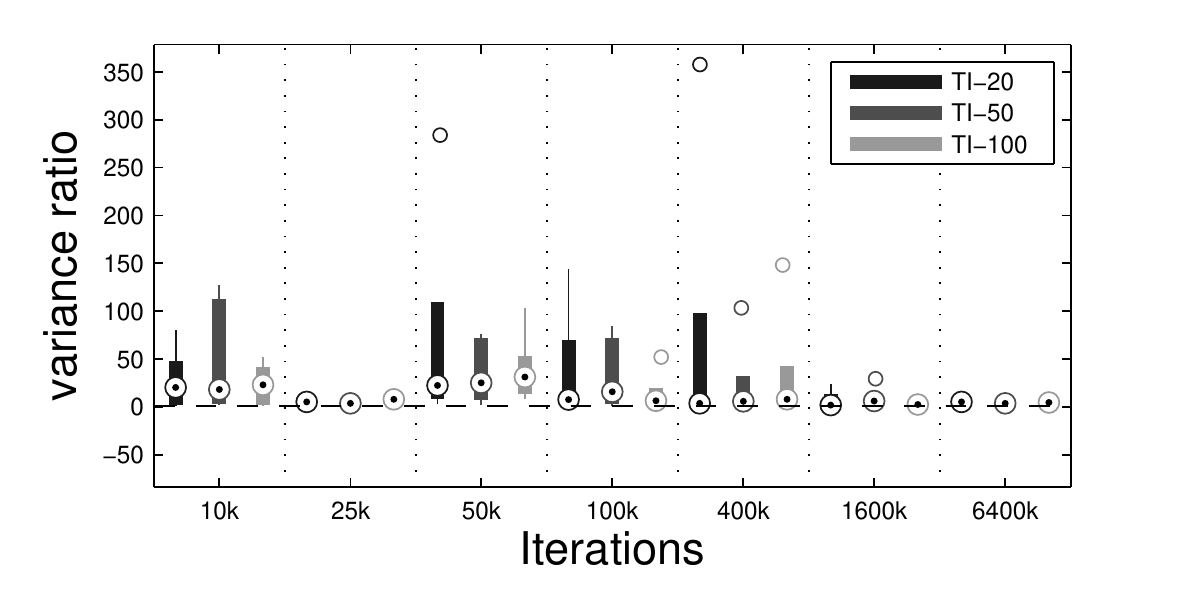}
	\includegraphics[width=0.45\textwidth]{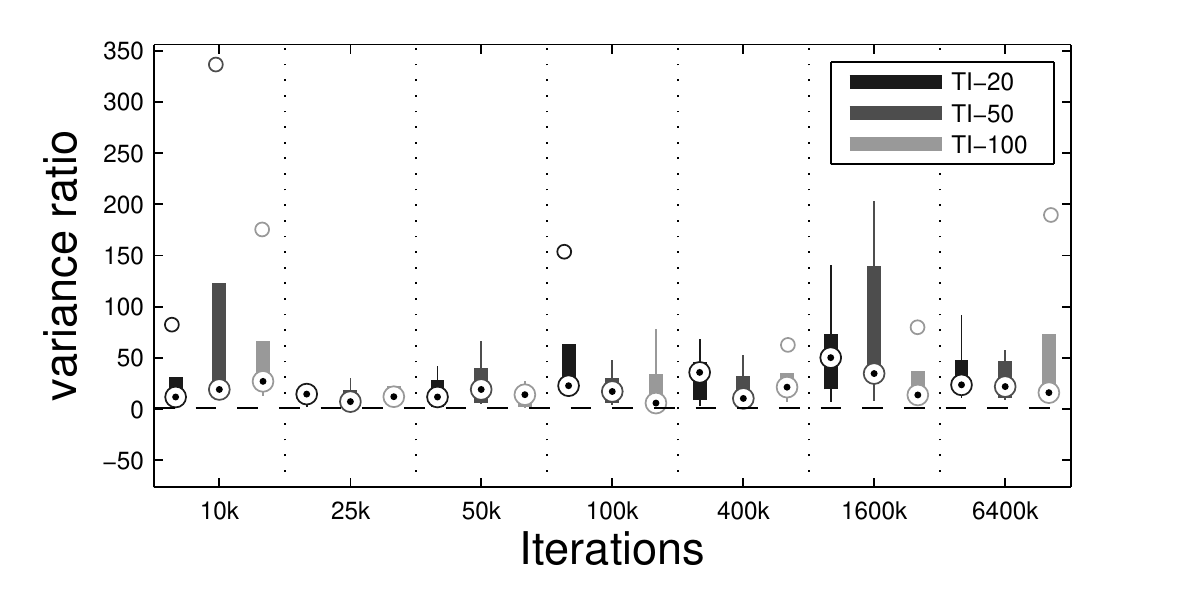} 
	\subcaption{Same as panel \subref{fig:netidiff_ti_seq_meanvar_c}, but for TI with $K=(20,50,100)$ inverse temperatures.}\label{fig:netidiff_ti_seq_meanvar_d}
	\end{minipage}
\caption{\label{fig:netidiff_ti_seq_meanvar} 
\textbf{Variance of log Bayes factor estimation: comparison between NET-DIFF and TI on the biopathway data.} The variance measures are obtained from five repeated simulations of the same data set. The mean measures correspond to the average from five different data instantiations, obtained from the biopathway of Figure~\ref{fig:goldstd1_a} (wildtype, left column), and from the biopathway of Figure~\ref{fig:goldstd1_b} (PRR7/PRR9 mutant, right column).}
\end{figure*}

\begin{figure}[tbph]
	\begin{minipage}[b]{0.5\linewidth}
	\centering 
\includegraphics[width=0.9\textwidth]{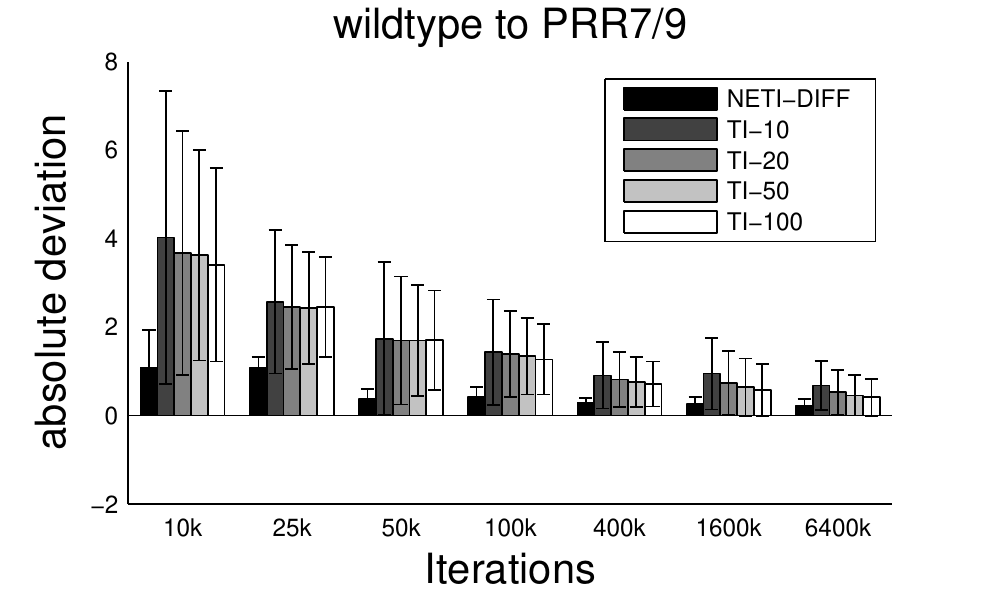} 
	\subcaption{}\label{fig:netidiff_ti_seq_biasvar_a}
	\end{minipage}
	\begin{minipage}[b]{0.5\linewidth}
	\centering 
	\includegraphics[width=0.9\textwidth]{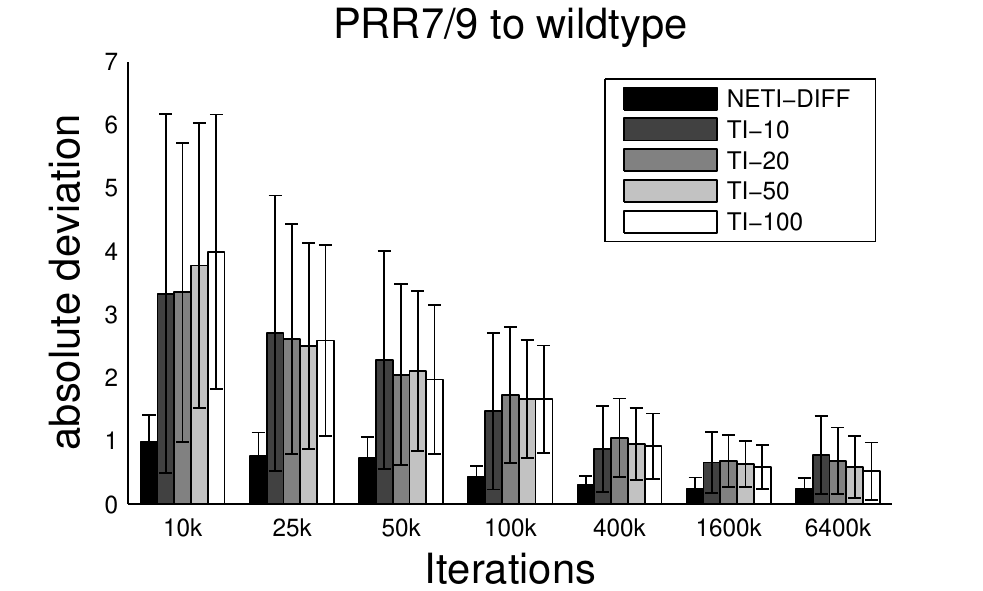}
	\subcaption{}\label{fig:netidiff_ti_seq_biasvar_b}
	\end{minipage}
\caption{\label{fig:netidiff_ti_seq_biasvar} 
\textbf{Mean absolute deviation of log Bayes factor estimation: comparison between NETI-DIFF and TI on the biopathway data.} 
Simulations were repeated for 5 independent data instantiations, obtained from the biopathway of Figure~\ref{fig:goldstd1_a} (wild type), and from the biopathway of Figure~\ref{fig:goldstd1_b} (PRR7/PRR9 mutant). 
Shown is the mean absolute deviation $\MAD$, defined in \eqname~(\ref{eq:MAD}). Vertical bar height: average over the five data instantiations. Error bars: standard deviation. The horizontal axis shows the total number of iterations $N_{iter}$. For each value of $N_{iter}$, the leftmost bar represents NETI-DIFF. The other bars with different grey shadings represent TI with different numbers of inverse temperatures, ranging from 10 to 100.
}
\end{figure}

% ========================================
\section{Results}
\label{sec:results}
% ========================================

In this section, we compare the efficiency and accuracy of three algorithms: standard thermodynamic integration (TI-standard) and optimal thermodynamic integration (TI-optimal) for computing the log marginal likelihood, and the proposed non-equilibrium thermodynamic integration for directly targeting the difference of the log marginal likelihood (NETI-DIFF).  

In TI-standard we  
compute,
based on \eqname~(\ref{eq:exPowerPost}),
 the expectation of the log likelihood  w.r.t. the power posterior, 
\begin{math}
\mathbb{E}_{\invT}[
\log p(\data|\bftheta,\model)]
\end{math},
for a set of a priori fixed inverse temperatures $\{\invT_i\}$, $i=1,\ldots,K$, spaced according to the power law of \eqname~(\ref{eq:powerLaw}). Following \cite{frielTIcorrection} we have set $K \in \{10,20,50,100\}$ and $\alpha=5$ in \eqname~(\ref{eq:powerLaw}). The log marginal likelihood is computed with the trapezoid rule (\eqname~\ref{eq:TI_trapezoid_standard}).

TI-optimal uses the two improvements proposed in \cite{frielTIcorrection}: the log marginal likelihood is computed with the improved numerical integration (\eqname~\ref{eq:TI_trapezoid_correction}), and the inverse temperatures are set iteratively according to an optimality criterion that aims to minimise the expected uncertainty; see  \cite{frielTIcorrection} for details.\footnote{Note that there is a typo in \eqname~(17) of \cite{frielTIcorrection}; $t = \frac{\hat{f}_{k+1}  - \hat{f}_{k}  + \hat{f}_{k}\hat{V}_{k} -\hat{f}_{k+1}\hat{V}_{k+1} }{\hat{V}_{k}-\hat{V}_{k+1}}$ must read: \\ 
	$t = \frac{\hat{f}_{k+1}  - \hat{f}_{k}  + t_{k}\hat{V}_{k} -t_{k+1}\hat{V}_{k+1} }{\hat{V}_{k}-\hat{V}_{k+1}}.$} 

Finally, NETI-DIFF is the algorithm proposed in the present article.

For each inverse temperature $\invT$ in TI-standard and TI-optimal, we discarded the first $20\%$ of the MCMC steps as burn-in 
(following \shortciteA{frielTIcorrection}). 
For NETI-DIFF, we discarded the first 1000 MCMC steps with the inverse temperature kept fixed at $\invT=0$, as burn-in.\footnote{Due to the non-equilibrium nature of NETI, not discarding any burn-in phase made little difference to the results.}
We recorded the total number of non-burn-in MCMC steps for all three algorithms, $N_{iter}$. As discussed in Appendix~\ref{sec:compconv} this is a measure of the total computational complexity. 
% }% endRed

We repeated the MCMC simulations $N_{simu}=5$ times from different initialisations.
Let $\mathcal{B}_i$ denote the log Bayes factor obtained from the $i$th MCMC simulation, 
and $\mathcal{B}_{true}$ the `true' log Bayes factor. For the Bayesian linear regression models 
applied to the Radiata and Radiocarbon data, a closed-form expression for $\mathcal{B}_{true}$
is available. For the Bayesian logistic regression model applied to the Pima Indians data,
and the hierarchical Bayesian model from Figure~\ref{fig:graphModelCheMA} for 
biopathway data, the log Bayes factor is not analytically tractable, and $\mathcal{B}_{true}$ 
was obtained from a very long  simulation, as in \cite{frielTIcorrection}.
We assessed the intrinsic estimation \emph{uncertainty} in terms of the variance:
\begin{equation} 
\label{eq:VAR}
\VAR  =  \frac{1}{N_{simu-1}} \sum_{i=1}^{N_{simu}} [\mathcal{B}_i - \overline{\mathcal{B}}]^2, \quad\quad
\overline{\mathcal{B}} =  \frac{1}{N_{simu}} \sum_{i=1}^{N_{simu}} \mathcal{B}_i
\end{equation}
and the \emph{accuracy} in terms of the mean absolute error:
\begin{equation}
\label{eq:MAD}
\MAD   =   \frac{1}{N_{simu}} \sum_{i=1}^{N_{simu}} |\mathcal{B}_i -  \mathcal{B}_{true}|
\end{equation}

\subsection{Radiata pine and Pima Indians}
\label{sec:RadiataPimaResults}
We start our empirical evaluation study with the analysis of the Radiate pine data (\Sect~\ref{sec:radiata}) and the Pima Indians data (\Sect~\ref{sec:pima}). These two data sets have been used in the literature before for the evaluation of the TI method proposed by \shortciteA{frielTIcorrection}, and in both cases the goal is to estimate the Bayes factor between two competing Bayesian regression models. For the Radiata pine data we compare two non-nested linear regression models. For the Pima Indians data we compare two logistic regression models, where the first model, $\mathcal{M}_1$, is nested in the second, $\mathcal{M}_2$ . We apply the NETI-DIFF approach with a sigmoid inverse temperature ladder, defined in \Sect~\ref{sec:Tladder}, and we instantiate NETI-DIFF such that in both applications the transition path runs from the first model, $\mathcal{M}_1$ ($\invT=0$), to the second, $\mathcal{M}_2$ ($\invT=1$). For the Pima Indians data this is the natural path, as $\mathcal{M}_1$ is nested within $\mathcal{M}_2$.

Figures~\ref{fig:RADIATA_1} and \ref{fig:PIMA_1} show the average absolute deviations (\eqname~\ref{eq:MAD}) between the analytically computed log Bayes factors and the estimated log Bayes factors for different total iteration numbers $N_{iter}$. 
Figure~\ref{fig:VARIANCES_1} compares the variance of the log Bayes factor estimates for the three different methods: TI-standard, TI-optimal, and NETI-DIFF. Figure~\ref{fig:VARIANCES_2} shows ratios of the variances obtained with TI-optimal and NETI-DIFF.

For the Radiata data, NETI-DIFF only achieves a slight reduction in the absolute deviation (Figure~\ref{fig:RADIATA_1}) and the variance 
(Figures~\ref{fig:VARIANCES_1}--\ref{fig:VARIANCES_2}) for the lowest number of iterations, $N_{iter}=64k$; otherwise NETI-DIFF and TI-optimal are on a par.
Note that the two alternative linear regression models applied to the Radiata data only share the intercept, while their sets of covariables are disjunct. 
This lack of model overlap presents the least favourable scenario for NETI-DIFF, and our results confirm that there is little room for improvement over standard TI. 

For the Pima Indians data, NETI-DIFF achieves a significant reduction in the absolute deviation (Figure~\ref{fig:PIMA_1}) and the variance 
(Figures~\ref{fig:VARIANCES_1}--\ref{fig:VARIANCES_2}) and clearly outperforms both TI methods: TI-standard and TI-optimal. The variance reduction ranges between ratios of 5 and 50. 
As opposed to the models applied to the Radiata data, the alternative logistic regression models applied to the Pima Indians data are nested, with the parameters of the less complex model forming a subset of those of the more complex one. Our results demonstrate that in this scenario, the new thermodynamic integration path of NETI-DIFF has potential for significant improvement over the established TI methods. 

We also investigated the effect of the inverse temperature ladder ('sigmoid' vs. 'power 5') and the starting point ($\mathcal{M}_1$ vs. $\mathcal{M}_2$). To this end, we systematically applied the proposed NETI-DIFF approach with all four combinations (two inverse temperature ladders times two starting points) to the two data sets: Radiata pine and Pima Indians. The  results can be found in Figure~\ref{fig:POWER_SIGMOID}. First, consider the Pima Indians data, where the two alternative models are nested, and the power inverse temperature scheme of \eqname~(\ref{eq:powerLaw}) has been applied.  There is a clear advantage of starting the thermodynamic integration at the less complex model over starting at the more complex model: the absolute errors are significantly higher in the latter case. This is not surprising. It is well known from standard TI for computing marginal likelihoods that 
for the power law of \eqname~(\ref{eq:powerLaw}),
the optimal transition path is from the prior to the posterior, with the majority of the inverse temperature points at the prior end. Applying this principle to NETI-DIFF, starting the transition path for the differential parameters (i.e. the parameters that are \emph{only} in the more complex model) at the prior, implies that the overall inverse temperature transition path has to lead from the less complex to the more complex model, in confirmation of our findings.  Interestingly, for the sigmoid temperature ladder from \Sect~\ref{sec:Tladder}, the difference between the two directions is substantially reduced, which is a natural consequence of the symmetry inherent in this scheme. There is no significant performance difference between the sigmoid and the power law inverse temperature paths when the models are nested (Pima Indians data, top row in Figure~\ref{fig:POWER_SIGMOID}). For the Radiata pine data on the other hand
(bottom row in Figure~\ref{fig:POWER_SIGMOID}), where the alternative models are not nested, the power law of \eqname~(\ref{eq:powerLaw}) is intrinsically suboptimal\footnote{
This is a consequence of the fact that due to the non-nested structure of the models, there is always a parameter for which the transition effectively  moves from the posterior to the prior, rendering the power law of \eqname~(\ref{eq:powerLaw}) suboptimal.
}, and the sigmoidal inverse temperature path of \Sect~\ref{sec:Tladder} is to be preferred.  

\subsection{Radiocarbon dating}
\label{sec:carbonResults}

Next, we consider model selection amongst different polynomial orders for polynomial regression on the Radiocarbon data. Since this is a linear model, the log Bayes factor is known and can be used for evaluating the accuracy of the different thermodynamic integration schemes. Besides comparing the proposed NETI-DIFF scheme with the established TI methods, we investigate the influence of the inverse temperature ladder and the transition path.
Due to the comparatively low computational costs, we have increased the number of discretisation points from $K \in \{10,20,50,100\}$ to $K \in \{20,50,100,200\}$.

Figure~\ref{fig:RADIOC_1} shows the absolute error (see \eqname~\ref{eq:MAD}) for NETI-DIFF and the better of the two established TI methods: TI-optimal. The task is to compute the log Bayes factor for the pairwise comparison of various polynomial orders, as indicated by the horizontal axis of each panel. It turns out that for TI-optimal, the accuracy of the estimate deteriorates with increasing difference of the model orders (black bars in the top panels of Figure~\ref{fig:RADIOC_1}), while NETI-DIFF is unaffected by model choice\footnote{NETI-DIFF is unaffected because it does not depend on the $K$ number of discretization points of the integral as the classical TI does. Instead, it continuously transforms one model into the other.}. In addition, NETI-DIFF considerably outperforms TI-optimal for the lower iteration numbers, as again seen from the top row in Figure~\ref{fig:RADIOC_1}.

The right column of Figure~\ref{fig:VARIANCES_1} compares the variances between NETI-DIFF and TI-optimal, and the right column of 
Figure~\ref{fig:VARIANCES_2} shows the corresponding variance ratios. 
It is seen that NETI-DIFF consistently outperforms TI-optimal, with the variance ratios ranging between 5 and 2000. It appears that for low iteration numbers $N_{iter}$,  the improvement is most pronounced when the alternative models differ substantially (polynomial order 1 versus 9), while for high iteration numbers $N_{iter}$, the clearest improvement is achieved when the alternative models are more similar (polynomial orders 4 versus 6).

The left panel of Figure~\ref{fig:RADIOC_2} compares the two inverse temperature ladders:  the power law of \eqname~(\ref{eq:powerLaw}) versus the sigmoidal form of \Sect~\ref{sec:Tladder}. Since the models are nested, we would expect the polynomial scheme to perform well, like for the Pima Indians data discussed above. Interestingly, the sigmoidal scheme achieves a better stabilization of the results w.r.t. model order, and a slightly better performance for the largest difference between the polynomial orders of the two alternative models considered. To shed more light on this trend, we have investigated the evolution of the standard deviation of the thermodynamic integral up to a given inverse temperature $\invT$. The results are shown in Figure~\ref{fig:RADIOC_3}. While the power law indeed achieves a lower standard deviation than the sigmoidal scheme at the low-inverse-temperature end (near the low-complex model), it contributes a larger proportion to the standard deviation at the high-inverse-temperature end (near the high-complex model). This suggests that the sparsity of inverse temperatures at the high-inverse-temperature end can be counterproductive due to insufficient sample size.

We finally investigated different model transition paths, with a comparison of three alternative schemes: (1) a staggered path from the low-complexity to the high-complexity model via a series of all intermediate models; (2) a transition via one intermediate model of medium complexity; and (3) a direct transition. The results are shown in the right panel of Figure~\ref{fig:RADIOC_2}. The differences are small without a clear trend. This suggests that NETI-DIFF is remarkably robust w.r.t. the choice of the model transition path.

\subsection{Biopathway}
\label{sec:BiopepaResults}
For the biopathway example, we considered two types of data. The first 
type was obtained from the wild type gene regulatory network shown in Figure~\ref{fig:goldstd1_a}; the second type was obtained from the mutant network shown in Figure~\ref{fig:goldstd1_b}. 
As we do not have a closed-form expression of the log Bayes factor
we chose, as a proxy, the average of the log Bayes factors 
obtained with the longest TI and NETI-DIFF simulations, 
which tended to be in reasonably good agreement.
Table~\ref{tab:verylongti_biopepa} shows the values of the log 
Bayes factor thus obtained, which confirms that  Bayesian model selection 
based on the hierarchical model of Figure~\ref{fig:graphModelCheMA} 
consistently identifies the true gene network.

\begin{table}
\begin{center}
\begin{tabular}{l c c c c c}
\hline
Data instance:     &  1 & 2 & 3 & 4 & 5 \\
\hline
\textit{\underline{wildtype}} $\rightarrow$ \textit{PRR7/9} & -27.8 & 
-29.3 & -26.1 & -25.4 & -28.4 \\
\textit{\underline{PRR7/9}} $\rightarrow$ \textit{wildtype} & 17.1  & 
9.8  & 14.4  & 5.8   & 4.0 \\
\hline
\end{tabular}
\end{center}
\caption{\label{tab:verylongti_biopepa}\textbf{Log Bayes factor for the 
biopathway data.} 
The table shows the log Bayes factor $\log p(\data|\model_2)/p(\data|\model_1)$, 
where $\model_1$ is the biopathway from Figure~\ref{fig:goldstd1_a} 
(wildtype), and $\model_2$ is the biopathway from Figure~\ref{fig:goldstd1_b} (PRR7/PRR9 mutant).
Top row: data obtained from $\model_1$; negative log Bayes factors select the true model.
Bottom row: data obtained from $\model_2$; positive log Bayes factors select the true model.
The five columns show values for different independent data instantiations.
The log Bayes factors were obtained by averaging the values obtained with NETI-DIFF and TI
for the largest number of iterations $N_{iter}$.
}
\end{table}

In a preliminary study, we compared the two inverse temperature ladders for NETI-DIFF: 
power law (see \eqname~(\ref{eq:powerLaw})) with power 5, as in \cite{frielTIcorrection}, versus the sigmoid transfer function of \Sect~\ref{sec:Tladder}.
We repeated the simulations on the 5 data sets of Table~\ref{tab:verylongti_biopepa}. From these data sets, we computed the 
mean of the variance $\VAR$, \eqname~(\ref{eq:VAR}), and the mean absolute error $\MAD$, \eqname~(\ref{eq:MAD}).
The results are shown in Figure~\ref{fig:netidiff_pow_sigm_biasvar}.
The trend is not as clear as in Figure~\ref{fig:POWER_SIGMOID}. However, the sigmoid inverse temperature ladder achieves 
more often a performance improvement over the power law (in terms of lower mean absolute error $\MAD$ and average 
variance $\VAR$) than the other way round, and we therefore adopted it for all subsequent studies.

The main question of interest is to compare TI and NETI-DIFF with respect to 
accuracy, estimation uncertainty and computational efficiency.
To improve the clarity of the presentation, we
only show the comparison between NETI-DIFF and TI-optimal, i.e. the TI scheme
with the improvements proposed by \shortciteA{frielTIcorrection}.
In what follows, we refer to "TI-optimal" simply as "TI".
The simulations were repeated for different total iteration 
lengths, $N_{iter}$, ranging from $N_{iter}=10,000$ to 
$N_{iter}=6,400,000$ MCMC steps. We repeated TI for different numbers of 
inverse temperatures, $K$, ranging from $K=10$ to 
$K=100$ (the same values as used in \cite{frielTIcorrection}). 

Figure~\ref{fig:netidiff_ti_seq_boxplot_r1} shows the distribution of estimated log Bayes factors obtained from $N_{simu}=5$ independent MCMC runs\footnote{These results were obtained from the first two data sets in the first column of Table~\ref{tab:verylongti_biopepa}.}.
The two columns refer to the different data types (from the wild type network, left column, and the mutant network, right column), and the rows (Panels~\ref{fig:netidiff_ti_seq_boxplot_r1_a}-\ref{fig:netidiff_ti_seq_boxplot_r1_d}) to the number of inverse temperatures used for TI (from $K=10$ to $K=100$; note that NETI-DIFF is unaffected by that choice).
The horizontal dashed lines show the `true' value, as described above.
As expected, the distribution width tends to decrease with increasing computational costs, $N_{iter}$, and for the highest value, TI and NETI-DIFF tend to be in close agreement, with distributions tightly focused on the `true' values.
However, for lower computational costs, $N_{iter} \leq 400k$, bias and uncertainty tend to be considerably lower for NETI-DIFF than for TI, irrespective of the number of inverse temperatures used for TI.

For a more systematic investigation, we repeated the MCMC simulations on ten independent data instantiations, for the ten data sets used in Table~\ref{tab:verylongti_biopepa}.  
Five data sets were obtained  from the biopathway of Figure~\ref{fig:goldstd1_a} (wildtype), and 
five data sets were obtained from the biopathway of Figure~\ref{fig:goldstd1_b} (PRR7/PRR9 mutant). 
For each data set, we computed the mean absolute deviation $\MAD$, defined in \eqname~(\ref{eq:MAD}), and the 
variance $\VAR$, as defined in \eqname~(\ref{eq:VAR}).

The top row in Figure~\ref{fig:netidiff_ti_seq_meanvar} shows the average variance $\overline{\VAR}$, averaged over all data instantiations. The second row shows the ratio of the average variance obtained with TI, divided by the average variance obtained with NETI-DIFF, averaged over all five data instantiations: 
$\overline{\VAR}(\mathrm{TI})/\overline{\VAR}(\mathrm{NETI-DIFF})$.
The third and fourth rows show the distribution of the variance ratios $\VAR(\mathrm{TI})/\VAR(\mathrm{NETI-DIFF})$ over the five different data instantiations, for different numbers of inverse temperatures (for TI), and different total interation numbers $N_{iter}$. For all ratios, values above 1 indicate a performance improvement with NETI-DIFF over TI. Our results indicate that NETI-DIFF consistently achieves a considerable variance reduction over TI. This reduction is particularly pronounced for small numbers of inverse temperatures, where it reaches up to three orders of magnitude. However, even for the highest number of inverse temperatures the variance reduction NETI-DIFF achieves over TI still varies between one and two orders of magnitude. This clear reduction in estimation uncertainty
is matched by a consistent reduction in the estimation error, as quantified in terms of $\MAD$ and shown in Figure~\ref{fig:netidiff_ti_seq_biasvar}. The reduction becomes stronger with decreasing iteration numbers $N_{iter}$ and decreasing numbers of inverse temperatures, which indicates that the performance improvement of NETI-DIFF over TI is particularly relevant in the regime of limited computational resources.

\section{Discussion}
\label{sec:discussion}

The objective of our work has been the direct targeting of the log Bayes factor via a modified thermodynamic integration path. This has been motivated by statistical physics, where the computation of a reaction free energy (mathematically equivalent to the log Bayes factor) is computationally more efficient than the computation of the difference of standard free energies (equivalent to the difference of log marginal likelihoods).
The modified transition path directly connects the posterior distributions of the two models involved. In this way, the high variance prior regime is avoided. We have carried out a comparative evaluation with the state-of-the-art TI method of \shortciteA{frielTIcorrection}. Our study confirms that a substantial variance reduction can be achieved when the models to be compared are nested. There is little room for improvement when comparing non-nested models with non-overlapping parameter sets.  However, even in this least favourable case, the performance achieved with the proposed method, referred to as NET-DIFF in the present manuscript, is still on a par with established TI methods. 
For inference in a complex systems described by coupled nonlinear differential equations (biopathway), we found that NETI-DIFF reduces the variance by up to two orders of magnitude over state-of-the-art TI methods. Our work has also revealed that NETI-DIFF achieves a 
considerable performance stabilisation with respect to a variation of the parameter prior.

When the task is model selection out of a set of cardinality $m$,
carrying out direct pairwise comparisons 
is of computational complexity $m^2$ and may not be
viable in practice. However, rather than reverting to the
standard TI scheme and computing the marginal likelihoods
\begin{equation}
p(\data|\model_1),\ldots,p(\data|\model_m)
\end{equation}
it appears more sensible to 
compute
the Bayes factors 
\begin{equation}
\label{the_individual_BFs}
\frac{p(\data|\model_1)}{p(\data|\model_0)},\ldots,\frac{p(\data|\model_m)}{p(\data|\model_0)}
\end{equation}
where $\model_0$ is a typical or representative model chosen from the 
set of models compared. 
The results for the Radiocarbon data, reported in Section~\ref{sec:carbonResults}, have demonstrated a remarkable robustness of the  proposed method w.r.t. a variation of the model transition path, meaning that there is no significant difference in efficiency and accuracy between the direct computation of
\begin{math}
\log \frac{p(\data|\model_1)}{p(\data|\model_2)},
\end{math}
and the indirect computation via 
\begin{math}
\log \frac{p(\data|\model_1)}{p(\data|\model_0)}
\end{math}
and 
\begin{math}
\log \frac{p(\data|\model_2)}{p(\data|\model_0)}
\end{math}.
This suggests that $1$-out-of-$m$ model selection can also be improved with the method we have proposed.
It is beyond the scope of this article to investigate this conjecture at greater depth, but it appears plausible 
that targeting Bayes factors
along an annealing path starting from a reference posterior distribution
associated with a reference model should
give smaller posterior variance than conventionally targeting marginal likelihoods
along an annealing path starting from the prior distribution.\\
If there are only those $m$ models, $\mathcal{M}_1,\ldots,\mathcal{M}_m$, then the $m$ Bayes factors in \eqname~(\ref{the_individual_BFs}) together with the (pre-defined) model prior probabilities $p(\mathcal{M}_i)$ ($i=1,\ldots,m$) and the normalisation condition fully specify the model posterior probabilities $p(\mathcal{M}_i|\data)$. With the definition:
\begin{equation}
\nonumber
b_{i,j} := \frac{p(\data|\mathcal{M}_i)}{p(\data|\mathcal{M}_j)} \cdot \frac{p(\mathcal{M}_i)}{p(\mathcal{M}_j)} = \frac{p(\data|\mathcal{M}_i)}{p(\data|\mathcal{M}_0)} \cdot \left(\frac{p(\data|\mathcal{M}_j)}{p(\data|\mathcal{M}_0)}\right)^{-1} \cdot \frac{p(\mathcal{M}_i)}{p(\mathcal{M}_j)} \;\;\;\;\;\; (i,j\in\{1,\ldots,m\})
\end{equation}
where the two Bayes factors on the right are known from \eqname~(\ref{the_individual_BFs}), we get:
\begin{equation}
\label{identical}
p(\mathcal{M}_i|\data) = \frac{p(\data|\mathcal{M}_i)\cdot p(\mathcal{M}_i)}{\sum_{j=1}^{m} p(\data|\mathcal{M}_j)\cdot p(\mathcal{M}_j) } = \frac{p(\data|\mathcal{M}_i)\cdot p(\mathcal{M}_i)}{\sum_{j=1}^{m} \frac{p(\data|\mathcal{M}_i)\cdot p(\mathcal{M}_i)}{b_{i,j}} } = \left(\sum_{j=1}^m b_{i,j}^{-1} \right)^{-1}
\end{equation}
\eqname~(\ref{identical}) is formally equivalent to \eqname~(4) in \cite{Berger_1987}. We have $m$ models with discrete prior probabilities $\pi_i = P(\mathcal{M}_i)>0$ and $\sum_{i=1}^m \pi_i=1$. We get, e.g., for model $\mathcal{M}_1$:
\begin{eqnarray}
\nonumber
p(\mathcal{M}_1|\data) & = & \left(1 + \sum_{j=2}^m b_{1,j}^{-1}   \right)^{-1} = \left(1 + \sum_{j=2}^m \frac{\pi_j}{\pi_1}   \cdot \frac{p(\data|\mathcal{M}_j)}{p(\data|\mathcal{M}_1)}   \right)^{-1} \\
\nonumber
& = &  \left(1 + \frac{1}{\pi_1 } \sum_{j=2}^m   \pi_j \cdot \frac{p(\data|\mathcal{M}_j)}{p(\data|\mathcal{M}_1)}  \right)^{-1} = \left(1 + \frac{1-\pi_1}{\pi_1} \sum_{j=2}^m   \frac{\pi_j}{1-\pi_1} \cdot \frac{p(\data|\mathcal{M}_j)}{p(\data|\mathcal{M}_1)}  \right)^{-1}\\
\nonumber
& = &  \left(1 + \frac{1-\pi_1}{\pi_1} \cdot \frac{\sum_{j=2}^m   p(\data|\mathcal{M}_j)  \cdot \frac{\pi_j}{1-\pi_1}}{p(\data|\mathcal{M}_1)}  \right)^{-1} 
 =  \left(1 + \frac{1-\pi_1}{\pi_1}   \cdot \frac{1}{B} \right)^{-1}
\end{eqnarray} where $B$ is the Bayes factor:
\begin{equation}
\label{new_bayes_fator}
B:=\frac{p(\data|\mathcal{M}_1)}{\sum_{j=2}^m p(\data|\mathcal{M}_j)\cdot g(\mathcal{M}_j)} = \frac{p(\data|\mathcal{H}_0)}{p(\data|\mathcal{H}_1)}  \;\;  \textnormal{with}\;\; g(\mathcal{M}_j):=\frac{\pi_j}{1-\pi_1} = \frac{\pi_j}{\sum_{j=2}^m \pi_j}
\end{equation}
and the hypotheses stand for: $\mathcal{H}_0:\mathcal{M}=\mathcal{M}_1$ and $\mathcal{H}_1:\mathcal{M}\in\{\mathcal{M}_2,\ldots,\mathcal{M}_m\}$ which are assumed to be true with the prior probabilities $\pi_1$ and $1-\pi_1$, respectively. \eqname~(\ref{new_bayes_fator}) corresponds to \eqname~(2) in \cite{Berger_1987}.\footnote{\cite{Berger_1987} study the Bayesian test problem: $\mathcal{H}_0: \theta=\theta_0$ vs. $\mathcal{H}_1: \theta\neq\theta_0$, where $\theta$ is a continuous parameter. In \cite{Berger_1987} the denominator of the Bayes factor $B$ in \eqname~(3) is given by: $P(\mathcal{D}|\mathcal{H}_1)=\int p(\mathcal{D}|\theta) g(\theta) d\theta$, where the prior $g(.)$ and the integral are over all parameters belonging to $\mathcal{H}_1$.
Here we can think of the test: $\mathcal{H}_0: \mathcal{M}=\mathcal{M}_1$ vs. $\mathcal{H}_1: \mathcal{M}\neq \mathcal{M}_1$. With the partition theorem we get for the joint probability:  $p(\data,\mathcal{H}_1)=p(\data,\{\mathcal{M}_2\cup\ldots\cup\mathcal{M}_m\})= \sum_{j=2}^m p(\data,\mathcal{M}_j)   =    \sum_{j=2}^m p(\data|\mathcal{M}_j) \cdot \pi_j$, and hence we have for the denominator of our Bayes factor: $p(\data|\mathcal{H}_1)=\frac{p(\data,\mathcal{H}_1)}{1-\pi_1}=\sum_{j=2}^m p(\data|\mathcal{M}_j)\cdot g(\mathcal{M}_j)$.}

% ========
One of the referees raised the interesting question of how the proposed method is applied to graphical Gaussian models and mixture models. 

We have included an additional section in the Appendix~\ref{app:ggm} where we discuss in detail how the proposed method can be applied to Graphical Gaussian models. We have also carried out an additional simulation study to illustrate the application of our method to Graphical Gaussian models. The key idea is to not apply the method to the configuration space of precision matrices directly, which would be cumbersome due to the constrained topology of this space (restriction to positive definite matrices).  Instead, we make use of the theorem that every multivariate normal density can be represented by a Gaussian belief network, and vice versa; see \cite{GeigerHeckGaussUAI}. This effectively defines an isomorphism between the space of Gaussian graphical models and the space of Gaussian belief networks. We exploit this isomorphism by defining the proposed NETI scheme in the space of Gaussian belief networks, as discussed in detail in Appendix~\ref{app:ggm}.

For mixture models, the proposed NETI method will not achieve any improvement over the standard thermodynamic integration scheme. The reason is that according to \eqname~(\ref{eq:ioio}), the modified thermodynamic integration path that we have proposed has the potential for a variance reduction if the two model likelihoods in the numerator and denominator share a substantial number of parameters. For mixture models, this is not the case, due to the intrinsic identifiability problem. In Appendix~\ref{sec:app_galaxy}, we demonstrate on an empirical simulation study that for a mixture model, the proposed new method and the established thermodynamic integration scheme are on a par.
% ===========

The focus of our study has been a comparison with the improved TI method proposed in \cite{frielTIcorrection}. Recently, a powerful new method for variance reduction in thermodynamic integration based on control variates, termed CTI (controlled thermodynamic integral), has been proposed \shortcite{Oates2015TI}. The idea is to add a zero-mean function from a given function family (e.g. a polynomial) to the integrand and then apply variational calculus to minimise the variance of the estimator. The resulting optimality equations depend on expectation values w.r.t. the unknown posterior distribution, which the authors approximate with samples from initial MCMC simulations.

On the Radiata data, CTI outperforms NETI-DIFF, due to the fact that NETI-DIFF offers little room for improvement on non-nested models with disjunct parameter sets, as discussed above. On the Pima Indians data, both NETI-DIFF and CTI achieve a significant variance reduction over the state-of-the-art TI method of \shortciteA{frielTIcorrection}. \shortciteA{Oates2015TI} applied their method with the standard trapezoid sum of \eqname~(\ref{eq:TI_trapezoid_standard}), CTI-1, and with the improved trapezoid sum of \eqname~(\ref{eq:TI_trapezoid_correction}), CTI-2. A comparison between Figure~\ref{fig:VARIANCES_2} in the present paper and Figure~3 in \shortciteA{Oates2015TI} shows that the performance of NETI-DIFF, which reduces the variance over state-of-the-art TI by a whole order of magnitude, lies between CTI-1 and CTI-2.
\shortciteA{Oates2015TI} argue that the linear curvature sum of \eqname~(\ref{eq:TI_trapezoid_standard}) is known to be biased, and the quadratic curvature rule of \eqname~(\ref{eq:TI_trapezoid_correction}) should be used. However, in \cite{STCO_2016} it was demonstrated that quadratic curvature can lead to an increase in the estimation error when vague prior distributions are used, and it is therefore not  always the automatic method of choice.

Current work in statistics is increasingly aiming to tackle more complex models, e.g. based on coupled nonlinear differential equations, like the biopathway model  discussed in Section~\ref{sec:Biopepa}. For data generated from an ordinary differential equation model of circadian regulation (Goodwin oscillator),
\shortciteA{Oates2015TI} found that CTI achieved  little improvement over state-of-the-art TI. The authors discuss that a potential problem CTI faces for complex models is multimodality of the posterior distributions, rendering the approximation of the posterior expectation values, which enter the optimality equations from variational calculus, less reliable. NETI-DIFF, on the other hand, does not rely on such estimates. In fact, our results, presented in Figure~\ref{fig:netidiff_ti_seq_meanvar}, suggest that NETI-DIFF achieves the most substantial variance reduction over state-of-the-art TI for the most complex, nonlinear biopathway model, reaching up to and  exceeding two orders of magnitude.   

We conclude that CTI and NETI-DIFF are \emph{not} competing methods, but rather conceptionally different approaches with the potential to complement each other. CTI aims to achieve variance reduction by adding control variates to the integrand; it requires a reliable estimation of posterior averages of quantities related to these control variates from initial MCMC runs. 
NETI-DIFF aims to achieve variance reduction by modifying the thermodynamic integration path; it works best for models with substantial parameter overlap. Both approaches can be combined, that is, the natural next step is to add control variates \emph{and} change the integration path, i.e. to target the log Bayes factor with the principles of CTI. This combination of NETI-DIFF and CTI has the potential to further extend the feasibility of Bayesian model selection to ever more complex models, and a closer investigation of such a hybrid approach poses a promising avenue for future research.

\section*{Acknowledgement}
Dirk Husmeier was supported by a grant from the Engineering and Physical Sciences Research Council (EPSRC) of the United Kingdom, grant reference EP/L020319/1. We would like to thank Marilyn Hurn for sending us the software implementation of TI-optimal \shortcite{frielTIcorrection} and for providing helpful explanations. 
We also thank two anonymous reviewers and the associate editor for insightful suggestions, which have improved the quality of our paper. 

\newpage
\bibliographystyle{spbasic}
\bibliography{references}

\section{Appendix}

\subsection{Proof of Jarzynski's theorem}
Using the definitions from Section~\ref{sec:rationale}, we get:
\begin{eqnarray*}
p(\data|\model_i) & = & 
\int p(\data|\bftheta,\model_i)p(\bftheta) d\bftheta = \frac{
\int \exp(-\EE_i[\bftheta]) p(\bftheta) d\bftheta
}{
\int p(\bftheta) d\bftheta
}
%\\
%& = & 
=\frac{\int \exp(-\EE_i(\bftheta)) p(\bftheta) d\bftheta}{
\int \exp(\EE_i[\bftheta]) \exp(-\EE_i[\bftheta]) p(\bftheta) d\bftheta}
\\
& = & 
\left(\int \exp(\EE_i[\bftheta])
\frac{\exp(-\EE_i[\bftheta]) p(\bftheta) d\bftheta}
{\int \exp(-\EE_i(\bftheta)) p(\bftheta) d\bftheta}
\right)^{-1}
%\\
%& = & 
=
\left(\int \exp(\EE_i[\bftheta])
p(\bftheta|\data,\model_i)
\right)^{-1}
\\
& = & 
\Big\langle \exp(\EE_i[\bftheta]) \Big\rangle_i^{-1}
\\
p(\data|\model_2) 
& = & 
\int \exp(-\EE_2[\bftheta]) p(\bftheta|\model_2) d\bftheta
%\\
%& = & 
=
\int \exp(-\{\EE_2[\bftheta]-\EE_1[\bftheta]\}) 
%\\
%& & 
\exp(-\{\EE_1[\bftheta]\}) p(\bftheta|\model_2) d\bftheta
\\
& = & 
\int \exp(-\Delta\EE[\bftheta]) \exp(-\{\EE_1[\bftheta]\}) p(\bftheta|\model_2) d\bftheta
\\
& = &
\int \exp(-\Delta\EE[\bftheta]) \exp(-\{\EE_1[\bftheta]\}) \frac{p(\bftheta|\model_2)}{p(\bftheta|\model_1)} p(\bftheta|\model_1)d\bftheta
\\
& = &
\int \exp(-\Delta\EET[\bftheta]) \exp(-\{\EE_1[\bftheta]\}) p(\bftheta|\model_1)d\bftheta
\\
\frac{p(\data|\model_2)}{p(\data|\model_1)}
& = & 
\int \exp(-\Delta\EET[\bftheta]) \frac{\exp(-\{\EE_1[\bftheta]\}) p(\bftheta|\model_1)}{p(\data|\model_1)} d\bftheta
%\\
%& = & 
\\
& = &
\int \exp(-\Delta\EET[\bftheta]) p(\bftheta|\data,\model_1) d\bftheta
\\
& = & 
\Big\langle\exp(-\Delta\EET[\bftheta]) \Big\rangle_1
\end{eqnarray*}

% ========================================
\subsection{Uncertainty quantification}
\label{sec:Var}
% ========================================
From \eqname~(\ref{eq:CCC}) we have:
\begin{eqnarray}
& & 
\frac{d}{d \invT}\mathbb{E}_{\invT}\left[
\log \left(\frac{p(\data|\bftheta,\model_2)}{p(\data|\bftheta,\model_1)}\right)
\right]
\nonumber \\
&=  & 
\frac{d}{d \invT}
\left(
\frac{1}{Z(\data|\invT,\model_1,\model_2)} \frac{d}{d\invT}  Z(\data|\invT,\model_1,\model_2)
\right)
\nonumber  \\
& = & 
\frac{1}{Z(\data|\invT,\model_1,\model_2)} \frac{d^2}{d\invT^2}  Z(\data|\invT,\model_1,\model_2)
-
\left(
\frac{1}{Z(\data|\invT,\model_1,\model_2)} \frac{d}{d\invT}  Z(\data|\invT,\model_1,\model_2)
\right)^2
\nonumber  \\
& = & 
\frac{1}{Z(\data|\invT,\model_1,\model_2)} \frac{d^2}{d\invT^2}  Z(\data|\invT,\model_1,\model_2)
-
\left\{
\mathbb{E}_{\invT}\left[
\log \left(\frac{p(\data|\bftheta,\model_2)}{p(\data|\bftheta,\model_1)}\right)
\right]
\right\}^2
\label{eq:CCC1}
\end{eqnarray}
For the first term we get:
\begin{eqnarray}
& & \frac{1}{Z(\data|\invT,\model_1,\model_2)} \frac{d^2}{d\invT^2}  Z(\data|\invT,\model_1,\model_2)
\nonumber \\
& = & 
\frac{1}{Z(\data|\invT,\model_1,\model_2)}
\int 
\frac{d^2}{d\invT^2} 
\left(\frac{p(\data|\bftheta,\model_2)}{p(\data|\bftheta,\model_1)}
\right)^{\invT}
p(\data|\bftheta,\model_1)
p(\bftheta|\model_1,\model_2)
d\bftheta  
\nonumber\\
& = & 
\int 
\left\{
\log 
\left(\frac{p(\data|\bftheta,\model_2)}{p(\data|\bftheta,\model_1)}
\right)\right\}^2
\frac{p(\data|\bftheta,\model_2)^{\invT}
	p(\data|\bftheta,\model_1)^{1-\invT}
	p(\bftheta|\model_1,\model_2)}
{Z(\data|\invT,\model_1,\model_2)}
d\bftheta  
\nonumber  \\
& = & 
\int 
p_{\invT}(\bftheta|\data,\model_1,\model_2)
\left\{
\log 
\left(\frac{p(\data|\bftheta,\model_2)}{p(\data|\bftheta,\model_1)}
\right)
\right\}^2
d\bftheta  
\nonumber\\
& = & 
\mathbb{E}_{\invT}\left[\left\{
\log \left(\frac{p(\data|\bftheta,\model_2)}{p(\data|\bftheta,\model_1)}\right)
\right\}^2
\right]
\label{eq:CCC2}
\end{eqnarray}
Combining \eqnames~(\ref{eq:CCC1}) and (\ref{eq:CCC2}), we get:
\begin{equation}
\frac{d}{d \invT}\mathbb{E}_{\invT}\left[
\log \left(\frac{p(\data|\bftheta,\model_2)}{p(\data|\bftheta,\model_1)}\right)
\right]
\nonumber 
= 
\var_{\invT}
\left[
\log \left(\frac{p(\data|\bftheta,\model_2)}{p(\data|\bftheta,\model_1)}\right)
\right]
\end{equation}
Define the following shorthand notation:
\begin{equation}
\Phi(\invT)
\; = \;
\left\{
\log \left(\frac{p(\data|\bftheta,\model_2)}{p(\data|\bftheta,\model_1)}\right)
\right\}_{\invT}
\end{equation}
which is an estimator of 
$
\mathbb{E}_{\invT}\left[
\log \left(\frac{p(\data|\bftheta,\model_2)}{p(\data|\bftheta,\model_1)}\right)
\right]
$
with sample size 1.
We can rewrite \eqname~(\ref{eq:ioio}) as:
\begin{eqnarray}
\log \left(\frac{p(\data|\model_2)}{p(\data|\model_1)}\right)
& \approx &
\int_0^1
\Phi(\invT)
d \invT
\label{eq:ioio2}
\;  \approx\;
\sum_n
\Phi(\invT_n)
\Delta \invT_n
\end{eqnarray}
For the variance we get:
\begin{eqnarray}
\var
\left\{
\log \left(\frac{p(\data|\model_2)}{p(\data|\model_1)}\right)
\right\} & \approx & 
\sum_n
\var \Big[\Phi(\invT_n)\Big]
[\Delta \invT_n]^2
\;\approx\;
\sum_n
\left(
\frac{\partial \Phi}{\partial \invT}
\right)_{\invT_n}\,
[\Delta \invT_n]^2
\nonumber
\\
& \approx &
\sum_n
\frac{\Delta \Phi(\invT_n)}{\Delta \invT_n}
[\Delta \invT_n]^2 
\nonumber
\\
& \approx &
\sum_n
\Delta \Phi(\invT_n)
\Delta \invT_n
\label{eq:varEmp}
\end{eqnarray}

\subsection{Pseudocode}
Table~\ref{pseudocode_mcmc_TI} shows the NETI-DIFF pseudocode for the Bayesian hierarchical model of Figure~\ref{fig:graphModelCheMA}. Pseudocode for standard MCMC, following a Metropolis-Hastings within Gibbs scheme, was provided in Table~1 of \cite{STCO_2016}. Table~\ref{pseudocode_mcmc_TI} shows the modification required to sample with the NETI-DIFF scheme from the tempered posterior distribution in \eqname~(\ref{eq:tempPost}).

 \begin{table*}
\centering
 \raisebox{-1cm}{\fbox{
   \begin{minipage}[c]{14.6cm}
	\footnotesize
    \vspace{0.3cm}
    {{\bf \underline{Initialization}:}} Consider response gene $i$ with gradient vector
    ${{\bf y}}_i=(y_{i,1},\ldots,y_{i,n})^{\top}$ and two competing regulator sets $\parents^{(1)}$ and $\parents^{(2)}$. Build the union $\parents:=\parents^{(1)}\cup\parents^{(2)}$ and apply the method described in Section~\ref{sec:Gibbs} to obtain the design matrices $\DmatA$ and $\DmatB$ for $\parents^{(1)}$ and $\parents^{(2)}$. Both design matrices include all regulators of the union $\parents$. In ${{\bf D}}^{(j)}$ all columns corresponding to regulators which are not in $\parents^{(j)}$ are set to zero ($j=1,2$). \\
    For the new union regulator set $\parents$ initialize the MCMC algorithm in iteration $\iter=0$ with the maximal reaction rate vector ${{\bf V }}_i^{(0)}={{\bf 1}}$, the Michaelis-Menten parameters ${{\bf K}}_i^{(0)}={{\bf 1}}$, the noise variance $\sigma_{i,(0)}^{2}=1$ and the parameter $\delta_{i,(0)}^{2}=1$. \\
    {{\bf \underline{MCMC iterations}:}} For $\iter=1,2,3,\ldots$ \\
    Given the current state ${{\bf V }}_i^{(\iter-1)}$, ${{\bf K}}_i^{(\iter-1)}$, $\sigma_{i,(\iter-1)}^{2}$, and $\delta_{i,(\iter-1)}^{2}$, successively:
    \begin {itemize}
    \item Re-sample the rate reaction parameter vector from its full conditional distribution \\ ${{\bf V }}_i^{(\iter)}\sim N_{\{{{\bf V }}_i^{(\iter)}\geq0\}}(\tilde{\bfmu},\tilde{\bfSigma})$, where\\
    $\tilde{\bfSigma} =  \Big({\delta_{i,(\iter-1)}^{-2}{\bf I}} +  
    \textcolor{red}{
    \invT{\DmatB_i}^{\top} {\DmatB_i} + (1-\invT){\DmatA_i}^{\top} {\DmatA_i}     }%endRed$
    \Big)^{-1}$,
    $\tilde{\bfmu} = \tilde{\bfSigma} \Big(\delta_{i,(\iter-1)}^{-2} {{\bf 1}} + \textcolor{red}{[\invT{\DmatB_i}+(1-\invT){\DmatA_i}]^{\top}
    } %endRED
     {{\bf y}}_i \Big)$, \\
     and 
     \textcolor{red}{
     $\DmatA_i=\DmatA_i({{\bf K}}_i^{(\iter-1)})$, $\DmatB_i=\DmatB_i({{\bf K}}_i^{(\iter-1)})$.
     } % endRed

    \item Re-sample the noise variance parameter from its full conditional distribution
    $\sigma_{i,(\iter)}^{2}\sim IG(\tilde{a}_{\sigma},\tilde{b}_{\sigma})$, where
    
      $\tilde{a}_{\sigma} = a_{\sigma} + \frac{1}{2} (n + \textcolor{red}{|\parentsA\cup\parentsB|}+1)$, and \\
      $\tilde{b}_{\sigma} = b_{\sigma} + \frac{1}{2} \Big[
      \textcolor{red}{
      \invT( {{\bf y}}_i - \DmatB_i{{\bf V }}_i^{(\iter)})^{\top}    ({{\bf y}}_i - \DmatB_i{{\bf V }}_i^{(\iter)})  
      + (1-\invT)( {{\bf y}}_i - \DmatA_i{{\bf V }}_i^{(\iter)})^{\top}    ({{\bf y}}_i - \DmatA_i{{\bf V }}_i^{(\iter)})
      }%endRed
      \\
      +  \delta_{i,(\iter-1)}^{2} ({{\bf V }}_i^{(\iter)} - {{\bf 1}})^{\top} ({{\bf V }}_i^{(\iter)} - {{\bf 1}})\Big]$ with ${{\bf D}}_i={{\bf D}}_i({{\bf K}}_i^{(r-1)})$.  \\

    \item Re-sample the $\delta_{i}$-hyperparameter from its full conditional distribution
    $\delta_{i,(\iter)}^{2}\sim IG(\tilde{a}_{\delta},\tilde{b}_{\delta})$, where
    
    $\tilde{a}_{\delta} = a_{\delta} + \frac{1}{2} (\textcolor{red}{|\parentsA\cup\parentsB|}  +1)$, and  \\
    $\tilde{b}_{\delta} = b_{\delta} + \frac{1}{2} \sigma_{i,(\iter)}^{2} ({{\bf V }}_i^{(\iter)} - {{\bf 1}})^{\top} ({{\bf V }}_i^{(\iter)} - {{\bf 1}}) $ 
    
    \item Perform a Metropolis-Hastings MCMC move that proposes to change the current value of the Michaelis-Menten parameters in ${{\bf K}}_i^{(\iter-1)}$ by sampling a realization ${{\bf u}}$ from a multivariate Gaussian distribution with expectation vector ${{\bf 0}}$ and covariance $\bfSigma=0.1 \cdot{{\bf I}}$. The newly proposed parameter vector ${{\bf K}}_i^{\star}:={{\bf K}}_i^{(\iter-1)}+{{\bf u}}$ is accepted with probability
    
    $A({{\bf K}}_i^{(\iter-1)},{{\bf K}}_i^{\star})=\min\left\{1,R({{\bf K}}_i^{(\iter-1)},{{\bf K}}_i^{\star})\right\}$, where
    
    $R({{\bf K}}_i^{(\iter-1)},{{\bf K}}_i^{\star})= \\
    \textcolor{red}{
    \frac{\exp\left\{\frac{-\invT}{2\sigma^2_{i,(\iter)}} ({{\bf y}}_i - {\DmatB_i}({{\bf K}}_i^{\star}) {{\bf V}}_i^{(\iter)})^{\top} ({{\bf y}}_i - {\DmatB_i}({{\bf K}}_i^{\star}) {{\bf V}}_i^{(\iter)})  \right\}}{\exp\left\{\frac{-\invT}{2\sigma^2_{i,(\iter)}} ({{\bf y}}_i - {\DmatB_i}({{\bf K}}_i^{(\iter-1)}) {{\bf V}}_i^{(\iter)})^{\top} ({{\bf y}}_i - {\DmatB_i}({{\bf K}}_i^{(\iter-1)}) {{\bf V}}_i^{(\iter)})  \right\}} 
    }$\\
    
    $\textcolor{red}{
    \frac{\exp\left\{\frac{-(1-\invT)}{2\sigma^2_{i,(\iter)}} ({{\bf y}}_i - {\DmatA_i}({{\bf K}}_i^{\star}) {{\bf V}}_i^{(\iter)})^{\top} ({{\bf y}}_i - {\DmatA_i}({{\bf K}}_i^{\star}) {{\bf V}}_i^{(\iter)})  \right\}}{\exp\left\{\frac{-(1-\invT)}{2\sigma^2_{i,(\iter)}} ({{\bf y}}_i - {\DmatA_i}({{\bf K}}_i^{(\iter-1)}) {{\bf V}}_i^{(\iter)})^{\top} ({{\bf y}}_i - {\DmatA_i}({{\bf K}}_i^{(\iter-1)}) {{\bf V}}_i^{(\iter)})  \right\}} 
   }  % endRed
   \frac{P_{\left\{{{\bf K}}_i^{\star}\geq0\right\}}({{\bf K}}_i^{\star})}{P_{\left\{{{\bf K}}_i\geq0\right\}}({{\bf K}}_i)}$\\ 
    If the move is accepted, set ${{\bf K}}_i^{(\iter)}={{\bf K}}_i^{\star}$; otherwise, leave the vector unchanged, ${{\bf K }}_i^{(\iter)}={{\bf K }}_i^{(\iter-1)}$.\\
   \end{itemize}
   {{\bf \underline{Output}:}} An MCMC sample from the joint posterior distribution:
   $({{\bf V }}_i^{(\iter)},{{\bf K}}_i^{(\iter)},\sigma_{i,(\iter)}^{2},\delta_{i,(\iter)}^{2})_{\iter=1,2,3,\ldots}$

  \end{minipage}}}
  \caption{\label{pseudocode_mcmc_TI} { {{\bf Pseudo code for a NETI-DIFF simulation}}  at inverse temperature $\invT$ for the biopathway model from Section~\ref{sec:Biopepa}. The differences to the standard MCMC scheme from Table~1 in \cite{STCO_2016} are marked in red fonts. 
  }}
 \end{table*}

\subsection{Application to Gaussian Graphical Models}
\label{app:ggm}

In this Appendix we show how the new method (NETI-DIFF) can be used to infer the Bayes factor between Gaussian graphical models (GGMs). We propose an indirect procedure which exploits that multivariate Gaussians can be represented as 'Gaussian belief networks' (\cite{GeigerHeckGaussUAI}). A Gaussian graphical model corresponds to an $\Ndim$-dimensional multivariate Gaussian distribution with mean vector ${{\bf m}}$ and covariance matrix ${{\boldsymbol \Sigma}}$ so that the density (PDF) is given by
\begin{equation}
\label{APP0}
p({{\bf x}}|{{\bf m}},{{\bf W}}) = \left(2\pi\right)^{-\Ndim/2} \cdot \det({{\bf W}}) \cdot \exp\{- \frac{1}{2} ({{\bf x}}-{{\bf m}})^{\transp} {{\bf W}} ({{\bf x}}-{{\bf m}})   \}
\end{equation}
where ${{\bf x}}=(x_1,\ldots,x_\Ndim)^{\transp}$ and ${{\bf W}}={{\boldsymbol \Sigma}}^{-1}$ is called the precision matrix. Each $0$ element of ${{\bf W}}$ indicates that the partial correlation between the corresponding variables is zero, e.g. ${{\bf W}}_{i,j}=0$ if the partial correlation between $x_i$ and $x_j$ is zero. We follow \cite{GeigerHeckGaussUAI} and identify this Gaussian distribution with a 'Gaussian belief network', i.e. we factorise the density in \eqname~(\ref{APP0}) with the chain rule: 
\begin{equation}
\label{APP1}
p({{\bf x}}|{{\bf m}},{{\bf W}})  = p(x_1) \cdot \prod_{i=1}^{\Ndim} p(x_i|x_1,\ldots,x_{i-1})
\end{equation}
where the conditional distributions are univariate Gaussians
\begin{equation}
\label{APP2}
x_i|x_1,\ldots,x_{i-1} \sim \mathcal{N}(m_i+\sum_{j=1}^{i-1} \beta_{j,i} (x_j-m_j),\sigma^2_i)
\end{equation}
${{\bf W}}_{j,i}=0$ implies that the 'regression coefficient' $\beta_{j,i}$ of the Gaussian belief network representation is zero, and vice-versa. Moreover, we have ${{\bf m}}=(m_1,\ldots,m_{\Ndim})^{\transp}$, and $\sigma_i^2$ is the conditional variance of $x_i$ given $x_1,\ldots,x_{i-1}$. From the parameters in \eqnames~(\ref{APP1}-\ref{APP2}) the precision matrix ${{\bf W}}={{\bf W}}(\Ndim)$ of the multivariate Gaussian distribution can be (re-)computed with the recursion:
\begin{equation}
\label{APP3}
{{\bf W}}(i+1) = \left(
\begin{matrix}
{{\bf W}}(i) + \frac{{{\boldsymbol \beta}}_i   {{\boldsymbol \beta}}_i^{\transp}}{\sigma_{i+1}^2}   \;\;\;\; & - \frac{{{\boldsymbol \beta}}_i}{\sigma_{i+1}^2}  \\
- \frac{{{\boldsymbol \beta}}_i^{\transp}}{\sigma_{i+1}^2} \;\;\;\;& \frac{1}{\sigma_{i+1}^2} 
\end{matrix}
\right)
\end{equation}
where ${{\bf W}}(1)=\frac{1}{\sigma_1^2}$ and ${{\boldsymbol \beta}}_i = (\beta_{1,i},\ldots,\beta_{i-1,i})^{\transp}$.

The most convenient way to compute the Bayes factor between two competing GGMs is to work with their Gaussian belief network representations.\footnote{When working directy with the precision matrix, one would have to guarantee that it stays positive-definite.} For a GGM with precision matrix ${{\bf W}}$, we impose a Wishart prior onto ${{\bf W}}$, and we represent the GGM in terms of the parameters ${{\bf m}}=(m_1,\ldots,m_\Ndim)^{\transp}$, ${\boldsymbol \sigma}^2=(\sigma_1^2,\ldots,\sigma_\Ndim^2)^{\transp}$, and
\begin{equation}
\nonumber
%\footnotesize
{{\bf B}} = 
\left(
\begin{matrix}
0 & 0 & 0 & \ldots & 0 \\
\beta_{1,2} & 0 & 0 & \ldots & 0 \\
\beta_{1,3} &\beta_{2,3} & 0 &  \ldots & 0 \\
\vdots & \vdots &  & \ddots & \vdots \\
\beta_{1,\Ndim} &\beta_{2,\Ndim} & \ldots &  \beta_{\Ndim - 1,\Ndim}  & 0 \\
\end{matrix}
\right)
\end{equation}
where $\beta_{j,i}=0$ if ${{\bf W}}_{j,i}=0$ ($j<i$). \\
Given two GGMs  $\mathcal{M}_1$ and $\mathcal{M}_2$ with precision matrices ${{\bf W}}^1$ and ${{\bf W}}^2$ we represent both as Gaussian belief networks with the regression coefficient matrices ${{\bf B}}^k$ whose elements are given by $\beta_{j,i}^k$ ($k=1,2$). We have $\beta_{j,i}^k=0$ if ${{\bf W}}^k_{j,i}=0$ ($k=1,2$) and $\beta_{j,i}^1=\beta_{j,i}^2$ if $\beta_{j,i}^1,\beta_{j,i}^2\neq 0$, so that all shared non-zero regression coefficients are equal. We assume that both GGMs share the mean vector ${{\bf m}}$, which we assume to be known, and the conditional variances $(\sigma_i^{2})^k = \sigma_i^2$. Let ${{\bf B}}$ denote the matrix of all regression coefficients which are non-zero in at last one of the GGMs. The elements of ${{\bf B}}$ are :
\begin{equation}
\label{APP4}
\beta_{j,i} =
\begin{cases}
\beta_{j,i}^1 & \text{if} \;\;  \beta_{j,i}^1\neq 0\;\; \text{and}\;\; \beta_{j,i}^2=0  \\
\beta_{j,i}^2 & \text{if} \;\;  \beta_{j,i}^1=0    \;\; \text{and}\;\; \beta_{j,i}^2\neq 0\\
\beta_{j,i}^1 & \text{if} \;\;  \beta_{j,i}^1=\beta_{j,i}^2
\end{cases}
\end{equation}
Given $n$ data points ${{\bf x}}_1,\ldots,{{\bf x}}_n$ the tempered posteriors take the form:
\begin{equation}
\nonumber
p_{\tau}({{\bf W}}|{{\bf x}}_1,\ldots,{{\bf x}}_n,\mathcal{M}_1,\mathcal{M}_2) \propto \left(\prod_{w=1}^n p({{\bf x}}_w|{{\bf m}},{{\bf W}}^1)\right)^{\tau}  \left( \prod_{w=1}^n p({{\bf x}}_w|{{\bf m}},{{\bf W}}^2)\right)^{1-\tau} p({{\bf W}})  
\end{equation}
where $\tau\in[0,1]$ and the three precision matrices ${{\bf W}}^1$, ${{\bf W}}^2$, and ${{\bf W}}$  can be computed with \eqname~(\ref{APP3}) from the conditional variances $\sigma_i^2$ and the regression parameters in ${{\bf B}}^1$, ${{\bf B}}^2$ and ${{\bf B}}$.\\
Sampling from the tempered posterior can be done with Metropolis-Hastings (MH) MCMC moves which we define in the space of the non-zero regression parameters in ${{\bf B}}$ and in the space of the logarithms of the conditional variances $\sigma_i^2$. We obtain a new candidate state ${{\bf B}}^{\star}$ and  $\sigma^2_{i,\star}$ by adding randomly sampled numbers to the non-zero elements of ${{\bf B}}$ and to $log(\sigma_i^2)$.\footnote{Those random numbers are uniformly distributed on a small interval $[-\eps,\eps]$ with center $0$.} From the new candidate matrix ${{\bf B}}^{\star}$ we extract the matrices ${{\bf B}}^{k,\star}$ ($k=1,2$) as follows: $\beta_{j,i}^{k,\star}=0$ if ${{\bf W}}^k_{i,j}$ is restricted to be zero and $\beta_{j,i}^{k,\star}=\beta_{j,i}^{\star}$ otherwise. The new precision matrices ${{\bf W}}^{\star}$, ${{\bf W}}^{1,\star}$, and ${{\bf W}}^{2,\star}$ can then be computed from ${{\bf B}}^{\star}$, ${{\bf B}}^{1,\star}$, and ${{\bf B}}^{2,\star}$ with \eqname~(\ref{APP3}), and the MH acceptance probability depends on the ratio of the tempered posteriors of the new precison matrix ${{\bf W}}^{\star}$ and the old precision matrix ${{\bf W}}$.\\
For a proof of concept we perform a simulation study: We consider the $\Ndim=7$ genes (1$\hat{=}$LHY, 2$\hat{=}$TOC1, 3$\hat{=}$PRR9, 4$\hat{=}$PRR7, 5$\hat{=}$GI, 6$\hat{=}$Y, and 7$\hat{=}$TOC1) of the Arabidopsis networks, shown in Figure~\ref{fig:goldstd1}, and we parametrize both graphs $\mathcal{M}_1$ and $\mathcal{M}_2$ as Gaussian belief networks. We set: ${{\bf m}}={{\bf 0}}$ and $\sigma_i^2=1$ for all $i$, and the non-zero regression coefficients appearing in both graphs are set to $\beta_{j,i}=1$, while the regression coefficients appearing only in the wildtype ($\mathcal{M}_1$) are set to $\beta\in\mathbb{R}$. The latter coefficients correspond to the edges 'PRR9-PRR7' ($\beta_{2,3}$) and 'PRR7-NI' ($\beta_{3,4}$) in Figure~\ref{fig:goldstd1}. $\beta$ is a tuning parameter for the strength of the two additional partial correlations in $\mathcal{M}_1$. For $\beta=0$ the partial correlations are zero and the nested mutant network $\mathcal{M}_2$ is the correct model. As prior on ${{\bf W}}$ we use a Wishart distribution with $\text{df}=10$ degrees of freedom and the identity matrix as precision matrix ${{\bf P}}={{\bf I}}_7$.  We  generate data sets with $n=100$ data points from $\mathcal{M}_1$, and we use NETI-DIFF (with 100k iterations, a sigmoidal temperature ladder and $\eps=0.1$) to compute the Bayes-factors. Figure~\ref{fig:app_new} shows the results. The Bayes factors are in favour of the true wildtype network ($\mathcal{M}_1$) if the additional regression coefficients have a sufficient size ($\beta\geq0.3$). For low values ($\beta\leq 0.2$) the Bayes factor is in favour of the mutant network ($\mathcal{M}_2$), which is actually the true network for $\beta=0$. Only for low positive values ($\beta=0.1$ and $\beta=0.2$) the wrong model is favoured over the true model. The latter can be explained by the prior. The Wishart prior with hyperparameters $\text{df}=10$ and ${{\bf P}}={{\bf I}}_7$ corresponds to $10$ pseudo data points from a GGM without any non-zero partial correlations, and hence, yield a higher penalty for the wildtype netwerk $\mathcal{M}_1$ than for the sparser mutant network $\mathcal{M}_2$.

\begin{figure}[t]
	\centering
	\includegraphics[width=0.9\textwidth]{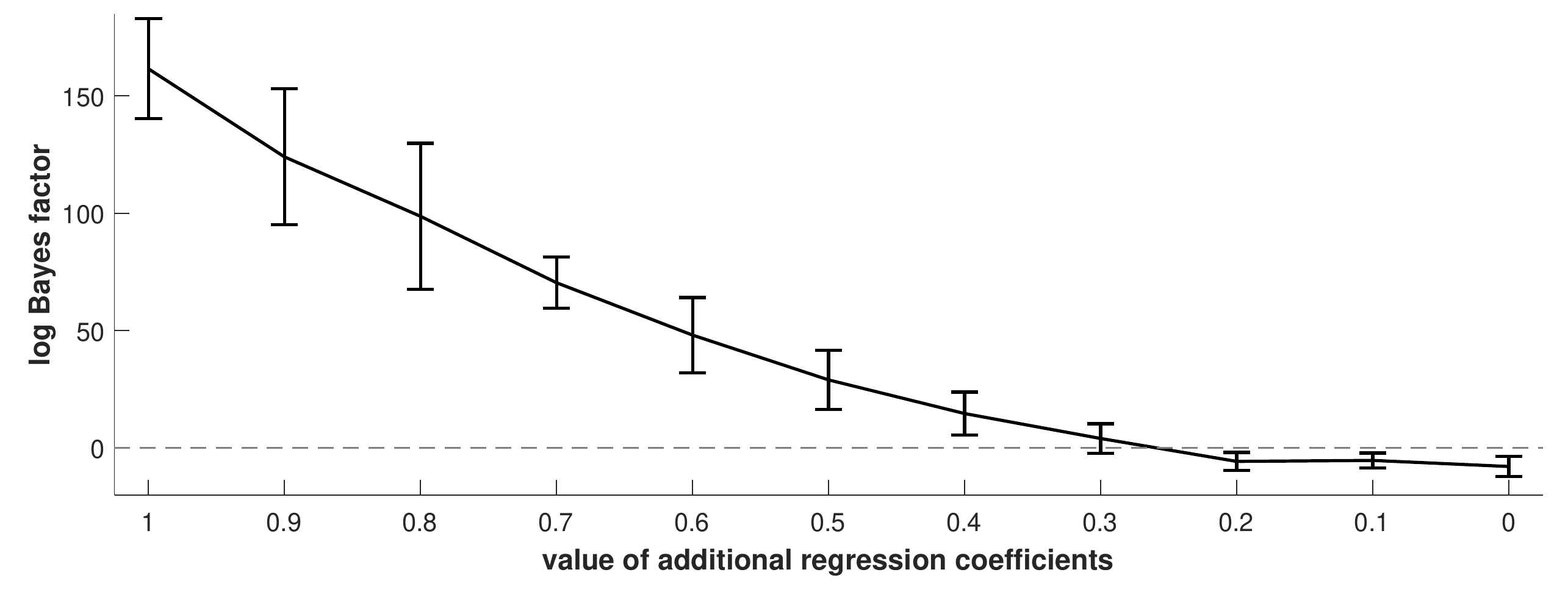}
	\vspace{-0.3cm}
	\caption{\label{fig:app_new} 
	%\footnotesize
	\textbf{Average log Bayes factor for GGMs: wildtype ($\mathcal{M}_1$) vs. mutant ($\mathcal{M}_2$) network.} The figure shows the average log Bayes factor obtained for 10 independent wildtype data instantiations with $n=100$ observations each. Bayes factors above the dotted reference line are in favour of the wildtype network ($\mathcal{M}_1$), which is the true network if the additional regression coefficients are greater than 0. The errorbars refer to standard deviations.}
\end{figure}

\subsection{Application to Mixture Models (Galaxy data)}
\label{sec:app_galaxy}

In this Appendix we show that the new method (NETI-DIFF) can also be used to compute the Bayes factor between mixture models with different numbers of mixture components. Like \cite{frielTIcorrection} we consider the Galaxy data from \cite{Richardson-Green}, which contain $n=82$ measurements $y_1,\ldots,y_{82}$ of galaxy velocities, and we compute the Bayes factor between two Bayesian Gaussian mixture models $\mathcal{M}_3$ with $K=3$ components and $\mathcal{M}_4$ with $K=4$ components. For our study we use exactly the same mixture model as \cite{frielTIcorrection} with the same prior distributions and the same hyperparameters. A latent allocation vector ${{\bf z}}=(z_1,\ldots,z_{82})^{\transp}$ allocates the individual data points to the $K$ mixture components, where $z_i=k$ if data point $y_i$ has been allocated to component $k$ ($k=1,\ldots,K$; $i=1,\ldots,82$). On the mixture weights $w_k:=P(z_i=k)$ we impose a Dirichlet prior:
\begin{equation}
\nonumber
(w_1,\ldots,w_K) \sim \text{DIR}(1,\ldots,1)
\end{equation} 
The data points within each component $k$ ($1\leq k\leq K$) are assumed to stem from a univariate Gaussian distribution with mean $\mu_k$ and variance $\sigma_k^2$, so that
\begin{equation}
\nonumber
y_i|(z_i=k)  \sim \text{N}(\mu_k,\sigma^2_k)
\end{equation} 
and for $\mu_k$ and $\sigma_k^2$ we use a Gaussian prior and an Inverse-Gamma prior:
\begin{equation}
\nonumber
\mu_k \sim \text{N}(0,1000) \quad \quad \quad \sigma_k^{-2} \sim \text{GAM}(1,1)
\end{equation} 
We define $\bftheta_K$ to be the set of all parameters of the mixture model $\mathcal{M}_K$ with $K$ components:
\begin{equation}
\nonumber
\bftheta_K = \{w_{{{\bf K}},1},\ldots,w_{{{\bf K}},K},\;\mu_{{{\bf K}},1},\ldots,\mu_{{{\bf K}},K},\;\sigma^2_{{{\bf K}},1},\ldots,\sigma^2_{{{\bf K}},K}   \}
\end{equation}

In the absence of limiting conditions, mixture models with different numbers of components (here: $\mathcal{M}_3$ and $\mathcal{M}_4$) do \emph{not} share any parameters, and the tempered NETI-DIFF posteriors take the form
\begin{equation}
\nonumber
p_{\tau}(\bftheta_3,\bftheta_4 |y_1,\ldots,y_m,\mathcal{M}_3,\mathcal{M}_4) \propto
p(y_1,\ldots,y_n|\bftheta_3)^{\tau}  p(y_1,\ldots,y_n|\bftheta_4)^{1-\tau} p(\bftheta_3|\mathcal{M}_3)  p(\bftheta_4|\mathcal{M}_4) 
\end{equation}
Because of this modular form, the parameters in the sets $\bftheta_3$ and $\bftheta_4$ can be sampled by disjunct MCMC sampling steps, which either re-sample subsets of the parameters $\tilde{\bftheta}_3 \subset \bftheta_3$ (or $\tilde{\bftheta}_4 \subset \bftheta_4$) from their full conditional distributions: 
\begin{eqnarray}
p_{\tau}(\tilde{\bftheta_3}|\tilde{\tilde{\bftheta_3}}, \bftheta_4,y_1,\ldots,y_m,\mathcal{M}_3,\mathcal{M}_4) & \propto &   p(y_1,\ldots,y_n|\bftheta_3)^{\tau} \;\;\;\;\cdot p(\bftheta_3|\mathcal{M}_3)   \nonumber \\
p_{\tau}(\tilde{\bftheta_4}|\tilde{\tilde{\bftheta_4}},\bftheta_3,y_1,\ldots,y_m,\mathcal{M}_3,\mathcal{M}_4) & \propto &   p(y_1,\ldots,y_n|\bftheta_4)^{1-\tau}  \cdot  p(\bftheta_4|\mathcal{M}_4)  \nonumber
\end{eqnarray}
where $\tilde{\tilde{\bftheta_K}}\cup \tilde{\bftheta_K} = \bftheta_K$,  
or via Metropolis Hastings sampling steps, whose acceptance probabilities are:
\begin{eqnarray}
A\left((\bftheta_3,\bftheta_4)\rightarrow (\bftheta_3^{\star},\bftheta_4)\right) & = & \min\{1,\left(\frac{ p(y_1,\ldots,y_n|\bftheta_3^{\star})}{p(y_1,\ldots,y_n|\bftheta_3)}\right)^{\tau}  \;\;\;\;\cdot\frac{p(\bftheta_3^{\star}|\mathcal{M}_3)}{p(\bftheta_3|\mathcal{M}_3)} \cdot \text{HR}\} \nonumber \\
A\left((\bftheta_3,\bftheta_4)\rightarrow (\bftheta_3,\bftheta_4^{\star})\right) & = & \min\{1,\left(\frac{p(y_1,\ldots,y_n|\bftheta_4^{\star})}{p(y_1,\ldots,y_n|\bftheta_4)}\right)^{1-\tau} \cdot \frac{p(\bftheta_4^{\star}|\mathcal{M}_4)}{p(\bftheta_4|\mathcal{M}_4)} \cdot \text{HR}\} \nonumber
\end{eqnarray}
where HR is the move-specific Hastings ratio and the $\star$ symbol indicates a new candidate parameter set. Since these are the standard equations for power posterior sampling, as used by the thermodynamic integration (TI) approach, the adaptation of the Metropolis-Hastings and Gibbs sampling steps of the power posterior sampling scheme for TI (\cite{frielTIcorrection}) is straightforward. At each temperature $\tau\in[0,1]$ NETI-DIFF updates the parameters in $\bftheta_3$ and in $\bftheta_4$ independently by performing the corresponding steps of the MCMC sampling scheme. The only difference is that the parameters in $\bftheta_4$ are subject to the complementary temperature $1-\tau$ rather than $\tau$, and we therefore implement NETI-DIFF with the sigmoid inverse temperature ladder from Section~\ref{sec:Tladder}. Moreover, we also take into account that NETI-DIFF has to perform twice as many sampling steps as TI, since NETI-DIFF re-samples the parameters of both models $\mathcal{M}_3$ and $\mathcal{M}_4$ within each iteration. Thus, NETI-DIFF iterations are approximately double as expensive as TI iterations, and we can perform only 50\% of the total number of iterations $N_{iter}$ with NETI-DIFF. \\
In our empirical study we compare the performance of NETI-DIFF with TI-standard and TI-optimal, and we implement both TI approaches with 100 discretisation points. We compute the Bayes factor between the mixture models $\mathcal{M}_3$ and $\mathcal{M}_4$ based on $N_{iter}=1000k$ and $N_{iter}=2000k$  iterations.\footnote{That is, we implement NETI-DIFF with $N_{iter}/2=500k$ (and $N_{iter}/2=1000k$) iterations, and for TI we take $N_{iter}/100=10k$ (and $N_{iter}/100=20k$) power posterior samples for each of the 100 inverse temperatures.} The results of our study are shown in Figure~\ref{fig:app_galaxy}. It can be seen that there are no significant differences between the performances. The NETI-DIFF estimates appear to be minimally less biased than the TI estimates, but on the other hand the NETI-DIFF estimates have a slightly increased standard deviation. This finding, that  NETI-DIFF does not lead to any improvement over the standard TI approach, is not surprising: Due to the fact that the two mixture models do not have any parameters in common, targeting the Bayes factor directly cannot have any advantages. For models with disjunct parameter spaces NETI-DIFF effectively just corresponds to two  simultaneously performed but independent non-equilibrium thermodynamic integration (NETI) approaches, where one model is subject to the complementary temperature transition from $\tau=1$ to $\tau=0$. Targeting the Bayes factor directly, as described in Section~\ref{sec:TInew}, can only lead to an improvement if the two models share parameters. In the direct transition paths between the two model posteriors, only those shared parameters constantly appear with the inverse temperature $1$ and do not undergo any temperature transitions (i.e. they are excluded from the annealing process). All non-shared parameters have to undergo the transitions from $\tau=0$ to $\tau=1$ or from $\tau=1$ to $\tau=0$, respectively.

\begin{figure}[t]
	\centering
	\includegraphics[width=1\textwidth]{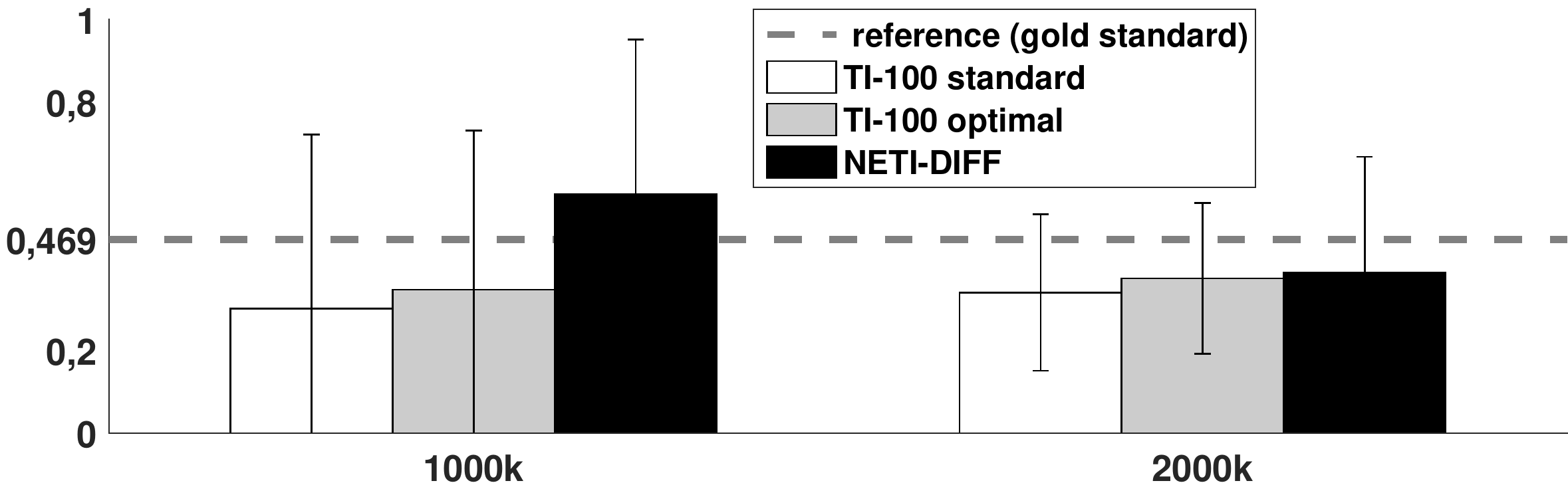}
	\vspace{-0.3cm}
	\caption{\label{fig:app_galaxy} 
	%\footnotesize
	\textbf{Bayes factor estimates for the Galaxy data.} Errorbar plot of the average log Bayes factor between the mixture models $\mathcal{M}_4$ with $K=4$ components and $\mathcal{M}_3$ with $K=3$ components. For each method the averages have been computed from 10 independent simulations. The dotted reference line indicates the (true) gold-standard Bayes factor of size $0.4685$, reported in \cite{frielTIcorrection}. The errorbars correspond to standard deviations, and the horizontal axis gives the total number of iterations ($N_{iter}=1000k$ and $N_{iter}=2000k$).}
\end{figure}

\subsection{Full conditional distributions of variance parameters}
\label{sec:fcd_var}

For linear models where the variance parameter $\sigma^2$ in \eqname~(\ref{reg_prior}) in Section~\ref{sec:Gibbs} is not known, a prior distribution has to be imposed on $\sigma^2$. A common choice is the conjugate Inverse-Gamma distribution with hyperparameters $a/2$ and $b/2$, symbolically $\sigma^{-2}\sim\text{GAM}(\frac{a}{2},\frac{b}{2})$. The tempered full conditional distribution of $\sigma^{-2}$ is then of closed-form and can be derived as follows:
\begin{eqnarray}
p_{\invT}(\sigma^{-2}|\data,\bftheta,\model_1,\model_2) & \propto &
p(\byi|\bftheta,\sigma^2,\model_2)^{\invT} \cdot p(\byi|\bftheta,\sigma^2,\model_1)^{1-\invT} \cdot p(\bftheta|\sigma^2,\model_1,\model_2) \cdot p(\sigma^2) \nonumber\\
& \propto &  N_n(\DmatB\bftheta,\sigma^2{{\bf I}})^{\invT} \cdot N_n(\DmatA\bftheta,\sigma^2{{\bf I}})^{1-\invT} \cdot N_p(\boldsymbol\mu_0,\sigma^2\delta^2 {{\bf I}}) \cdot GAM(\sigma^{-2}) \nonumber\\
& \propto & \left(\frac{1}{\sigma^2}\right)^{\invT \cdot \frac{ n}{2}}  \exp\left(
-\frac{1}{2\sigma^2} \cdot
\left[
\DmatA\Vmat-\byi
\right]^{\transp}
\left[
\DmatA\Vmat-\byi
\right] \cdot (1-\invT)
\right)\cdot \nonumber \\
& & \left(\frac{1}{\sigma^2}\right)^{(1-\invT) \cdot \frac{ n}{2}} \exp\left(
-\frac{1}{2\sigma^2} \cdot
\left[
\DmatB\Vmat-\byi
\right]^{\transp}
\left[
\DmatB\Vmat-\byi
\right] \cdot \invT
\right) \cdot \nonumber \\  & & 
 \left(\frac{1}{\sigma^2}\right)^{\frac{ p}{2}}
\exp\left(
-\frac{1}{2\sigma^2\delta^2} \cdot
[\Vmat-\boldsymbol\mu_0]^{\transp}
[\Vmat-\boldsymbol\mu_0]
\right)\cdot \nonumber \\ 
& & 
 \left(\sigma^{-2}\right)^{a/2-1} \exp\left(-\sigma^{-2} \cdot \frac{b}{2}     \right)\nonumber
\\
& = & 
\left(\sigma^{-2}\right)^{\tilde{a} -1} \cdot \exp\left(-\sigma^{-2} \cdot \tilde{b} \right)  \nonumber
\end{eqnarray}
where $p$ is the length of the regression coefficient vector $\bftheta$ and 
\begin{eqnarray}
\footnotesize
\tilde{a} & = & \frac{1}{2} (a+n+p) 
\nonumber 
\\
\tilde{b} &  =  &\frac{1}{2} \left(b+ (1-\invT)
\left[
\DmatA\Vmat-\byi
\right]^{\transp}
\left[
\DmatA\Vmat-\byi
\right] + \invT
\left[
\DmatB\Vmat-\byi
\right]^{\transp}
\left[
\DmatB\Vmat-\byi
\right] + \delta^{-2}
[\Vmat-\boldsymbol\mu_0]^{\transp}
[\Vmat-\boldsymbol\mu_0]    \right)
\nonumber
\end{eqnarray}
\normalsize
Comparing this with the identity:
\begin{equation}
\nonumber
\text{GAM}(\sigma^{-2}|\tilde{a},\tilde{b}) = \frac{\tilde{b}^{\tilde{a}}}{\Gamma(\tilde{a})} \cdot \left(\sigma^{-2}\right)^{\tilde{a}-1} \exp \{-  \sigma^{-2}\cdot \tilde{b}   \}  \propto \left(\sigma^{-2}\right)^{\tilde{a}-1} \exp \{-  \sigma^{-2}\cdot \tilde{b} \}
\end{equation}
we get the full conditional distribution 
\begin{equation}
\nonumber
\sigma^{-2}|(\data,\bftheta,\model_1,\model_2,\invT)\sim\text{GAM}(\tilde{a},\tilde{b})
\end{equation}
Hence, we can also sample $\sigma^{-2}$ directly from the tempered full conditional distributon in a Gibbs sampling scheme, and $\sigma^2 = 1/\sigma^{-2}$.

%
% Runtimes from Matlab script 'WF_NETI_Runtime/Runtime_all.m' 
%
\begin{table}[tb]
  \center
  \begin{tabular}{l c   c c c}
	\hline
                                    &  $n$ & NETI-DIFF   & TI-standard & TI-optimal  \\
    \hline 
    Radiata pine                    & 42   & 81.6 (4.1)  &  85.6 (4.4) & 85.6 (4.4)  \\
    Pima Indians                    & 532  & 9.3 (0.1)   &  9.5 (0.3)  & 10.3 (0.1)  \\  
    Radiocarbon                     & 343  & 19.5 (0.5)  &  -          & 35.4 (0.8)  \\
		Biopepa                         & 143  & 54.1 (1.2)  &  53.6 (2.0) & 58.7 (2.5)  \\ 
  \hline
  \end{tabular}
  \caption{\label{tab:runtime}
    \textbf{Runtimes in seconds for different data sets and 100000 iterations.} The data size is given by $n$ observations. The number of variables can differ depending on the model. Time values are averages over 10 runs and different models on an Intel Core i7 6700HQ processor. The standard deviations are indicated in brackets. The runtime for Biopepa is for a single response out of seven possible responses. 
  }
  \end{table}

\begin{figure*}[tb]
	\begin{center}
	\begin{minipage}[b]{0.49\linewidth}
	\centering 
	\includegraphics[width=1\textwidth]{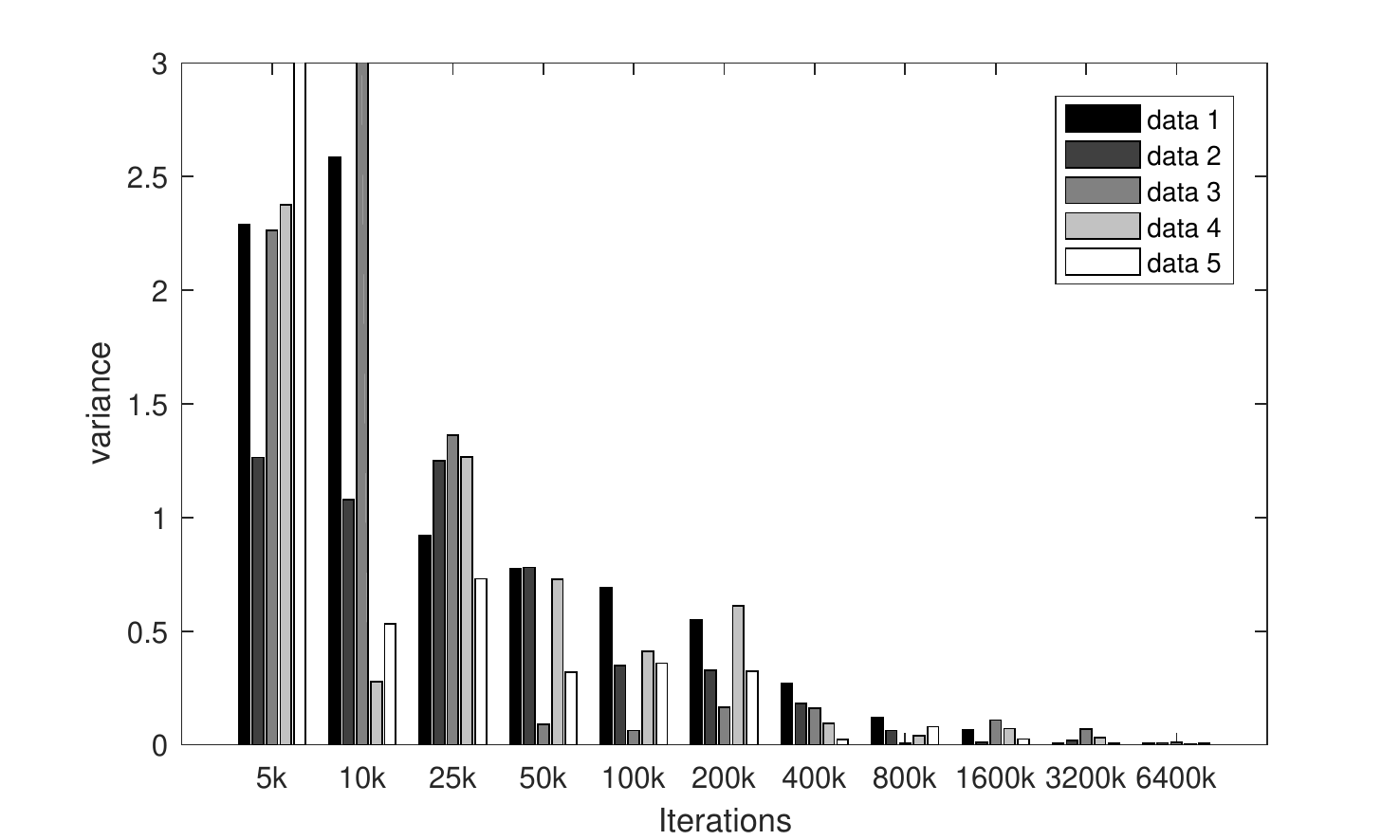}
	\subcaption{NETI-DIFF with power law ladder. }\label{fig:NETIDIFF_convergence_a}
	\end{minipage}
	\begin{minipage}[b]{0.49\linewidth}
	\centering 
	\includegraphics[width=1\textwidth]{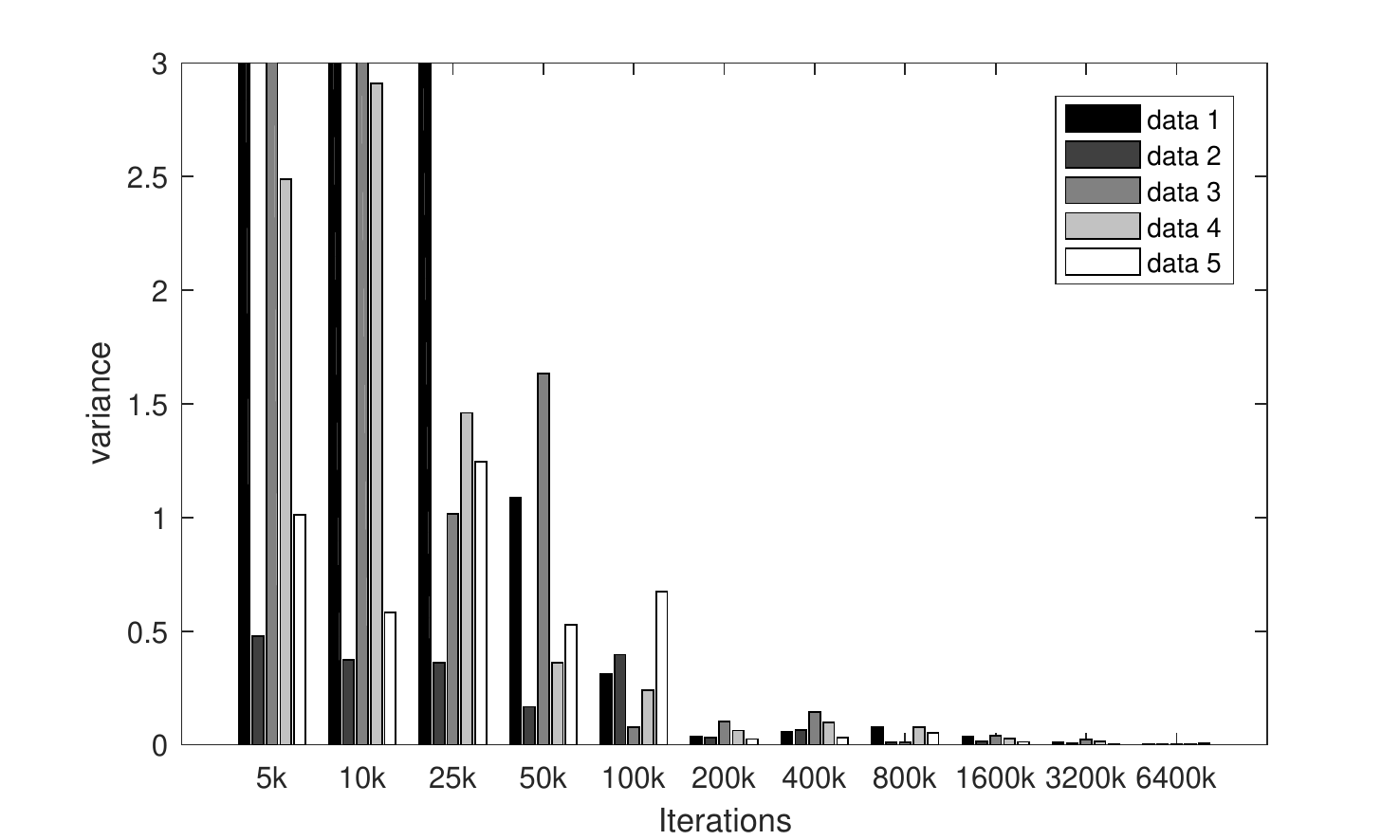}
	\subcaption{NETI-DIFF with sigmoidal ladder.}\label{fig:NETIDIFF_convergence_b}
	\end{minipage}
		\caption{\label{fig:NETIDIFF_convergence}
		\textbf{Convergence of NETI-DIFF for the Biopepa data.} This figure gives the variance of marginal log likelihood estimators for NETI-DIFF and the two ladder types (panel~\subref{fig:NETIDIFF_convergence_a} and \subref{fig:NETIDIFF_convergence_b}). The variance is calculated from five repetitions and for five different data instances shown in the legend. This figure complements Figure~\ref{fig:netidiff_pow_sigm_biasvar}, where the average variances, averaged over all five data instantiations, are shown. 
		}
	\end{center}
\end{figure*}

\subsection{Computational run times and convergence diagnostics}
\label{sec:compconv}

It is important to assess the convergence of the NETI simulations accurately. However, conventional convergence diagnostics for MCMC, like the Gelman-Rubin potential scale reduction factor, are not applicable here.  The reason is that the combination of the NETI scheme, described in Section~\ref{sec:NETI}, and the new thermodynamic integration path, described in Section~\ref{sec:TInew}, continuously transform one model into another via a series of non-equilibrium configurations. We need to point out that any samples taken during this transformation are of no interest in themselves; the only quantity of interest is the log Bayes factor, computed according to \eqname~(\ref{eq:ioio}). The estimate of the log Bayes factor from \eqname~(\ref{eq:ioio}) is a random variable that is subject to the intrinsic stochasticity of the MCMC sampler. A natural convergence diagnostic is the variance of this estimator: for an infinite simulation time, the variance should go to zero as the estimate should not depend on the particular idiosyncrasies of any MCMC trajectory.  We have investigated this conjecture in Figures~\ref{fig:VARIANCES_1}-\ref{fig:VARIANCES_2}, \ref{fig:netidiff_pow_sigm_biasvar_c}-\subref{fig:netidiff_pow_sigm_biasvar_d} and \ref{fig:netidiff_ti_seq_meanvar_a}. Since Figures~\ref{fig:netidiff_pow_sigm_biasvar_c}-\subref{fig:netidiff_pow_sigm_biasvar_d} and \ref{fig:netidiff_ti_seq_meanvar_a} provide average variances over five independent data instantiations, we have included Figure~\ref{fig:NETIDIFF_convergence} that shows the individual variances for each data set separately.  All these figures demonstrate that the variance approaches zero as the simulation time, regarding the number of MCMC steps, is increased. Figure~\ref{fig:VARIANCES_1} quantifies the improvement in convergence that the proposed method achieves over the established schemes, in the form of a faster decrease of the variance with increasing simulation times. 

The figures mentioned above, e.g. Figures~\ref{fig:VARIANCES_1} and \ref{fig:netidiff_ti_seq_meanvar_a},  monitor convergence in terms of iteration numbers. For a fair comparison between different methods, we also need to take into consideration the computational costs per iteration shown in Table~\ref{tab:runtime}: The computational run times of the three algorithms compared are approximately equal; if there is any difference at all, it appears to be in favour of the proposed NETI scheme. From this, we can conclude that monitoring inference uncertainty as a function of MCMC iteration numbers, as carried out throughout our paper, provides an appropriate quantification of computational complexity.
% ==================
\end{document}